\documentclass[10pt,twocolumn,english,amsmath,amssymb,aps,prl]{revtex4-1}
\usepackage[T1]{fontenc}
\usepackage[latin9]{inputenc}
\usepackage{color}
\usepackage{babel}
\usepackage{amsmath}
\usepackage{amssymb}
\usepackage{graphicx}
\usepackage[unicode=true,pdfusetitle,
 bookmarks=true,bookmarksnumbered=false,bookmarksopen=false,
 breaklinks=false,pdfborder={0 0 1},backref=false,colorlinks=true]
 {hyperref}
\begin{document}
\title{Strain-tuned quantum criticality in electronic Potts-nematic systems}
\author{Anzumaan R. Chakraborty}
\author{Rafael M. Fernandes}
\affiliation{School of Physics and Astronomy, University of Minnesota, Minneapolis,
Minnesota 55455, USA}
\begin{abstract}
Motivated by recent observations of threefold rotational symmetry
breaking in twisted moiré systems, cold-atom optical lattices, quantum
Hall systems, and triangular antiferromagnets, we phenomenologically
investigate the strain-temperature phase diagram of the electronic
3-state Potts-nematic order. While in the absence of strain the quantum
Potts-nematic transition is first-order, quantum critical points (QCP)
emerge when uniaxial strain is applied, whose nature depends on whether
the strain is compressive or tensile. In one case, the nematic amplitude
jumps between two non-zero values while the nematic director remains
pinned, leading to a symmetry-preserving meta-nematic transition that
terminates at a quantum critical end-point. For the other type of
strain, the nematic director unlocks from the strain direction and
spontaneously breaks an in-plane twofold rotational symmetry, which
in twisted moiré superlattices triggers an electric polarization.
Such a piezoelectric transition changes from first to second-order
upon increasing strain, resulting in a quantum tricritical point.
Using a Hertz-Millis approach, we show that these QCPs share interesting
similarities with the widely studied Ising-nematic QCP. The existence
of three minima in the nematic action also leaves fingerprints in
the strain-nematic hysteresis curves, which display multiple loops.
At non-zero temperatures, because the upper critical dimension of
the 3-state Potts model is smaller than three, the Potts-nematic transition
is expected to remain first-order in 3D, but to change to second-order
in 2D. As a result, the 2D strain-temperature phase diagram displays
two first-order transition wings bounded by lines of critical end-points
or tricritical points, reminiscent of the phase diagram of metallic
ferromagnets. We discuss how our results can be used to unambiguously
identify spontaneous Potts-nematic order.
\end{abstract}
\date{\today}
\maketitle

\section{Introduction}

Electronic nematicity, which consists of the electronically-driven
breaking of the discrete rotational symmetry of a system \citep{Kivelson1998},
has been observed in various correlated electronic materials, including
three families of unconventional superconductors: cuprates \citep{Kivelson2003,Hinkov2008,Vojta2009},
heavy-fermion compounds \citep{Okazaki2011,Ronning2017,Seo2020},
and iron-based materials \citep{Chu2012,Fernandes2014,Bohmer2016,Bohmer2022}.
In all those cases, the underlying tetragonal lattice renders the
electronic nematic order parameter Ising-like \citep{Fradkin2010},
as the system must select between two nearest-neighbor (or next-nearest-neighbor)
bonds of the square lattice, which are related by a $90^{\circ}$
rotation. The selected bond will either expand or contract, since
nematic order necessarily triggers a lattice distortion \citep{Fernandes2014}.
Conversely, application of uniaxial strain along one of the bond directions
completely lifts the degeneracy between the two bonds, leading to
a smearing of the nematic phase transition. The situation is analogous
to the case of an Ising ferromagnet in the presence of a longitudinal
magnetic field, since strain acts as a conjugate field to the nematic
order parameter. Due to the ubiquituous presence of residual and random
strain in crystals \citep{Carlson2006,Carlson2011,Meese2022}, this
property can make it experimentally challenging to distinguish whether
an anisotropic property is due to spontaneous nematic order, nematic
order induced by strain (perhaps associated with an enhanced nematic
susceptibility), or simply strain \citep{XiaoyuWang2022}. More broadly,
the intrinsic coupling between electronic nematicity and uniaxial
strain gives rise to a rich phenomenology \citep{Schmalian2016,Paul2017,Carvalho2019,Massat2022}.

\begin{figure}[h]
\centering{}\includegraphics[width=1\columnwidth]{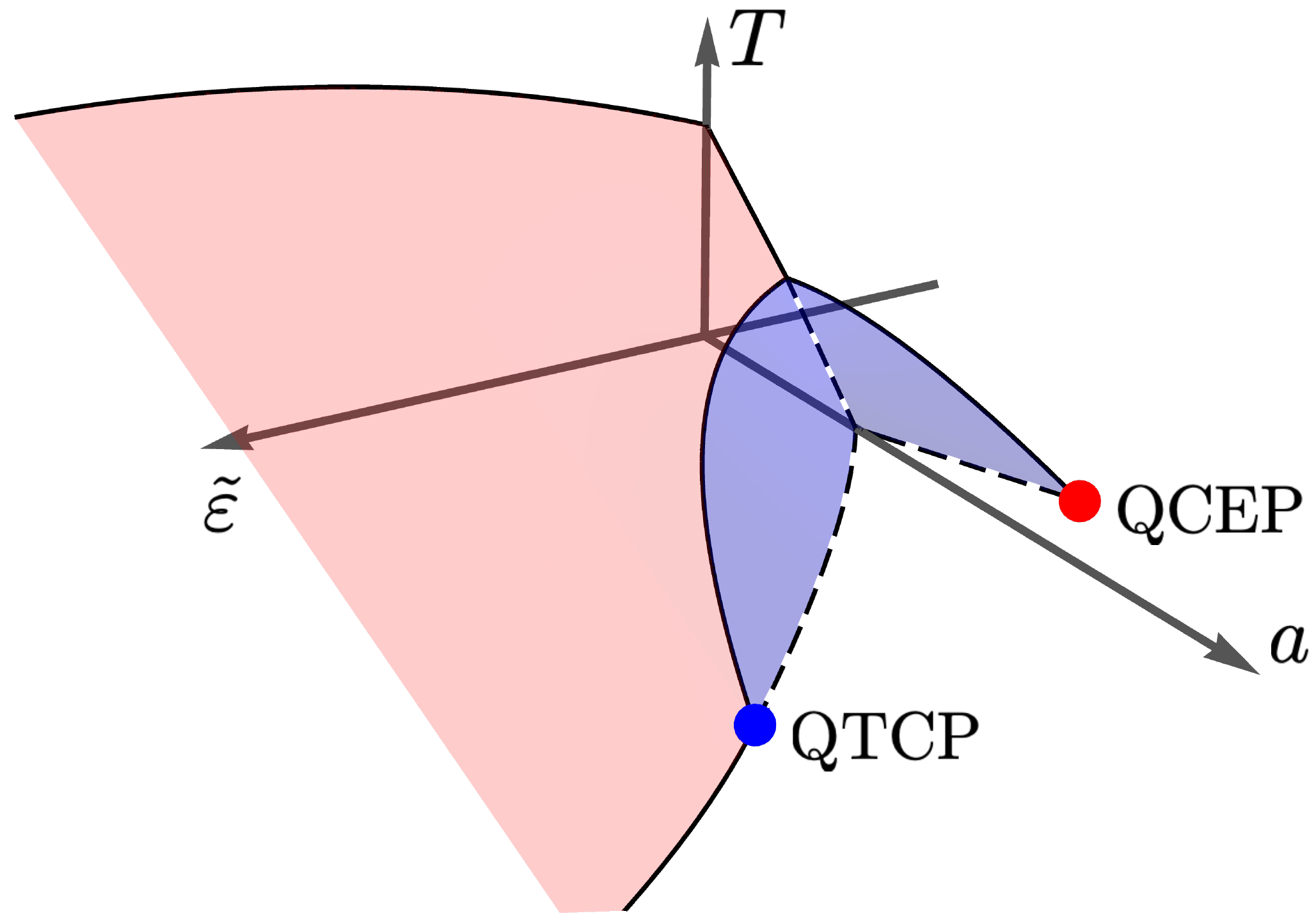}\caption{{\footnotesize{}Qualitative $(\tilde{\varepsilon},a,T)$ phase diagram
of a 2D Potts-nematic system, displaying first-order transition wings
(blue surfaces). Here, $a$ is a non-thermal tuning parameter like
doping or pressure; $\tilde{\varepsilon}$ is linearly proportional
to the uniaxial strain applied along one of the high-symmetry directions
of the threefold rotationally-symmetric lattice, but its sign depends
on Landau coefficients of the Potts-nematic action. For $\tilde{\varepsilon}<0$,
the isolated wing of first-order transitions is bounded by a line
of classical end-points that terminates at a quantum critical end-point
(QCEP). No symmetries are broken across this meta-nematic transition.
For $\tilde{\varepsilon}>0$, the wing is bounded by a line of classical
tricritical points terminating at a quantum tricritical point (QTCP),
and is thus surrounded by a surface of continuous transitions (red
surface). The in-plane two-fold rotational symmetry is broken spontaneously
across these transitions, giving rise to a piezoelectric phase in
twisted moiré systems. \label{fig_phase_diagram_T}}}
\end{figure}

Recently, electronic nematic order has also been observed in systems
whose underlying lattices have threefold rotational symmetry (i.e.
triangular, honeycomb, and kagome), such as the hexagonal (111) surface
of bismuth subjected to large magnetic fields \citep{Feldman2016},
the trigonal lattice of the doped topological insulator Bi$_{2}$Se$_{3}$
\citep{Tamegai2019,Cho2020}, the triangular antiferromagnet Fe$_{1/3}$NbS$_{2}$
\citep{Little2020}, a triangular optical lattice of cold $^{87}$Rb
atoms \citep{Jin2021}, and the trangular moiré superlattices of twisted
bilayer graphene (TBG) \citep{Kerelsky2019,Jiang2019,Choi2019,Cao2021},
twisted double-bilayer graphene (TDBG) \citep{Rubio2022}, twisted
trilayer graphene \citep{Zhang_Vafek2022}, and heterobilayer transition
metal dichalcogenides \citep{Jin2021_stripe}. More broadly, Potts-nematicity
has been proposed to emerge in diverse settings, from frustrated magnets
\citep{Mulder2010,Orth2022,Li2022,Nedic2022,Strockoz2022} to interacting
moiré systems \citep{Dodaro2018,Venderbos2018,Kozii2019,Xu2020,Fernandes_Venderbos,Kang2020,Bernevig_TBGVI,Chichinadze2020,Nori2020,Wang_Kang2021,Kontani2022,Savary2022,Matty2022}
and kagome metals \citep{Grandi2023}. In contrast to the case of
lattices with fourfold rotational symmetry, the nematic order parameter
here has a 3-state Potts character \citep{Hecker2018,Fernandes2019},
corresponding to selecting one among three nearest-neighbor bonds
related by a $120^{\circ}$ (or $60^{\circ}$) rotation. The linear
coupling between such a Potts-nematic order parameter and in-plane
strain has been recently explored in different contexts \citep{Hecker2018,How2019,Kuntsevich2019,Fernandes_Venderbos,Kostylev2020,Little2020,Cao2021,Kimura2022,Hecker2022}.
An interesting result is that application of uniaxial strain along
one of the bond directions may not fully lift the degeneracy between
the three bonds. Consequently, unlike the case of a tetragonal lattice,
a nematic-related transition -- dubbed nematic-flop transition in
Ref. \citep{Fernandes_Venderbos}-- can take place in a triangular
lattice even in the presence of uniaxial strain. The situation is
analogous to a 3-state Potts ferromagnet in the presence of an external
magnetic field that points along one of the three allowed magnetic
moment directions \citep{Straley1973ThreestatePM,blankschtein1980effects}.
If a ``positive'' field is applied, i.e. a field that favors one
of the moment directions, no additional symmetries can be spontaneously
broken. However, if a ``negative'' field is applied, i.e. a field
that penalizes one of the moment directions, there is a residual Ising
symmetry associated with the two remaining moment directions. Such
a symmetry is spontaneously broken in the vicinity of the zero-field
ferromagnetic transition.

Another peculiarity of the 3-state Potts model is that its upper critical
dimension is $d_{u}^{\mathrm{Potts}}\lesssim3$ (for a review, see
Ref. \citep{review_Potts}), whereas in the Ising model, $d_{u}^{\mathrm{Ising}}=4$.
Most importantly, the character of the 3-state Potts transition is
fundamentally different for dimensions above and below $d_{u}^{\mathrm{Potts}}$.
For $d\geq3$, a mean-field description works and the transition is
first-order, due to the existence of a cubic invariant in the Landau
free-energy expansion. However, for $d=2$, the 3-state Potts transition
is second-order. This has important consequences for two-dimensional
systems subjected to a 3-state Potts nematic instability, such as
twisted moiré systems. At high enough temperatures, $d=2$ and one
expects a second-order nematic transition. However, at $T=0$, since
$d+z>d_{u}^{\mathrm{Potts}}$ for the expected values of the dynamic
critical exponent $z$ (i.e. $z=1$ for an insulator and $z=3$ for
a metal \citep{Lohneysen2007}), the nematic transition should be
first-order. This not only implies the absence of a Potts-nematic
quantum critical point (QCP), but it also indicates that, as the nematic
transition temperature is suppressed by a non-thermal tuning parameter,
a tricritical point should emerge.

In this paper, we use a phenomenological model to study the Potts-nematic
phase diagram in the presence of uniaxial strain. The nematic order
parameter is parametrized as a two-component ``vector'' $\boldsymbol{\Phi}=\phi\left(\cos2\theta,\,\sin2\theta\right)$,
where $\phi$ is the magnitude and the director angle $\theta$ is
restricted to $3$ possible values. The tuning parameters are the
temperature $T$ and a non-thermal control parameter $a$, such as
doping, which suppresses the Potts-nematic transition temperature
to zero. Fig. \ref{fig_phase_diagram_T} summarizes our main findings
for a 2D system whose underlying lattice has threefold rotational
symmetry. At $T=0$, since the system is above the 3-state-Potts upper
critical dimension, it undergoes a first-order quantum nematic phase
transition upon changing the non-thermal parameter $a$, where the
threefold rotational symmetry $C_{3z}$ is broken. The fate of the
transition upon application of uniaxial strain along one of the nematic-bond
directions depends on the sign of $\tilde{\varepsilon}$, which is
linearly proportional to the strain $\varepsilon$, which in turn
can be either compressive ($\varepsilon<0$) or tensile ($\varepsilon>0$).

For $\tilde{\varepsilon}<0$, upon increasing the strain magnitude,
a first-order transition line transition extends to larger $a$ values,
ending at a quantum critical end-point (QCEP), analogously to the
case of the liquid-gas transition of water. The magnitude of the nematic
order parameter $\phi$ jumps across the first-order transition line,
whereas the nematic director angle $\theta$ remains pinned by the
strain direction, signaling a symmetry-preserving \emph{quantum meta-nematic
transition}. Beyond the QCEP, there is only a crossover signaled by
the Widom line.

For $\tilde{\varepsilon}>0$, while a first-order transition line
extending to larger values of $a$ also appears upon increasing the
strain magnitude, the situation is completely different. The first
key difference is that the director angle $\theta$ spontaneously
unpins from the strain direction across the transition, selecting
one among two possible angles, which in turn are related by twofold
rotations with respect to in-plane axes (denoted by $C'_{2}$). Therefore,
across this first-order Ising transition, the $C'_{2}$ symmetry is
broken, resulting in the emergence of an out-of-plane ferroelectric
polarization in the case of twisted moiré systems. Because this ferroelectricity
only appears in the presence of strain of a particular type (compressive
or tensile), we dub this a \emph{quantum piezoelectric transition}.
The second key difference with respect to the case of $\tilde{\varepsilon}<0$
is that, upon applying stronger strain, the first-order transition
line ends at a quantum tricritical point (QTCP), beyond which a line
of piezoelectric QCPs emerges. A Hertz-Millis type of calculation
for both the piezoelectric QCPs and the meta-nematic QCEP in the case
of metallic systems reveals that they behave very similarly to an
Ising-nematic QCP \citep{Oganesyan2001,Metzner2003,Garst2010,Metlitski2010,Schattner2016,Lederer2017,Klein_Chubukov2018},
not only possessing the same dynamical critical exponent $z=3$, but
also cold spots at the Fermi surface.

We also compute the upper and lower spinodal lines associated with
the different first-order transition lines and employ a generalized
Stoner-Wohlfarth approach \citep{Stoner1948} to show that the asymmetry
between the effects of compressive and tensile strains is manifested
in the hysteresis curves of $\boldsymbol{\Phi}\left(\varepsilon\right)$.
In particular, because there are three action minima available, rather
than the usual two, the hysteresis curves can show multiple loops
depending on the initial conditions. These characteristic features
of the hysteresis curves provide concrete criteria to unambigously
determine experimentally whether a twofold anisotropic signal observed
in a system with threefold rotational symmetry is due to spontaneous
nematic order or induced nematic order by strain.

The extension of the results to non-zero temperatures depends on the
dimensionality of the system. For $d=3$, the system is always above
the 3-state-Potts upper critical dimension, implying that the phase
diagram at non-zero temperatures should be similar to that at $T=0$.
However, for $d=2$, the Potts-nematic transition should generally
become second-order at high enough temperatures, thus displaying the
typical critical exponents of the 2D 3-state Potts model \citep{review_Potts}.
Consequently, a tricritical point at $T\neq0$ should exist for unstrained
systems, connecting the first-order quantum phase transition to the
second-order transition at high $T$. As illustrated in Fig. \ref{fig_phase_diagram_T},
this tricritical point is expected to directly connect to the QCEP
and the QCTP, giving rise to two first-order transition wings. This
shape of the phase diagram resembles that of an itinerant ferromagnet
\citep{Belitz2005,Brando2016}, although the mechanisms by which the
$T=0$ transition becomes first-order are unrelated \citep{Belitz1999,Chubukov2004,Maslov2009}.
An important difference is that, in the Potts-nematic case, the first-order
transition wing on the $\tilde{\varepsilon}<0$ side is isolated,
bounded by the line of critical endpoints, whereas the wing on the
$\tilde{\varepsilon}>0$ side is bounded by a line of tricritical
points, and thus exists inside a much broader wing signaling the second-order
transition to the piezoelectric phase. 

The paper is organized as follows: in Sec. \ref{sec:Zero-temperature-phase-diagram}
we apply mean-field theory to determine the $T=0$ phase diagram of
the Potts-nematic model, focusing on the emergence of the meta-nematic
QCEP and the piezoelectric QTCP in Sec. \ref{subsec:metanematic}
and Sec. \ref{subsec:piezoelectric}, respectively. The Potts-nematic
hysteresis curves are analyzed in Sec. \ref{sec:Hysteresis-and-spinodal},
whereas Sec. \ref{sec:Non-zero-temperature-phase} presents a qualitative
analysis of the $T>0$ phase diagram. Conclusions are presented in
Sec. \ref{sec:Conclusions}.

\section{Zero-temperature phase diagram \label{sec:Zero-temperature-phase-diagram}}

\subsection{Mean-field solution of the Potts-nematic model}

The ``in-plane'' nematic order parameter can always be parametrized
as $\boldsymbol{\Phi}\equiv\left(\phi_{1},\,\phi_{2}\right)=\phi\left(\cos2\theta,\,\sin2\theta\right)$,
where $\phi>0$ is the magnitude and $0\leq\theta\leq\pi$ is the
nematic director angle \citep{Fradkin2010,Fernandes_Venderbos}. By
construction, the order parameter satisfies $\boldsymbol{\Phi}(\theta)=\boldsymbol{\Phi}(\theta+\pi)$,
as is the case for the classical nematic order parameter. Physically,
the components $\phi\cos2\theta$ and $\phi\sin2\theta$ transform
as the expectation values of the electronic quadrupolar moments $\rho_{x^{2}-y^{2}}$
and $\rho_{xy}$, as well as the strain components $\varepsilon_{xx}-\varepsilon_{xy}$
and $\varepsilon_{xy}$. Here, $\varepsilon_{ij}=\left(\partial_{i}u_{j}+\partial_{j}u_{i}\right)/2$
is the strain tensor and $\boldsymbol{u}$, the displacement vector.
The allowed values of $\theta$ are constrained by the symmetries
of underlying crystal lattice, as well as by the presense of in-plane
uniaxial strain. Hereafter, we focus on lattices that are invariant
under threefold rotations with respect to the $z$ axis ($C_{3z}$
operation) and twofold rotations with respect to at least one in-plane
axis ($C'_{2}$ operation). These include the triangular lattice as
well as any other lattice with point groups $\mathrm{D}_{\mathrm{6h}}$,
$\mathrm{D}_{\mathrm{3h}}$, $\mathrm{D_{3d}}$, $\mathrm{D}_{\mathrm{6}}$,
and $\mathrm{D_{3}}$. Note that the latter two describe certain twisted
moiré superlattices, like TBG and TDBG. We assume that external uniaxial
strain is applied along a direction that makes an angle $\alpha$
with respect to the $x$ axis, and consider both tensile ($\varepsilon>0$)
and compressive ($\varepsilon<0$) strain. In this case, the nematic
action is given by \citep{Hecker2018,Fernandes_Venderbos,Xu2020,Cao2021}:
\begin{multline}
\mathcal{S}_{\text{nem}}\left[\boldsymbol{\Phi}\left(q\right)\right]=\frac{1}{2}\int_{q}\phi_{-q}\chi_{0}^{-1}\left(q\right)\phi_{q}\\
+\int_{x}\left[\frac{\lambda}{3}\,\phi^{3}\cos6\theta+\frac{u}{4}\,\phi^{4}-\gamma\varepsilon\phi\cos\left(2\theta-2\alpha\right)\right]\label{eq:S_nem}
\end{multline}

Here, $x=(\tau,\mathbf{x})$ denotes imaginary time $\tau\in[0,\beta]$
and spatial variable $\mathbf{x}$, whereas $q=(\Omega_{n},\boldsymbol{q})$
consists of bosonic Matsubara frequencies $\Omega_{n}=2\pi nT$ and
momentum $\mathbf{q}$. The inverse nematic susceptibility is given
by $\chi_{0}^{-1}(q)=a+\boldsymbol{q}^{2}+\Omega_{n}^{2}$, where
$a$ is a non-thermal tuning parameter. The coupling constants $\lambda$
and $u>0$ describe the non-harmonic terms of the action, whereas
$\gamma$ is the elasto-nematic coupling. 

For $\varepsilon=0$, the action (\ref{eq:S_nem}) maps onto the 3-state
Potts model -- or, equivalently, the $3$-state clock model. Indeed,
for $\lambda>0$, the nematic director is pinned to the three high-symmetry
directions $\theta=(2n+1)\pi/6$, whereas for $\lambda<0$, the three
allowed values are $\theta=2n\pi/6$, with $n=0,1,2$. In the presence
of strain, there are important changes in the problem. In this paper,
we consider strain applied along one of the high-symmetry directions
of the lattice. In this case, we can set without loss of generality
$\alpha=2m\pi/6$, with $m=0,1,2$, since the action (\ref{eq:S_nem})
is invariant under $\varepsilon\rightarrow-\varepsilon$ and $\alpha\rightarrow\alpha+\pi/2$.
This invariance reflects the fact that, for the nematic order parameter,
compressive strain applied along one axis has the same effect as tensile
strain applied along an orthogonal axis. Shifting the director angle
such that it is measured with respect to the strain direction, $\tilde{\theta}=\theta-\alpha$,
and considering the case of static and homogeneous fields, the action
``density'' becomes:

\begin{equation}
S_{\text{nem}}\left(\boldsymbol{\Phi}\right)=\frac{a}{2}\,\phi^{2}+\frac{u}{4}\,\phi^{4}+\frac{\lambda}{3}\,\phi^{3}\cos6\tilde{\theta}-\gamma\varepsilon\phi\cos2\tilde{\theta}
\end{equation}

Upon defining the rescaled quantities $\tilde{\phi}\equiv\frac{u}{\left|\lambda\right|}\,\phi$,
$\tilde{S}_{\text{nem}}\equiv\frac{u^{3}}{\lambda^{4}}\,S_{\text{nem}}$,
$\tilde{a}=\frac{u}{\lambda^{2}}\,a$, and $\tilde{\varepsilon}\equiv\frac{\gamma u^{2}}{\lambda^{3}}\,\varepsilon$,
we can rewrite the action in a more convenient form:

\begin{equation}
\tilde{S}_{\text{nem}}=\frac{\tilde{a}}{2}\,\tilde{\phi}^{2}+\frac{1}{4}\,\tilde{\phi}^{4}+\mathrm{sgn}\lambda\left(\frac{1}{3}\,\tilde{\phi}^{3}\cos6\tilde{\theta}-\tilde{\varepsilon}\tilde{\phi}\cos2\tilde{\theta}\right)\label{eq:Snem_tilde}
\end{equation}

Regardless of whether the system is 2D or 3D, at $T=0$ the effective
dimensionality $d+z\geq3$, implying that the system is above the
upper critical dimension of the 3-state Potts model. As a result,
a mean-field solution is appropriate; setting $\partial\tilde{S}_{\text{nem}}/\partial\tilde{\phi}=0$
and $\partial\tilde{S}_{\text{nem}}/\partial\tilde{\theta}=0$, and
assuming $\tilde{\phi}\neq0$, we find the mean-field equations:

\begin{align}
\tilde{a}\tilde{\phi}+\mathrm{sgn}\lambda\,\tilde{\phi}^{2}\cos6\tilde{\theta}+\tilde{\phi}^{3} & =\mathrm{sgn}\lambda\,\tilde{\varepsilon}\cos2\tilde{\theta}\label{eq:meanfield_1}\\
\sin6\tilde{\theta} & =\frac{\tilde{\varepsilon}}{\tilde{\phi}^{2}}\sin2\tilde{\theta}\label{eq:meanfield_2}
\end{align}

To make the notation less cumbersome, hereafter we drop the tilde
of all quantities except for $\tilde{\varepsilon}$; the latter is
to emphasize that the relevant quantity is the combination $\tilde{\varepsilon}=\left(\gamma u^{2}/\lambda^{3}\right)\varepsilon$,
whose overall sign depends not only on whether the applied strain
$\varepsilon$ is compressive or tensile, but also on the signs of
the nemato-elastic coupling $\gamma$ and on the cubic Landau coefficient
$\lambda$. We first review the well-known results in the case of
no applied strain, $\tilde{\varepsilon}=0$ \citep{Hecker2018,Fernandes_Venderbos}.
Eq. (\ref{eq:meanfield_2}) gives the extrema $\theta_{0}=p\pi/6$,
with $p=0,\cdots,5$. Computing the second derivative of the action
at the extrema, we find $\left(\frac{\partial^{2}S_{\text{nem}}}{\partial\theta^{2}}\right)_{\theta_{0}}=-12\phi^{3}(-1)^{p}\mathrm{sgn}\lambda$.
Therefore, the minima (maxima) of the action are given by $\theta_{0}=p\pi/6$
with even (odd) $p$ if $\lambda<0$ and odd (even) $p$ if $\lambda>0$.
Meanwhile, Eq. (\ref{eq:meanfield_1}) becomes:

\begin{equation}
a-\phi_{0}+\phi_{0}^{2}=0
\end{equation}
which gives:

\begin{equation}
\phi_{0,\pm}=\frac{1}{2}\left(1\pm\sqrt{1-4a}\right)
\end{equation}

Clearly, a $\phi_{0}\neq0$ solution can only exist if $a<a_{\mathrm{us}}\equiv1/4$,
which sets the upper spinodal of the first-order Potts-nematic transition.
The first-order transition takes place for $a\equiv a_{c,0}$ such
that $S_{\text{nem}}\left(\phi_{0,+},\theta_{0}\right)=0$, which
gives $a_{c,0}=2/9$. The jump in the nematic order parameter at the
transition is thus given by $\Delta\phi_{0}=1/3$. 

The mean-field solution for non-zero strain has been in part discussed
in Refs. \citep{Fernandes2019,Cao2021} and, more broadly, in the
literature of the 3-state Potts model under the presence of a magnetic
field \citep{Straley1973ThreestatePM,blankschtein1980effects,review_Potts}.
The second mean-field equation (\ref{eq:meanfield_2}) can be rewritten
as (where, we recall, the director angle $\theta$ is measured with
respect to the direction strain is applied):

\begin{equation}
\sin2\theta\left[\cos^{2}2\theta-\frac{1}{4}\left(1+\frac{\tilde{\varepsilon}}{\phi^{2}}\right)\right]=0\label{eq:aux_meanfield2}
\end{equation}

\begin{figure*}[t]
\begin{centering}
\includegraphics[width=0.6\paperwidth]{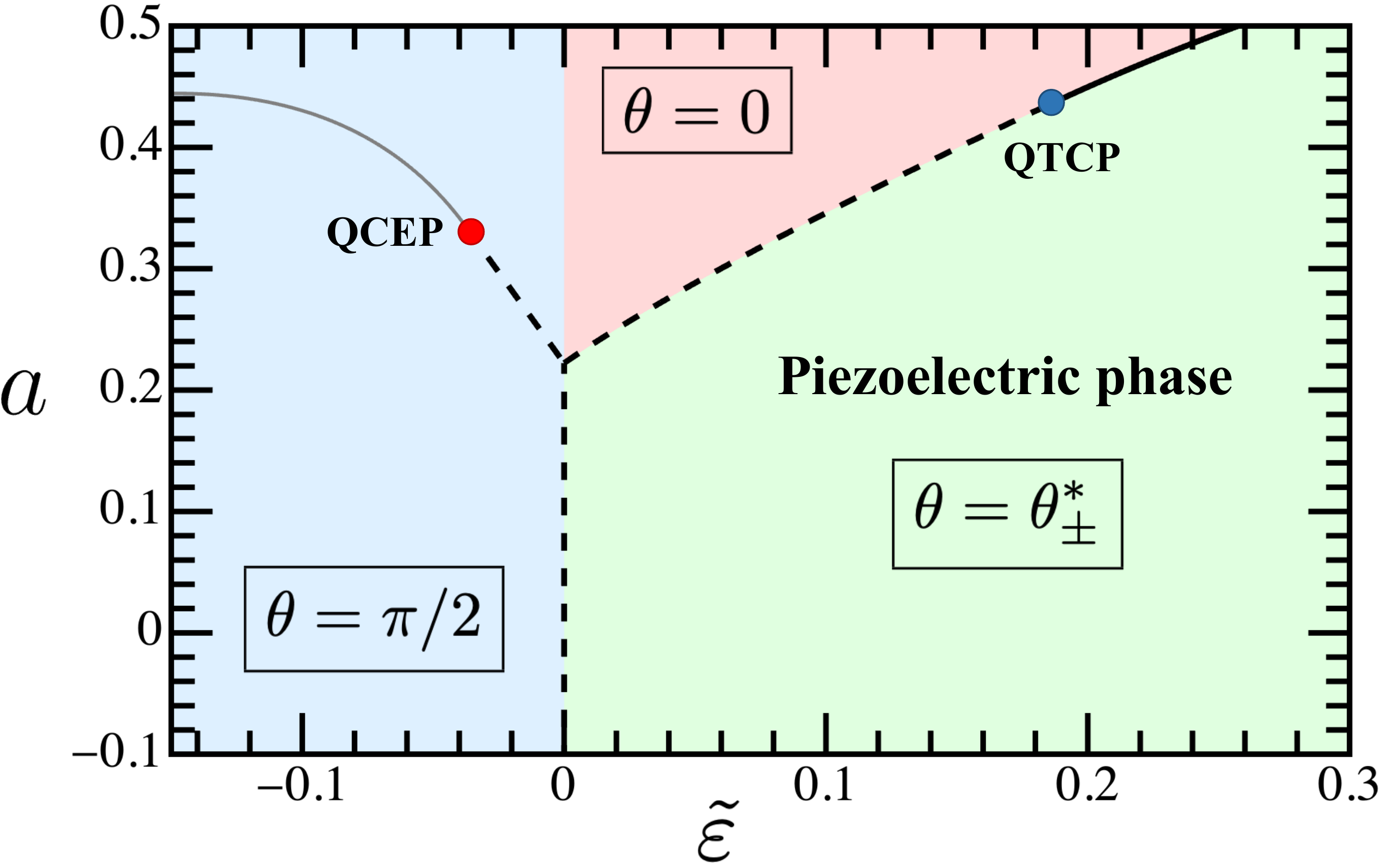}
\par\end{centering}
\caption{{\small{}Zero-temperature $(\tilde{\varepsilon},a)$ phase diagram
of a Potts-nematic system. The red and blue regions correspond to
the phases in which the director angle $\theta$ is fixed to the high-symmetry
directions $\theta_{+}=0$ and $\theta_{-}=\pi/2$, respectively,
whereas the green region corresponds to the piezoelectric phase, in
which $\theta=\theta_{\pm}^{*}$ is not fixed {[}see Eq. (\ref{eq:theta_star}){]}.
For $\tilde{\varepsilon}<0$, there is a first-order meta-nematic
transition line (dashed black) that ends at a QCEP. Beyond the QCEP,
there is a crossover signaled by the Widom line (solid gray). For
$\tilde{\varepsilon}>0$, the first-order piezoelectric transition,
which spontaneously breaks in-plane two-fold rotational symmetry,
ends at a QTCP, after which it becomes second-order (solid black line).\label{fig:phase_diagram_T=00003D0}}}
\end{figure*}

This equation always admits two solutions: $\theta_{+}=0$, corresponding
to a nematic director parallel to the strain direction, and $\theta_{-}=\pi/2$,
denoting a nematic director perpendicular to the strain direction.
Note that, by definition, $\mathbf{\Phi}\rightarrow-\mathbf{\Phi}$
upon a rotation of $90^{\circ}$ of the director angle $\theta$.
In both cases, the mean-field equation (\ref{eq:meanfield_1}) that
determines the $\phi_{\pm}$ values corresponding to $\theta_{\pm}$
is:

\begin{equation}
a\phi_{\pm}+\phi_{\pm}^{3}\pm\mathrm{sgn}\lambda\,\left(\phi_{\pm}^{2}-\tilde{\varepsilon}\right)=0\label{eq:thetaplus_1}
\end{equation}
whereas the action evaluated at these extrema is given by:

\begin{equation}
S_{\pm}=\frac{a}{2}\,\phi_{\pm}^{2}+\frac{1}{4}\,\phi_{\pm}^{4}\pm\mathrm{sgn}\lambda\left(\frac{1}{3}\,\phi_{\pm}^{3}-\tilde{\varepsilon}\phi_{\pm}\right)\label{eq:thetaplus_2}
\end{equation}

To check which of these solutions (if any) is a minimum of the action,
we evaluate the second derivative:

\begin{equation}
\left(\frac{\partial^{2}S_{\mathrm{nem}}}{\partial\theta^{2}}\right)_{\theta_{\pm}}=\pm4\mathrm{sgn}\lambda\,\phi_{\pm}\tilde{\varepsilon}\left(1-\frac{3\phi_{\pm}^{2}}{\tilde{\varepsilon}}\right)
\end{equation}

It follows that, when $\tilde{\varepsilon}<0$, the $\theta_{-}$
($\theta_{+}$) solution is always a local action minimum for $\lambda>0$
($\lambda<0$). Meanwhile, when $\tilde{\varepsilon}>0$, the situation
is more involved. Far enough from the Potts-nematic transition of
the unstrained system, where the nematic order parameter induced by
the strain is expected to be small, $\phi^{2}\ll\tilde{\varepsilon}$,
we find that the $\theta_{+}$ ($\theta_{-}$) solution is a local
minimum of the action for $\lambda>0$ ($\lambda<0$). Once the nematic
order parameter increases such that $\phi^{2}>\tilde{\varepsilon}/3$,
however, this solution switchs to a local maximum. This indicates
that another solution is available. Indeed, the mean-field equation
for $\theta$, Eq. (\ref{eq:aux_meanfield2}), admits two additional
solutions:

\begin{equation}
\theta_{\pm}^{*}=\pm\frac{1}{2}\arccos\left(\frac{1}{2}\sqrt{1+\frac{\tilde{\varepsilon}}{\phi^{2}}}\right)\label{eq:theta_star}
\end{equation}
provided that the argument is smaller than $1$, i.e. $\phi^{2}>\tilde{\varepsilon}/3$.
This is the same condition for which the $\theta_{+}$ ($\theta_{-}$)
solution becomes a local maximum of the action for $\lambda>0$ ($\lambda<0$).
Note that, in the director space spanned by the angle $\theta$, the
points $\theta_{+}^{*}$ and $\theta_{-}^{*}$ are related by a twofold
rotation with respect to the horizontal axis (as well as the vertical
axis), which indicates that selecting one of the two solutions will
break a spatial symmetry of the system. We will return to this point
later. Using Eq. (\ref{eq:theta_star}), it is straightforward to
obtain the mean-field equation for the corresponding nematic amplitude
$\phi_{\pm}^{*}$

\begin{equation}
a\phi_{\pm}^{*}+\left(\phi_{\pm}^{*}\right)^{3}\mp\mathrm{sgn}\lambda\,\phi_{\pm}^{*}\sqrt{\tilde{\varepsilon}+\left(\phi_{\pm}^{*}\right)^{2}}=0\label{eq:thetastar_1}
\end{equation}
as well as the values of the action evaluated at these solutions:

\begin{equation}
S_{\pm}^{*}=\frac{a}{2}\left(\phi_{\pm}^{*}\right)^{2}+\frac{1}{4}\left(\phi_{\pm}^{*}\right)^{4}\mp\frac{\mathrm{sgn}\lambda}{3}\left[\tilde{\varepsilon}+\left(\phi_{\pm}^{*}\right)^{2}\right]^{3/2}\label{eq:thetastar_2}
\end{equation}

Eq. (\ref{eq:thetastar_1}) can be solved in a straightforward way:

\begin{equation}
\phi_{\pm}^{*}=\sqrt{\frac{1}{2}-a\pm\sqrt{\tilde{\varepsilon}-a+\frac{1}{4}}}\label{eq:phi_star}
\end{equation}

It turns out that the $\phi_{-}^{*}$ solution is either a saddle-point
of the action or does not satify the condition $\phi^{2}<\tilde{\varepsilon}/3$.
Consequently, $\phi_{+}^{*}$ is the desired solution, yielding:

\begin{equation}
\theta_{\pm}^{*}=\pm\frac{1}{2}\arccos\left(\frac{1}{2}\left[1+\frac{\tilde{\varepsilon}}{\frac{1}{2}-a+\sqrt{\tilde{\varepsilon}-a+\frac{1}{4}}}\right]^{1/2}\right)\label{eq:theta_star_final}
\end{equation}

We therefore obtain three different viable solutions for $\tilde{\varepsilon}\neq0$:
$\theta_{+}=0$, $\theta_{-}=\pi/2$, and $\theta_{\pm}^{*}$ given
by Eq. (\ref{eq:theta_star_final}). Following our analysis above,
either $\theta_{-}$ or $\theta_{+}$ is expected to be the global
minimum for $\tilde{\varepsilon}<0$, depending on whether $\lambda>0$
or $\lambda<0$, respectively. On the other hand, for $\tilde{\varepsilon}>0$,
two different minima are expected for distinct ranges of $a$: $\theta_{+}$
or $\theta_{-}$ (for $\lambda>0$ and $\lambda<0$, respectively)
and $\theta_{\pm}^{*}$. 

The full phase diagram can be directly obtained by comparing the actions
evaluated at the three solutions, Eqs. (\ref{eq:thetaplus_2}) and
(\ref{eq:thetastar_2}), after solving for the corresponding nematic
amplitude in Eqs. (\ref{eq:thetaplus_1}) and (\ref{eq:theta_star_final}).
The resulting phase diagram is shown in Fig. \ref{fig:phase_diagram_T=00003D0};
as anticipated, it is analogous to the phase diagram of the ferromagnetic
3-state Potts-model in the presence of a magnetic field \citep{Straley1973ThreestatePM,blankschtein1980effects}.
For concreteness, we consider the case in which $\lambda>0$. For
$\tilde{\varepsilon}<0$, we indeed find that the $\theta_{-}=\pi/2$
solution is the global minimum for any value of $a$. This does not
mean, however, that the system does not undergo a phase transition.
As denoted by the dashed line in Fig. \ref{fig:phase_diagram_T=00003D0},
for small enough $\left|\tilde{\varepsilon}\right|$ and close enough
to the nematic transition at zero strain, $a_{c}=2/9$, the system
undergoes a symmetry-preserving first-order transition in which the
nematic amplitude $\phi_{-}$ jumps, while the nematic director angle
$\theta$ remains fixed. This is expected, since the nematic order
parameter undergoes a first-order transition in the absence of strain
and $\tilde{\varepsilon}$ acts as a conjugate field to the nematic
order parameter. We dub this a meta-nematic quantum phase transition,
in analogy to the $T=0$ meta-magnetic transition that takes place
in a metallic ferromaget subjected to an external field. The first-order
line ends in a critical end-point, similarly to the liquid-gas transition
of water. This side of the phase diagram is further discussed in Sec.
\ref{subsec:metanematic}.

The $\tilde{\varepsilon}>0$ side of the phase diagram is qualitatively
different. As displayed in Fig. \ref{fig:phase_diagram_T=00003D0},
for $a\gg a_{c}$, the global minimum is at $\theta_{+}=0$. However,
upon approaching the nematic transition point of the unstrained system,
$a_{c,0}=2/9$, the nematic director angle that minimizes the action
switches to $\theta_{\pm}^{*}$. In contrast to the transition on
the $\tilde{\varepsilon}<0$ side of the phase diagram, this is not
a symmetry-preserving transition, since the spatial symmetry that
relates the two nematic director angles $\theta_{+}^{*}$ and $\theta_{-}^{*}$
is spontaneously broken. For small enough $\tilde{\varepsilon}$,
this transition is first-order whereas for large enough $\tilde{\varepsilon}$,
it becomes second-order. Therefore, there is a tricritical point,
marked in the figure, for intermediate values of $\tilde{\varepsilon}$.
We will analyze this side of the phase diagram in more detail in Sec.
\ref{subsec:piezoelectric}.

The change in the nematic director angle upon decreasing $a$ for
$\tilde{\varepsilon}>0$, discussed also in Ref. \citep{Fernandes_Venderbos},
can be understood directly from the action in Eq. (\ref{eq:Snem_tilde}).
For $\lambda>0$, the cubic term is minimized for $\theta=\pi/6,\,\pi/2,\,5\pi/6$
and maximized for $\theta=0,\,\pi/3,\,2\pi/3$. The linear term, on
the other hand, is minimized by $\theta=\pi/2$ and maximized by $\theta=0$,
for $\tilde{\varepsilon}<0$, and minimized by $\theta=0$ and maximized
by $\theta=\pi/2$ for $\tilde{\varepsilon}>0$. Therefore, in the
regime $\tilde{\varepsilon}<0$, both the linear and cubic terms can
be simultaneously minimized by the same nematic director angle, $\theta=\pi/2$.
In contrast, in the regime $\tilde{\varepsilon}>0$, the minimum of
the cubic term is the maximum of the linear term and vice versa. For
large enough values of $a$, where the amplitude of the nematic order
parameter is small, the linear term wins over the cubic one. Once
the system approaches its intrinsic nematic instability, the nematic
ampitude increases and the two terms eventually give comparable contributions
to the action. This frustration between the minima and maxima of the
cubic and linear terms is lifted by a compromise value for the nematic
director $\theta$. Indeed, Eq. (\ref{eq:theta_star}) for $\theta_{\pm}^{*}$
interpolates between $0$ when $\phi^{2}=\tilde{\varepsilon}/3$,
which mimizes the linear term, to $\pi/6$ and $5\pi/6$ when $\phi^{2}\gg\tilde{\varepsilon}/3$,
which mimizes the cubic term.

\subsection{Meta-nematic quantum critical endpoint \label{subsec:metanematic}}

\begin{figure}[h]
\centering{}\includegraphics[width=1\columnwidth]{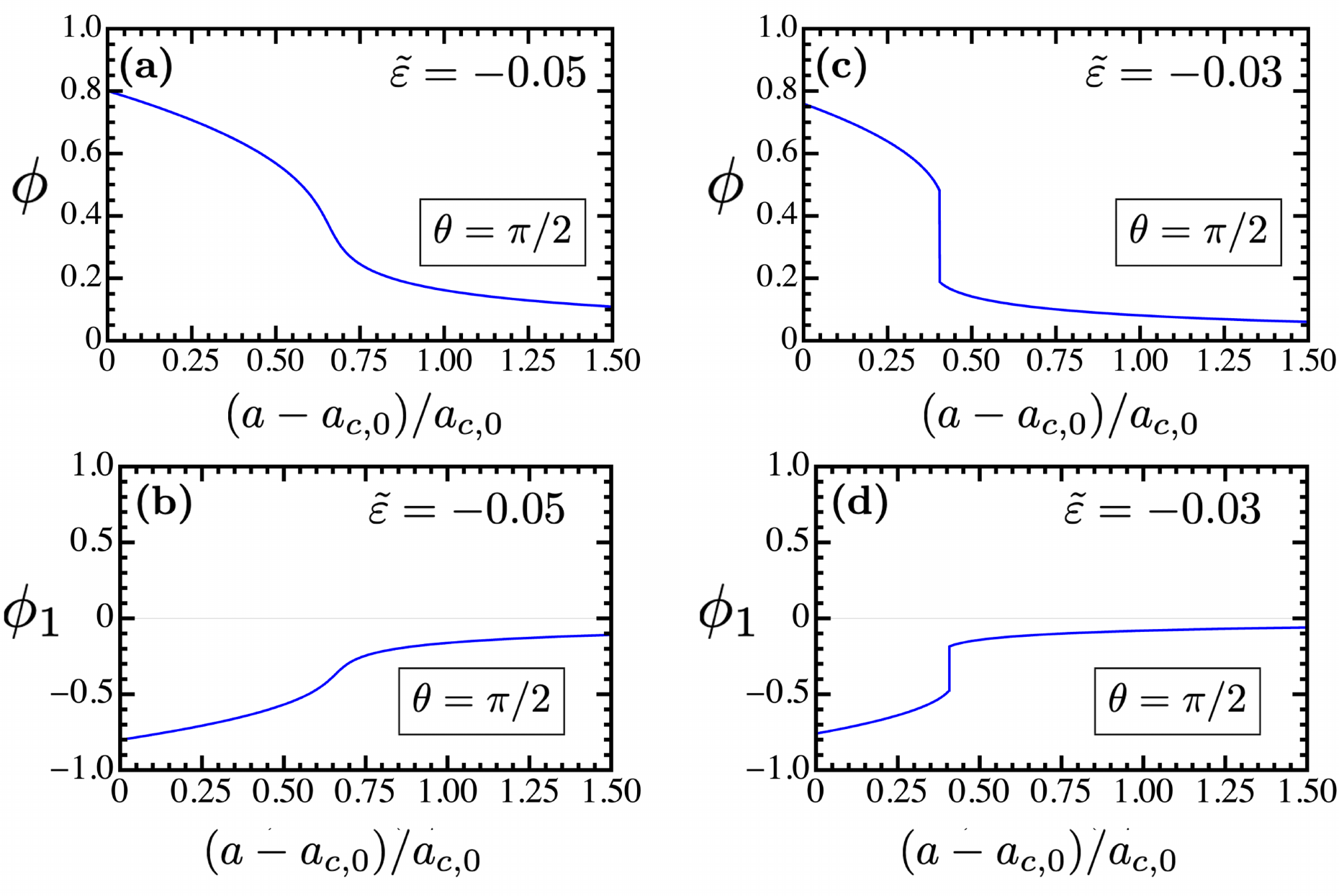}\caption{{\small{}Behavior of the nematic order parameter for fixed $\tilde{\varepsilon}<0$
across the crossover Widom line {[}panels (a) and (b){]} and the meta-nematic
transition line {[}panels (c) and (d){]}, as defined in the phase
diagram of Fig. \ref{fig:phase_diagram_T=00003D0}. Representative
values of $\tilde{\varepsilon}$ are chosen from the left and right
side of $\tilde{\varepsilon}_{\text{QCEP}}=-1/27\approx-0.037$, respectively.
The top panels show the magnitude of the nematic order parameter $\phi$
and the bottom panels, the nematic component projected along the strain
direction, $\phi_{1}=\phi\cos2\theta$; in all cases, the nematic
director angle remains fixed at $\pi/2$. The horizontal axis corresponds
to the non-thermal tuning parameter $a$ measured relative to unstrained
Potts-nematic transition point $a_{c,0}=2/9$. \label{fig:phi_metanematic}}}
\end{figure}

To gain further insight into the $\tilde{\varepsilon}<0$ region of
the phase diagram, we substitute the value of the nematic director
angle that minimizes the action, $\theta_{-}=\pi/2$, in Eq. (\ref{eq:Snem_tilde})
-- recall that we are considering $\lambda>0$. We then obtain an
action that depends only on the magnitude $\phi$:

\begin{equation}
S_{\text{nem}}^{(\tilde{\varepsilon}<0)}\left(\phi\right)=\frac{a}{2}\,\phi^{2}+\frac{1}{4}\,\phi^{4}-\frac{1}{3}\,\phi^{3}+\tilde{\varepsilon}\phi
\end{equation}

To proceed, we recall that, for zero strain, $\tilde{\varepsilon}=0$,
the system undergoes a first-order transition at $a_{c,0}=2/9$ in
which the nematic order parameter jumps by $\Delta\phi_{0}=1/3$.
Therefore, it is convenient to introduce the shifted nematic order
parameter $\delta\phi\equiv\phi-\Delta\phi_{0}$, as it effectively
removes the cubic term above. We find: 
\begin{equation}
S_{\text{nem}}^{(\tilde{\varepsilon}<0)}\left(\delta\phi\right)=\frac{A}{2}(\delta\phi)^{2}+\frac{1}{4}(\delta\phi)^{4}-H\,\delta\phi\label{eq:S_nem_negative}
\end{equation}
where we dropped a constant term and defined:

\begin{align}
A & \equiv a-\frac{1}{3}\\
H & \equiv-\tilde{\varepsilon}-\frac{1}{3}\left(a-a_{c,0}\right)
\end{align}

Eq. (\ref{eq:S_nem_negative}) is nothing but the Ising model in the
presence of an external field $H$, widely employed to describe symmetry-preserving
phase transitions, such as the Mott transition \citep{Terletska2011,Furukawa2015}
and certain magnetic transitions \citep{Wang_Fernandes2022}. It consists
of a first-order transition line parametrized by $H=0$, below which
the order parameter $\phi$ jumps between two non-zero values, signaling
a meta-nematic transition. The first-order transition line -- and
thus the jump in $\phi$ -- terminates at the so-called critical
end-point, given by $A=H=0$, which in our case is a quantum critical
end-point (QCEP), since the system is at $T=0$. This allows us to
obtain the location of the QCEP in the $\left(\tilde{\varepsilon},a\right)$
phase diagram,

\begin{align}
a_{\text{QCEP}} & =\frac{1}{3}\\
\tilde{\varepsilon}_{\text{QCEP}} & =-\frac{1}{27}
\end{align}
as well as the equation describing the first-order transition line:

\begin{equation}
a_{c}\left(\tilde{\varepsilon}\right)=\frac{2}{9}-3\tilde{\varepsilon}\,,\quad\mathrm{for}\;\,\tilde{\varepsilon}_{\text{QCEP}}<\tilde{\varepsilon}<0\,.
\end{equation}

The behavior of the magnitude of the nematic order parameter, $\phi$,
and of the nematic component projected along the strain direction,
$\phi_{1}=\phi\cos2\theta$, is shown in Fig. \ref{fig:phi_metanematic}
as a function of the non-thermal tuning parameter $a$ for fixed values
of strain $\tilde{\varepsilon}<0$. For $\tilde{\varepsilon}<\tilde{\varepsilon}_{\text{QCEP}}$,
as shown in Figs. \ref{fig:phi_metanematic}(a)-(b), the nematic order
parameter evolves continuously and displays a crossover behavior at
a characteristic $a$ value corresponding to the Widom line located
at the left of the QCEP.\textcolor{red}{{} }On the other hand, for $\tilde{\varepsilon}>\tilde{\varepsilon}_{\text{QCEP}}$,
the nematic order parameter undergoes a jump between two non-zero
values, signaling a symmetry-preserving meta-nematic transition, as
shown in Figs. \ref{fig:phi_metanematic}(c)-(d).

To characterize the properties of the QCEP, we calculate, the dynamical
critical exponent $z$. For an insulator, the bare dynamics of the
nematic susceptibility $\chi_{0}\left(q\right)$ is unchanged by the
coupling to the electrons, resulting in $z=1$. For a metal, we employ
a Hertz-Millis approach \citep{Hertz1976,Millis2002,Lohneysen2007}
to compute the one-loop polarization bubble $\Pi\left(q\right)$ that
renormalizes the nematic susceptibility, $\chi^{-1}\left(q\right)=\chi_{0}^{-1}\left(q\right)-\Pi\left(q\right)$.
As discussed elsewhere \citep{Xu2020,Fernandes_Venderbos}, for a
single-band system, the interaction (with coupling constant $g$)
between the nematic field $\boldsymbol{\Phi}$ and the electronic
quadrupolar charge density is given by the Hamiltonian:

\begin{equation}
H_{I}=g\phi\sum_{\mathbf{k}}\cos\left(2\beta_{\mathbf{k}}-2\theta\right)c_{\mathbf{k+q}/2,\sigma}^{\dagger}c_{\mathbf{k-q}/2,\sigma}^{\phantom{}}
\end{equation}
where $\tan\beta_{\mathbf{k}}\equiv k_{y}/k_{x}$ and the annihilation
operator $c_{\mathbf{k},\sigma}$ refers to an electron with momentum
$\mathbf{k}$ and spin $\sigma$. Spin indices are implicitly summed.
Recall that, in our notation, $\theta$ is measured with respect to
the strain direction $\alpha$. Therefore, it is convenient to define
$\tilde{\beta}_{\mathbf{k}}\equiv\beta_{\mathbf{k}}-\alpha$. The
coupled nematic-electronic action is then given by
\begin{align}
\mathcal{S}\left[\boldsymbol{\Phi},\psi,\psi^{\dagger}\right] & =\mathcal{S}_{\text{nem}}\left[\boldsymbol{\Phi}\right]+\int_{k}\left(-i\omega_{n}+\xi_{\mathbf{k}}\right)\psi_{k\sigma}^{\dagger}\psi_{k\sigma}^{\phantom{}}\label{eq:S_electronic}\\
 & +g\int_{k,q}\phi_{q}\cos\left(2\tilde{\beta}_{\mathbf{k}}-2\tilde{\theta}_{q}\right)\psi_{k+q/2,\sigma}^{\dagger}\psi_{k-q/2,\sigma}^{\phantom{}}\nonumber 
\end{align}
where we reintroduced the tilde in $\tilde{\theta}$ for the sake
of clarity. Here, $\psi$, $\psi^{\dagger}$ are Grassmann variables,
$\xi_{0,\mathbf{k}}=\epsilon_{\mathbf{k}}-\mu$ is the electronic
dispersion, and $k=(\omega_{n},\text{\textbf{k}})$, where $\omega_{n}=2\pi(n+1/2)k_{B}T$
is the fermionic Matsubara frequency. In the $\tilde{\varepsilon}<0$
side of the phase diagram, the nematic director $\tilde{\theta}$
is fixed at $\tilde{\theta}_{-}=\pi/2$. Therefore, in terms of $\delta\phi\equiv\phi-\Delta\phi_{0}$,
the interacting action becomes:

\begin{equation}
\mathcal{S}_{\text{int}}=-g\int_{k,q}\delta\phi_{q}\cos\left(2\tilde{\beta}_{\mathbf{k}}\right)\psi_{k+q/2}^{\dagger}\psi_{k-q/2}^{\phantom{}}\label{eq:Sint_metanematic}
\end{equation}

Moreover, the electronic dispersion is renormalized to $\xi_{\mathbf{k}}=\xi_{0,\mathbf{k}}-\frac{g}{3}\cos\left(2\tilde{\beta}_{\mathbf{k}}\right)$,
signaling the Fermi surface distortion caused by the non-zero nematic
order parameter. The coupling in Eq. (\ref{eq:Sint_metanematic})
is analogous to the case of a metal in the presence of an \emph{Ising-nematic}
QCP \citep{Oganesyan2001,Metzner2003,Garst2010,Metlitski2010}. The
lowering from 3-state Potts symmetry to Ising symmetry is due to the
external strain pinning the nematic director. The residual Ising degree
of freedom is not associated with any symmetry of the system, but
a consequence of the fact that the transition in the absence of the
conjugate field is first-order. The situation is analogous to the
QCEP of a metallic ferromagnet in the presence of a magnetic field
\citep{Millis2002}.

It is now straightforward to compute the polarization bubble. To leading
order in $g$, it is given by: \emph{
\begin{equation}
\Pi(q)=-2g^{2}\int_{k}\cos^{2}\left(2\tilde{\beta}_{\mathbf{k}}\right)G_{0,k+q/2}G_{0,k-q/2}
\end{equation}
}where $G_{0,k}^{-1}=i\omega_{n}-\xi_{0,\mathbf{k}}+\frac{g}{3}\cos\left(2\tilde{\beta}_{\mathbf{k}}\right)$
is the fermionic propagator for the distorted band dispersion. We
find: 

\begin{align}
\delta\Pi(q) & =-\frac{g^{2}}{2E_{F}}\,f_{1}\left(\frac{g}{3E_{F}},\cos2\tilde{\beta}_{\mathbf{q}}\right)\frac{|\Omega|}{v_{F}|\mathbf{q}|}\nonumber \\
 & -\frac{g^{2}}{2E_{F}}f_{2}\left(\frac{g}{3E_{F}},\cos2\tilde{\beta}_{\mathbf{q}}\right)\Big(\frac{\Omega}{v_{F}|\mathbf{q}|}\Big)^{2}
\end{align}
 where $E_{F}$ and $v_{F}$ are the Fermi energy and the Fermi velocity
of the undistorted Fermi surface, $\delta\Pi\left(q\right)\equiv\Pi\left(q\right)-\Pi\left(\mathbf{q},\Omega=0\right)$,
and we defined the functions:
\begin{align}
f_{1}(\tilde{g},x) & =\frac{[(1+\tilde{g}^{2})x-2\tilde{g}]^{2}}{\pi(1-\tilde{g}^{2})^{5/2}(1-\tilde{g}x)^{5/2}}\label{eq:aniso_pol_1}\\
f_{2}(\tilde{g},x) & =\frac{4\tilde{g}(2+\tilde{g}^{2})x-9\tilde{g}^{2}-(2+\tilde{g}^{2})(2x^{2}-1)}{\pi(1-\tilde{g}^{2})^{5/2}(1-\tilde{g}x)^{3}}\label{eq:aniso_pol_2}
\end{align}

Thus, as in the case of an Ising-nematic QCP, the Hertz-Millis dynamical
critical exponent is $z=3$, since $f_{1}(x)\geq0$, except for the
cold spots defined by $f_{1}(x_{\mathrm{cs}})=0$. From Eq. (\ref{eq:aniso_pol_1}),
we find that the cold spots are located at
\begin{equation}
\tilde{\beta}_{\mathbf{q}}=\beta_{\mathbf{q}}-\alpha=\frac{1}{2}\arccos\left(\frac{6gE_{F}}{9E_{F}^{2}+g^{2}}\right)
\end{equation}

Due to the Fermi surface distortion caused by the non-zero nematic
order parameter, the cold spots shift away from the value $\tilde{\beta}_{\mathbf{q}}=\pm\pi/4$,
which is recovered in the limit $g\Delta\phi\to0$. Moreover, because
$f_{2}(x_{\mathrm{cs}})>0$, at the cold spots the dynamical critical
exponent is given by $z=2$.

\subsection{Piezoelectric quantum tricritical point \label{subsec:piezoelectric}}

We now move to the $\tilde{\varepsilon}>0$ side of the $\left(\tilde{\varepsilon},a\right)$
phase diagram. As discussed above, there are two different minima:
$\theta_{+}=0$ far above $a_{c}$ and $\theta_{\pm}^{*}$, as given
by Eq. (\ref{eq:theta_star}), far below $a_{c}$. Our numerical results
showed that the transition between the two corresponding phases is
first-order for small strain but second-order for large strain. To
understand this behavior analytically, we start from Eq. (\ref{eq:Snem_tilde})
and substitute $\phi=\sqrt{\frac{\tilde{\varepsilon}}{-1+4\cos^{2}2\theta}}$
(recall that we are considering $\lambda>0$). Near the QTCP, we can
expand the action to leading order in $\theta$. Dropping a constant
term, we obtain:

\begin{equation}
S_{\text{nem}}^{(\tilde{\varepsilon}>0)}\left(\theta\right)=\frac{\mathcal{A}}{2}\,\theta^{2}+\frac{\mathcal{U}}{4}\,\theta^{4}+\frac{\mathcal{W}}{6}\,\theta^{6}\label{eq:Snem_tricritical}
\end{equation}
where we defined:

\begin{align}
\mathcal{A} & \equiv\frac{16\tilde{\varepsilon}\left(3a-2\sqrt{3\tilde{\varepsilon}}+\tilde{\varepsilon}\right)}{27}\\
\mathcal{U} & \equiv\frac{64\tilde{\varepsilon}\left(18a-13\sqrt{3\tilde{\varepsilon}}+10\tilde{\varepsilon}\right)}{81}\\
\mathcal{W} & \equiv\frac{64\tilde{\varepsilon}\left(1512a-1183\sqrt{3\tilde{\varepsilon}}+1224\tilde{\varepsilon}\right)}{1215}
\end{align}

The nematic order parameter in this case is given by:

\begin{equation}
\phi=\sqrt{\frac{\tilde{\varepsilon}}{3}}\left(1+\frac{8\theta^{2}}{3}\right)\label{eq:phi_tricritical}
\end{equation}

\begin{figure}[h]
\centering{}\includegraphics[width=1\columnwidth]{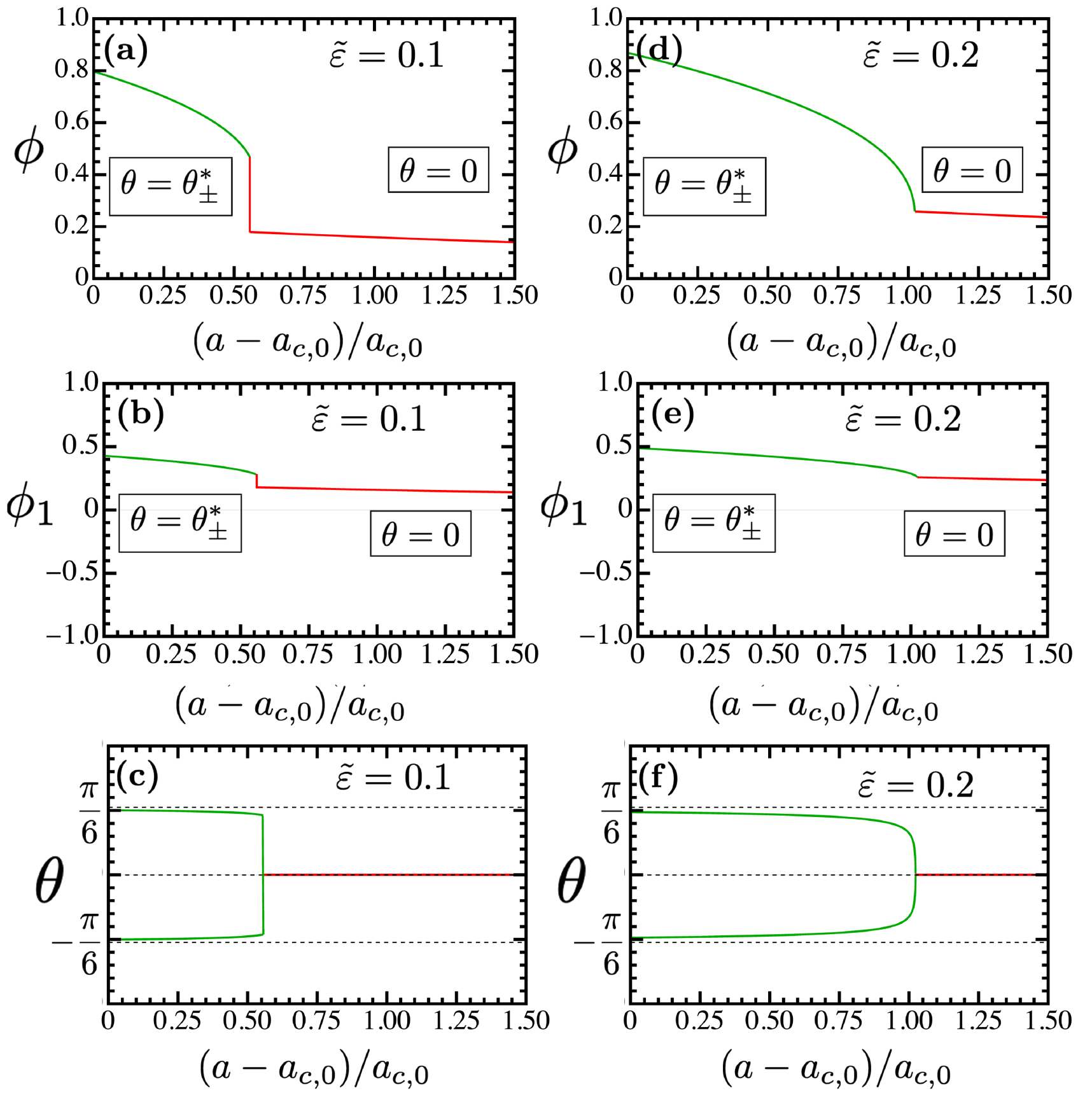}\caption{{\small{}Behavior of the nematic order parameter for fixed $\tilde{\varepsilon}>0$
across the first-order transition line {[}panels (a), (b), and (c){]}
and the second-order transition line {[}panels (d), (e), and (f){]},
as defined in the phase diagram of Fig. \ref{fig:phase_diagram_T=00003D0}.
Representative values of $\tilde{\varepsilon}$ are chosen from the
left and right side of $\tilde{\varepsilon}_{\text{QTCP}}=3/16\approx0.19$,
respectively. The top panels show the magnitude of the nematic order
parameter $\phi$; the middle panels show the nematic component projected
along the strain direction, $\phi_{1}=\phi\cos2\theta$; the bottom
panels show the nematic director angle $\theta$. The horizontal axis
corresponds to the non-thermal parameter $a$ measured relative to
unstrained Potts-nematic transition $a_{c,0}=2/9$. The red curve
corresponds to the $\theta=0$ phase whereas the green curve corresponds
to the $\theta=\theta_{\pm}^{*}$ phase. \label{fig:phi_piezoelectric}}}
\end{figure}

Before analyzing the behavior of Eq. (\ref{eq:Snem_tricritical}),
let us discuss the nature of the phase transition from the $\theta_{-}=0$
phase to the $\theta_{\pm}^{*}$ phase. In contrast to the $\tilde{\varepsilon}<0$
case discussed in Sec. \ref{subsec:metanematic}, here the emergent
Ising degree of freedom $\theta$ is related to a symmetry of the
system, namely, twofold rotations with respect to an in-plane axis,
$C'_{2}$. Indeed, as pointed out in Ref. \citep{Fernandes_Venderbos},
when the director moves away from the high-symmetry directions $p\pi/6$
(with $p=1,\cdots,5$), which is the case only in the $\theta_{\pm}^{*}$
phase, the twofold rotational symmetry $C'_{2}$ is spontaneously
broken -- in addition to the threefold rotational symmetry $C_{3z}$
that is explicitly broken by the external strain. 

More formally, focusing on a lattice with point group $\mathrm{D_{6h}}$,
strain applied along a high symmetry direction lowers the point group
symmetry to $\mathrm{D_{2h}}$. The onset of the $\theta_{\pm}^{*}$
phase breaks the in-plane twofold rotational symmetry, further lowering
the point group symmetry to $\mathrm{C_{2h}}$. In the phase diagram
of Fig. \ref{fig:phase_diagram_T=00003D0}, starting from the $\tilde{\varepsilon}=0$
axis slightly above the nematic transition point ($a>a_{c,0}$) and
then increasing $\tilde{\varepsilon}$ (i.e. $\tilde{\varepsilon}>0$),
the sequence of point-group-symmetry lowering is $\mathrm{D_{6h}\rightarrow D_{2h}\rightarrow C_{2h}}$.
Importantly, while the first symmetry-breaking is explicit and caused
by any non-zero $\tilde{\varepsilon}$, the second one is spontaneous
and requires a threshold value for $\tilde{\varepsilon}$. In contrast,
upon decreasing $\tilde{\varepsilon}$ (i.e. $\tilde{\varepsilon}<0$),
there is only the explicit symmetry breaking $\mathrm{D_{6h}\rightarrow D_{2h}}$
caused by a non-zero $\tilde{\varepsilon}$. Following the same steps
for the other point groups considered here, we find the following
sequences of symmetry lowering upon increasing $\tilde{\varepsilon}$:
$\mathrm{D_{3h}\rightarrow C_{2v}\rightarrow C_{s}}$, $\mathrm{D_{3d}\rightarrow C_{2h}\rightarrow S_{2}}$,
$\mathrm{D_{6}\rightarrow D_{2}\rightarrow C_{2}}$, and $\mathrm{D_{3}\rightarrow C_{2}\rightarrow C_{1}}$.

This result becomes even more interesting in the case of lattices
described by the point groups $\mathrm{D_{6}}$ and $\mathrm{D_{3}}$,
which lack any mirror symmetries. These are the groups that describe
the symmetries of twisted bilayer graphene and twisted double-bilayer
graphene. In these cases, spontaneous breaking of the in-plane twofold
rotational symmetry in the $\theta_{\pm}^{*}$ phase results in the
condensation of an electric polarization $P_{z}$ pointing out of
the plane. This can be seen by analyzing how the Potts-nematic order
parameter couples to $P_{z}$ in these groups. Following Ref. \citep{Samajdar2021},
the nemato-electric action is given by:

\begin{equation}
S'=\frac{\Upsilon}{6}\,P_{z}\phi^{3}\sin6\theta=\frac{\Upsilon}{6}\,P_{z}\phi^{3}\sin6\tilde{\theta}
\end{equation}
where $\Upsilon$ is a coupling constant. Expanding for small $\theta$,
we find:

\begin{equation}
S'\approx\Upsilon\left(\frac{\tilde{\varepsilon}}{3}\right)^{3/2}P_{z}\,\theta\label{eq:S_piezo}
\end{equation}

Therefore, a non-zero $\theta$ necessarily triggers a non-zero out-of-plane
electric polarization, which allows us to identify the $\theta_{\pm}^{*}$
phase with a ferroelectric phase. However, because this phase is only
accessible in the presence of externally applied uniaxial strain,
we dub it a \emph{piezoelectric }phase. We emphasize that the onset
of piezoelectricity is a specific property of $\mathrm{D_{6}}$ and
$\mathrm{D_{3}}$ lattices only.

The shape of the piezoelectric transition line in the $\left(\tilde{\varepsilon},a\right)$
phase diagram, as well as the character of the transition, can be
directly obtained from minimization of Eq. (\ref{eq:Snem_tricritical}).
The QTCP takes place for $\mathcal{A}=\mathcal{U}=0$, yielding:
\begin{align}
a_{\text{QTCP}} & =\frac{7}{16}\\
\tilde{\varepsilon}_{\text{QTCP}} & =\frac{3}{16}
\end{align}

For $\tilde{\varepsilon}>\tilde{\varepsilon}_{\text{QTCP}}$, the
piezoelectric transition is second-order, since $\mathcal{U}>0$.
In this regime, the transition line is given by $\mathcal{A}=0$,
which corresponds to:
\begin{equation}
a_{c}\left(\tilde{\varepsilon}\right)=\frac{2\sqrt{3\tilde{\varepsilon}}-\tilde{\varepsilon}}{3}\,,\quad\mathrm{for}\;\,\tilde{\varepsilon}>\tilde{\varepsilon}_{\text{QTCP}}\,.\label{eq:ac_piezo}
\end{equation}

On the other hand, for $0<\tilde{\varepsilon}<\tilde{\varepsilon}_{\text{QTCP}}$,
$\mathcal{U}<0$ and the piezoelectric transition is first-order.
Minimizing Eq. (\ref{eq:Snem_tricritical}), we find that the first-order
transition takes place when the following condition is met:

\begin{equation}
\mathcal{A}=\frac{3\mathcal{U}^{2}}{16\mathcal{W}}
\end{equation}
\begin{widetext}which corresponds to:

\begin{equation}
a_{c}\left(\tilde{\varepsilon}\right)=\frac{-1278\tilde{\varepsilon}+1606\sqrt{3\tilde{\varepsilon}}+\sqrt{15\tilde{\varepsilon}}\sqrt{25344\tilde{\varepsilon}-12312\sqrt{3\tilde{\varepsilon}}+4487}}{2214}\,,\quad\mathrm{for}\;\,0<\tilde{\varepsilon}<\tilde{\varepsilon}_{\text{QTCP}}\,.
\end{equation}

Note that this is an approximate expression valid only close to $\tilde{\varepsilon}_{\text{QTCP}}$.
\end{widetext} In Fig. \ref{fig:phi_piezoelectric}, we show the
behavior of different components of the nematic order parameter --
the magnitude $\phi$, the projection $\phi_{1}=\phi\cos2\theta$,
and the angle $\theta$ -- as a function of the non-thermal tuning
parameter for two different values of $\tilde{\varepsilon}>0$. For
$\tilde{\varepsilon}<\tilde{\varepsilon}_{\text{QTCP}}$, all components
change discontinuously across the piezoelectric transition {[}Figs.
\ref{fig:phi_piezoelectric}(a)-(c){]}, whereas for $\tilde{\varepsilon}>\tilde{\varepsilon}_{\text{QTCP}}$,
all components change continuously {[}Figs. \ref{fig:phi_piezoelectric}(d)-(f){]}.
We note that, for large enough $\tilde{\varepsilon}>0$, the derivative
of the second-order transition line with respect to $\tilde{\varepsilon}$
changes, as described by Eq. (\ref{eq:ac_piezo}), resulting in a
reentrance of the $\theta_{+}=0$ phase as a function of strain for
fixed $a$. This behavior is not shown in the phase diagram of Fig.
\ref{fig:phase_diagram_T=00003D0} because it only happens for very
large strain values, $\tilde{\varepsilon}>3$ (for comparison, recall
that the nematic order parameter jump across the unstrained Potts-nematic
transition is $\Delta\phi_{0}=1/3$).

We finish this section by discussing the properties of the line of
piezoelectric QCPs described by Eq. (\ref{eq:ac_piezo}). As in the
case of the QCEP discussed in the previous section, the dynamical
critical exponent is $z=1$ in the case of an insulator. For a metallic
system, we start from the action (\ref{eq:S_electronic}), substitute
$\phi_{q}=\sqrt{\frac{\tilde{\varepsilon}}{3}}$ and expand for small
$\theta$ to obtain:
\begin{equation}
\mathcal{S}_{\text{int}}=\sqrt{\frac{4\tilde{\varepsilon}}{3}}\,g\int_{k,q}\theta_{q}\sin\left(2\tilde{\beta}_{\mathbf{k}}\right)\psi_{k+q/2}^{\dagger}\psi_{k-q/2}^{\phantom{}}\label{eq:Sint_piezo}
\end{equation}
where, as before, $\tilde{\beta}_{\mathbf{q}}=\beta_{\mathbf{q}}-\alpha$.
Note that the electronic dispersion is also renormalized due to the
external strain, $\xi_{\mathbf{k}}=\xi_{0,\mathbf{k}}+\sqrt{\frac{\tilde{\varepsilon}}{3}}\,g\cos\left(2\tilde{\beta}_{\mathbf{k}}\right)$.
Like the QCEP case studied in Sec. \ref{subsec:metanematic}, the
form factor in Eq. (\ref{eq:Sint_piezo}) is that of an Ising-nematic
QCP. Interestingly, the two Ising-nematic form factors in Eqs. (\ref{eq:Sint_metanematic})
and (\ref{eq:Sint_piezo}) are ``orthogonal'' in the nematic space,
corresponding to the longitudinal and transverse modes of a hypothetical
XY nematic order parameter \citep{Oganesyan2001,Garst2010}. This
is a consequence of the fact that, for $\tilde{\varepsilon}<0$, the
nematic director angle is pinned and the nematic amplitude is fluctuating,
whereas for $\tilde{\varepsilon}>0$ it is $\theta$ that fluctuates.

To one-loop order, the polarization bubble is given by: \emph{
\begin{equation}
\Pi(q)=-\frac{8}{3}\,\tilde{\varepsilon}g^{2}\int_{k}\sin^{2}\left(2\tilde{\beta}_{\mathbf{k}}\right)G_{0,k+q/2}G_{0,k-q/2}
\end{equation}
}which evaluates to:
\begin{align}
\delta\Pi(q) & =-\frac{\tilde{\varepsilon}g^{2}}{2E_{F}}\,f_{3}\left(\frac{g}{E_{F}}\sqrt{\frac{\tilde{\varepsilon}}{3}},\cos2\tilde{\beta}_{\mathbf{q}}\right)\frac{|\Omega|}{v_{F}|\mathbf{q}|}\nonumber \\
 & -\frac{\tilde{\varepsilon}g^{2}}{2E_{F}}\,f_{4}\left(\frac{g}{E_{F}}\sqrt{\frac{\tilde{\varepsilon}}{3}},\cos2\tilde{\beta}_{\mathbf{q}}\right)\left(\frac{\Omega}{v_{F}|\mathbf{q}|}\right)^{2}
\end{align}
 where we defined the functions:
\begin{align}
f_{3}(\tilde{g},x) & =\frac{4}{3\pi}\frac{1-x^{2}}{\sqrt{1-\tilde{g}^{2}}(1+\tilde{g}x)^{5/2}}\\
f_{4}(\tilde{g},x) & =\frac{4}{3\pi}\frac{3\tilde{g}^{2}+4\tilde{g}x+(2-\tilde{g}^{2})(2x^{2}-1)}{(1-\tilde{g}^{2})^{3/2}(1+\tilde{g}x)^{3}}
\end{align}

Thus, within a Hertz-Millis approximation for the dynamical critical
exponent $z$, we find $z=3$, except for the cold spots parametrized
by $f_{3}\left(x_{\mathrm{cs}}\right)=0$, for which $z=2$. The last
result follows from the fact that, since $\tilde{g}$ is small, $f_{4}\left(x_{\mathrm{cs}}\right)>0$.
Moreover, note that $f_{3}\left(x\right)\geq0$ for any $x$. Interestingly,
in contrast to the QCEP case, here the cold spots are the same as
in the case of the undistorted Fermi surface, $\tilde{\beta}_{\mathbf{q}}=0,\text{ }\pi/2$.
This can be understood geometrically by noting that the semi-major
axes of the elliptical Fermi surface coincide with its cold spots.
As a result, cold-spot fermions at $\beta_{\mathbf{k}}=0$ ($\beta_{\mathbf{k}}=\pi/2$)
will necessarily exchange underdamped collective bosons with momentum
direction $\tilde{\beta}_{\mathbf{q}}=\pi/2$ ($\tilde{\beta}_{\mathbf{q}}=0$).

\section{Hysteresis and spinodal lines \label{sec:Hysteresis-and-spinodal}}

The phase boundaries in the phase diagram of Fig. \ref{fig:phase_diagram_T=00003D0}
were obtained by determining the global minimum of the action. In
the case of first-order transitions, however, the action also has
local minima, which correspond to metastable phases. While they are
formally inaccessible in true equilibrium, they can be probed via
hysteresis measurements in which the order parameter $\boldsymbol{\Phi}$
is measured upon cycling the conjugate field $\tilde{\varepsilon}$.
The interesting aspect of the Potts-nematic state is that the action
has three discrete minima rather than two, which should lead to more
complex hysteresis loops as compared to the standard Ising-nematic
case.

\begin{figure}[h]
\centering{}\includegraphics[width=1\columnwidth]{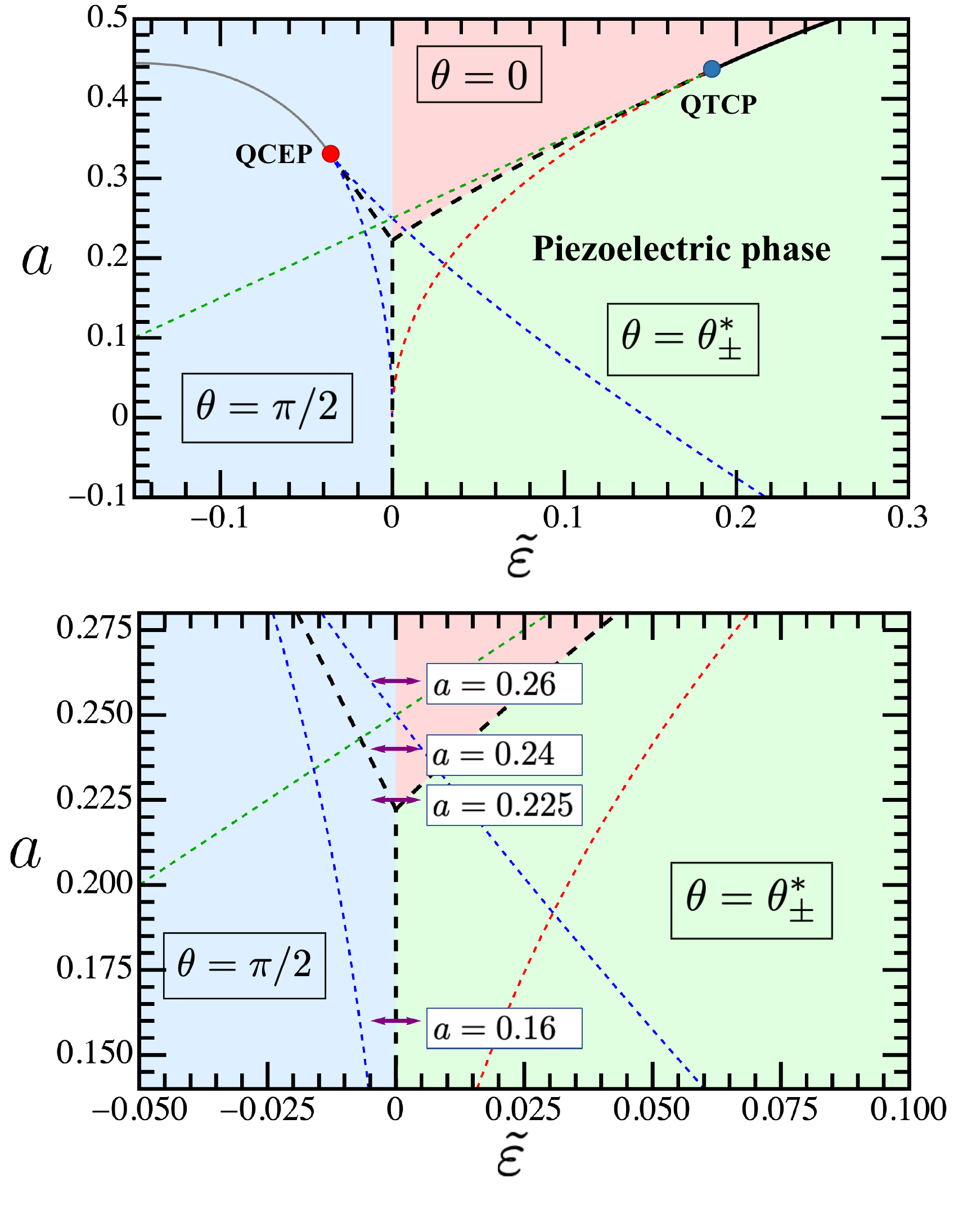}\caption{The zero temperature phase diagram of Fig. \ref{fig:phase_diagram_T=00003D0}
is shown with the spinodal lines included (top panel). The two blue
dashed lines on the left side ($\tilde{\varepsilon}<0$) are the upper
and lower spinodals corresponding to the two phases associated with
the meta-nematic transition. They coincide at the QCEP. The green
and red dashed curves on the right side ($\tilde{\varepsilon}>0$)
are the upper and lower spinodals of the $\theta=\theta_{\pm}^{*}$
(piezoelectric) and $\theta=0$ phases, respectively. They coincide
at the QTCP. The bottom panel is a zoom of the top panel; the purple
arrows show the values of $a$ for which the hysteresis curves $\boldsymbol{\Phi}\left(\tilde{\varepsilon}\right)$
in Fig. \ref{fig:hysteresis} are shown. \label{fig:phase_diagram_w_spinodals}}
\end{figure}

To calculate these hysteresis curves, we first derive the upper and
lower spinodals associated with the first-order transition lines in
Fig. \ref{fig:phase_diagram_T=00003D0}. As in Sec. \ref{sec:Zero-temperature-phase-diagram},
we consider $\lambda>0$ and drop the tilde of the rescaled variables
(except for $\tilde{\varepsilon}$). The spinodals are curves on the
$(\tilde{\varepsilon},a)$-plane that bound the regions of metastability
of the different phases. We consider first the phase $\theta_{-}=\pi/2$;
it corresponds to a local minimum as long as the following metastability
conditions are met
\begin{equation}
\partial_{\phi}S_{\text{nem}}(\theta=\pi/2)=\phi^{3}-\phi^{2}+a\phi+\tilde{\varepsilon}=0\label{eq:extremize_pi/2}
\end{equation}

\begin{equation}
\begin{cases}
\partial_{\phi}^{2}S_{\text{nem}}(\theta=\pi/2)=3\phi^{2}-2\phi+a>0\\
\partial_{\theta}^{2}S_{\text{nem}}(\theta=\pi/2)=3\phi^{3}-\tilde{\varepsilon}\phi>0
\end{cases}\label{eq:hessian_pi/2}
\end{equation}

Since $\partial_{\theta}\partial_{\phi}S_{\text{nem}}$ vanishes,
positive-definiteness of the Hessian matrix of second derivatives
$\left(\partial_{i}\partial_{j}S_{\text{nem}}\right)$ is ensured
by Eq. (\ref{eq:hessian_pi/2}). It is convenient to define the cubic
discriminant of Eq. (\ref{eq:extremize_pi/2}), $\mathcal{D}_{\pi/2}=a^{2}-4a^{3}+4\tilde{\varepsilon}-27\tilde{\varepsilon}^{2}-18a\varepsilon$.
When $\mathcal{D}_{\pi/2}<0$, Eq. (\ref{eq:extremize_pi/2}) has
only one real $\phi$ solution, whereas when $\mathcal{D}_{\pi/2}>0$,
there are three real $\phi$ solutions. In the latter case, $\mathcal{D}_{\pi/2}>0$,
the largest and smallest values of $\phi$ that solve Eq. (\ref{eq:extremize_pi/2})
correspond to the two solutions associated with the meta-nematic transition.
On the other hand, in the former case, $\mathcal{D}_{\pi/2}<0$, the
single real solution indicates that there is no meta-nematic transition,
as is the case to the left of the QCEP. This suggests that $\mathcal{D}_{\pi/2}=0$
gives the spinodals associated with the meta-nematic transition. There
is, however, one subtlety: by construction, $\phi$ must be positive.
Therefore, it is not enough to ensure the existence of a real solution,
but of a real and positive solution. It turns out that, when $\tilde{\varepsilon}<0$,
the real solutions of Eq. (\ref{eq:extremize_pi/2}) in both cases
($\mathcal{D}_{\pi/2}>0$ and $\mathcal{D}_{\pi/2}<0$) are always
positive. However, when $\tilde{\varepsilon}>0$, the single solution
in the $\mathcal{D}_{\pi/2}<0$ case is negative, whereas only one
among the two positive solutions in the $\mathcal{D}_{\pi/2}>0$ case
is an action minimum. Taking these conditions into account and solving
the equation $\mathcal{D}_{\pi/2}=0$ for $a$, we find the equations
describing the spinodals of the meta-nematic transition. The three
solutions of $\mathcal{D}_{\pi/2}=0$ can be written as:

\begin{align}
a_{n}(\tilde{\varepsilon}) & =\frac{1}{12}\Big(1-\mathrm{e}^{2ni\pi/3}\,b(\tilde{\varepsilon})+\mathrm{e}^{-2ni\pi/3}\,\frac{216\tilde{\varepsilon}-1}{b(\tilde{\varepsilon})}\Big)\,,\\
b(\tilde{\varepsilon}) & =\big[108\tilde{\varepsilon}(54\tilde{\varepsilon}-5)+24\sqrt{3\tilde{\varepsilon}(27\tilde{\varepsilon}+1)^{3}}-1\big]^{\frac{1}{3}}
\end{align}
with $n=0,1,2$. Then, the upper spinodal is given by:

\begin{equation}
a_{\text{us}}^{\text{meta}}(\tilde{\varepsilon})=\left\{ \begin{array}{ll}
a_{1}(\tilde{\varepsilon}) & \,,\quad\mathrm{for}\;\,\tilde{\varepsilon}_{\text{QCEP}}<\tilde{\varepsilon}<\tilde{\varepsilon}_{*}\\
a_{0}(\tilde{\varepsilon}) & \,,\quad\mathrm{for}\;\,\tilde{\varepsilon}>\tilde{\varepsilon}_{*}
\end{array}\right.
\end{equation}
with $\tilde{\varepsilon}_{*}=1/216$ defined such that $b\left(\tilde{\varepsilon}_{*}\right)=0$.
For the lower spinodal, we obtain:

\[
a_{\text{ls}}^{\text{meta}}(\tilde{\varepsilon})=a_{2}(\tilde{\varepsilon})\,,\quad\mathrm{for}\;\,\tilde{\varepsilon}_{\text{QCEP}}<\tilde{\varepsilon}<0\,
\]

Here, the subscripts ``us'' and ``ls'' denote upper spinodal and
lower spinodal, respectively. In particular, $a_{\text{us}}^{\text{meta}}(\tilde{\varepsilon})$
giv es the limit of metastability of the $\theta_{-}$ phase below
the meta-nematic transition line, whereas $a_{\text{ls}}^{\text{meta}}(\tilde{\varepsilon})$
gives the limit of metastability of the $\theta_{-}$ phase above
the meta-nematic transition line. These spinodal lines are shown by
the blue dashed lines in the phase diagram of Fig. \ref{fig:phase_diagram_w_spinodals}. 

We now analyze the metastability of the $\theta_{+}=0$ phase. The
metastability conditions are given by:
\begin{equation}
\partial_{\phi}S_{\text{nem}}(\theta=0)=\phi^{3}+\phi^{2}+a\phi-\tilde{\varepsilon}=0\label{eq:extremize_0}
\end{equation}
\begin{equation}
\begin{cases}
\partial_{\phi}^{2}S_{\text{nem}}(\theta=0)=3\phi^{2}+2\phi+a>0\\
\partial_{\theta}^{2}S_{\text{nem}}(\theta=0)=-3\phi^{3}+\tilde{\varepsilon}\phi>0.
\end{cases}\label{eq:hessian_0}
\end{equation}

Applying a similar analysis as in the $\theta_{-}=\pi/2$ case, we
find that $\theta_{+}=0$ ceases to be a local minimum when the second
condition of Eq. (\ref{eq:hessian_0}) fails. Plugging in $\phi^{2}=\tilde{\varepsilon}/3$
into Eq. (\ref{eq:extremize_0}), we find: 
\begin{equation}
a_{\text{ls}}^{\text{piezo}}(\tilde{\varepsilon})=\frac{2\sqrt{3\tilde{\varepsilon}}-\tilde{\varepsilon}}{3}\,,\quad\mathrm{for}\;\,0<\tilde{\varepsilon}<\tilde{\varepsilon}_{\text{QTCP}}\label{eq:a_ls_piezo}
\end{equation}
which corresponds to the lower spinodal of the first-order piezoelectric
phase transition, shown by the dashed red line in the phase diagram
of Fig. \ref{fig:phase_diagram_w_spinodals}. To obtain the upper
spinodal of this transition, we need to analyze the metastability
of the $\theta=\theta_{\pm}^{*}$ phase. Eq. (\ref{fig:phi_metanematic})
gives the nematic magnitude $\phi_{+}^{*}$ in the $\theta_{\pm}^{*}$
phase. For $\tilde{\varepsilon}<\tilde{\varepsilon}_{\text{QTCP}}$,
the condition $(\phi_{+}^{*})^{2}\ge\tilde{\varepsilon}/3$ required
for $\theta_{\pm}^{*}$ in Eq. (\ref{eq:theta_star}) to exist is
always satisfied. Moreover, the $\phi_{+}^{*}$ solution exists as
long as the argument of the square root in Eq. (\ref{fig:phi_metanematic})
is positive, $a<\tilde{\varepsilon}+1/4$. This therefore defines
the limit of metastability of the $\theta_{\pm}^{*}$ phase, which
corresponds to the upper spinodal of the first-order piezoelectric
transition. It is shown by the dashed green line in Fig. \ref{fig:phase_diagram_w_spinodals}(a)
and given by:
\begin{equation}
a_{\text{us}}^{\text{piezo }}(\tilde{\varepsilon})=\tilde{\varepsilon}+\frac{1}{4}\,,\quad\mathrm{for}\;\,\tilde{\varepsilon}<\tilde{\varepsilon}_{\text{QTCP}}
\end{equation}

Interestingly, for $\tilde{\varepsilon}>\tilde{\varepsilon}_{\text{QTCP}}$,
the condition $(\phi_{+}^{*})^{2}\ge\tilde{\varepsilon}/3$ would
imply an upper spinodal $a_{\text{us}}^{\text{piezo }}(\tilde{\varepsilon})=\frac{2\sqrt{3\tilde{\varepsilon}}-\tilde{\varepsilon}}{3}$,
which is identical to what the lower spinodal would be in this strain
range, see Eq. (\ref{eq:a_ls_piezo}). The coincidence between the
upper and lower spinodals implies that the transition is actually
second-order. Indeed, these would-be spinodals have the same expression
as the one describing the second-order transition line, Eq. (\ref{eq:ac_piezo}).

\begin{figure*}[t]
\begin{centering}
\includegraphics[width=0.8\paperwidth]{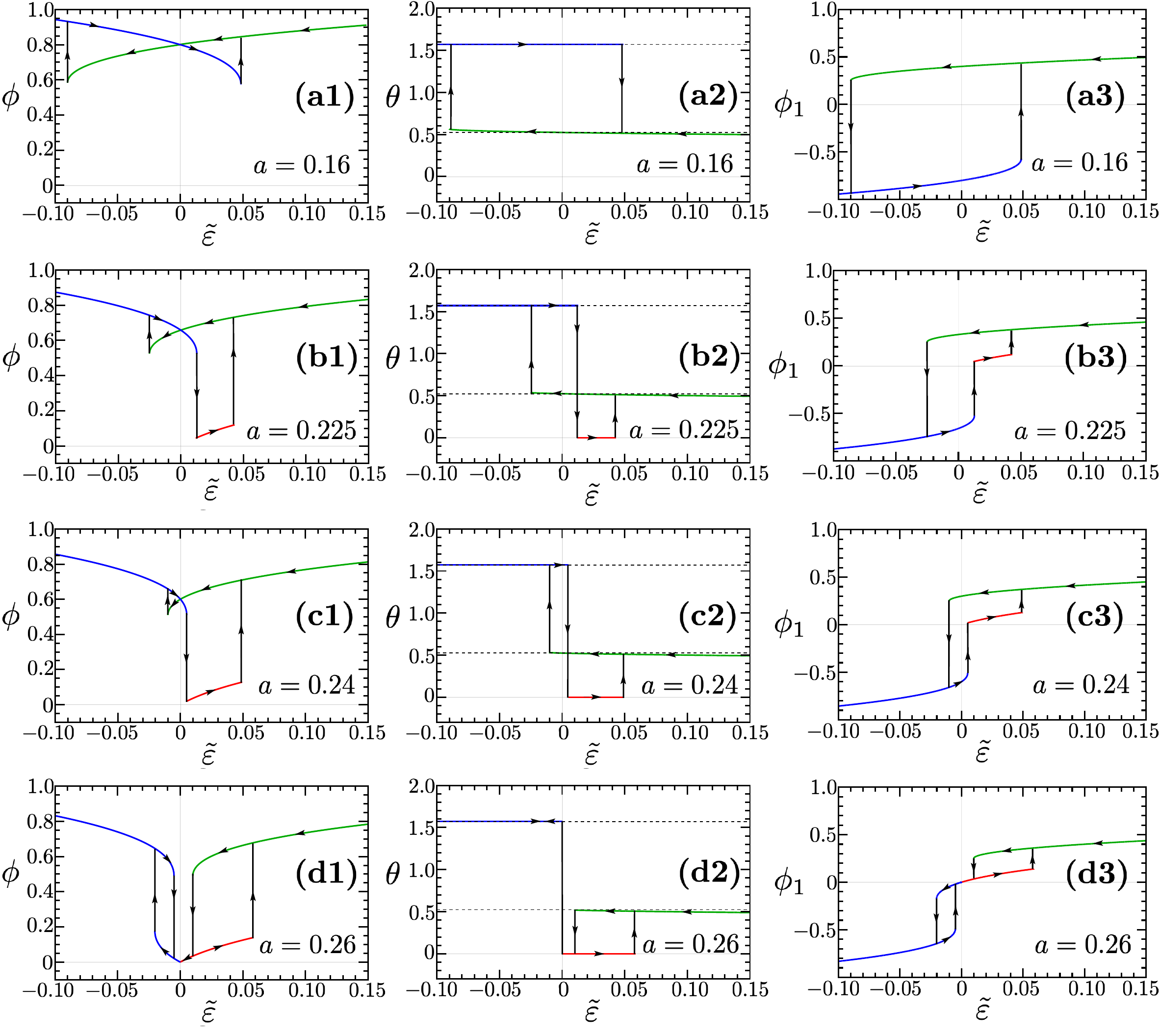}
\par\end{centering}
\caption{Hysteresis curves of the nematic order parameter as a function of
uniaxial strain $\tilde{\varepsilon}$ for four representive values
of $a$ marked by the purple arrows in Fig. \ref{fig:phase_diagram_w_spinodals}.
Panels (a) correspond to $a=0.16$; (b), to $a=0.225$; (c), to $a=0.24$;
and (d), to $a=0.26$. Left, middle, and right panels, which are identified
by the numbers 1, 2, and 3, respectively, correspond to the nematic
magnitude $\phi$, the nematic director angle $\theta$, and the nematic
component projected along the direction of the strain, $\phi_{1}=\phi\cos2\theta$.
Red, green, and blue colors denote the action minimum $\theta=0$,
$\theta=\theta_{\pm}^{*}$, and $\theta=\pi/2$, respectively. \label{fig:hysteresis}}
\end{figure*}

We are now in position to analyze the hysteresis curves $\boldsymbol{\Phi}\left(\tilde{\varepsilon}\right)$
as the strain $\tilde{\varepsilon}$ is cycled. We employ the Stoner-Wohlfarth
approach \citep{Stoner1948}: starting deep in one of the ordered
states, we assume that the system remains in this state until it is
no longer a local minimum of the action, i.e. until its spinodal line
is crossed, at which point the system moves to another minimum. In
the Ising-nematic case, this last step is straightforward, as there
is only one minimum available in the action landscape after the spinodal
line is crossed. However, in the Potts-nematic case, there can be
two local minima. To decide which of the two minima the system chooses,
we employ a ``gradient-descent criterion.'' Specifically, we introduce
a ``time'' variable $s$, promoting $\boldsymbol{\Phi}$ to a dynamical
field $\boldsymbol{\Phi}(s)$, and define a generalized gradient-descent
equation: 
\begin{equation}
\dot{\phi}_{i}=-\eta_{ij}\frac{\partial S_{\text{nem}}}{\partial\phi_{j}},\label{eq:langevin}
\end{equation}
where, we recall, $\boldsymbol{\Phi}\equiv\left(\phi_{1},\,\phi_{2}\right)=\phi\left(\cos2\theta,\,\sin2\theta\right)$.
Here, repeated indices are implicitly summed, $\dot{\phi}_{i}=\partial\phi_{i}/\partial s$,
and $\eta_{ij}$ is a positive-definite matrix. The last condition
ensures that Eq. (\ref{eq:langevin}) remains purely diffusive, such
that $\boldsymbol{\Phi}$ approaches a local minimum of $S_{\text{nem}}$
as $s\to\infty$. We set $\eta_{ij}=\eta\delta_{ij}$ with $\eta>0$,
in which case solutions to Eq. (\ref{eq:langevin}) are trajectories
of steepest descent. We rescale $s\to s'=s/\eta$ and redefine $\dot{\phi}_{i}=\partial\phi_{i}/\partial s'$
to obtain the autonomous system 
\begin{equation}
\begin{cases}
\dot{\phi}_{1} & =-a\phi_{1}-\phi_{1}(\phi_{1}^{2}+\phi_{2}^{2})-(\phi_{1}^{2}-\phi_{2}^{2})+\tilde{\varepsilon}\\
\dot{\phi}_{2} & =-a\phi_{2}-\phi_{2}(\phi_{1}^{2}+\phi_{2}^{2})+2\phi_{1}\phi_{2}
\end{cases}\label{eq:gradient_descent}
\end{equation}

The procedure we adopt once the system approaches a spinodal at $\tilde{\varepsilon}^{*}$,
for a fixed $a$ value, is as follows: let $\tilde{\varepsilon}^{+}$
and $\tilde{\varepsilon}^{-}$ be strain values near $\tilde{\varepsilon}^{*}$
and within the unstable and stable sides of the spinodal curve, respectively.
Let $\boldsymbol{\Phi}_{\text{min}}^{(\tilde{\varepsilon}^{-})}$
be the local minimum of the action at $\tilde{\varepsilon}=\tilde{\varepsilon}^{-}$,
which disappears once $\tilde{\varepsilon}=\tilde{\varepsilon}^{+}$.
We then set $\tilde{\varepsilon}=\tilde{\varepsilon}^{+}$ in Eq.
(\ref{eq:gradient_descent}) and choose several initial conditions
$\boldsymbol{\Phi}(0)$ in a narrow neighborhood of $\boldsymbol{\Phi}_{\text{min}}^{(\tilde{\varepsilon}^{-})}$,
$|\boldsymbol{\Phi}(0)-\boldsymbol{\Phi}_{\text{min}}^{(\tilde{\varepsilon}^{-})}|\lesssim10^{-4}$,
letting the system evolve until a new local minima is encountered.
We found that $\boldsymbol{\Phi}(s)$ approaches the same minimum
for all initial conditions we investigated, which suggests that the
outcome is insensitive to any initial condition within a small vicinity
of $\boldsymbol{\Phi}_{\text{min}}^{(\tilde{\varepsilon}^{-})}$. 

We applied this procedure to four representative fixed values of $a$
in the phase diagram of Fig. \ref{fig:phase_diagram_w_spinodals},
marked by the purple arrows in the bottom panel. They each correspond
to one of the four regions bounded by the values of $a$ in which
two different spinodal lines intersect, namely: 
\begin{equation}
\begin{cases}
a_{1} & =-39+16\sqrt{6}\approx0.1918\\
a_{2} & =\frac{-45+13\sqrt{13}}{8}\approx0.2340\\
a_{3} & =a_{\text{us}}^{\text{Potts}}=0.25.
\end{cases}
\end{equation}

Here, $a_{1}$ corresponds to the crossing between the blue and red
dashed spinodal lines; $a_{2}$ corresponds to the lower crossing
between the blue and green dashed spinodal lines; and $a_{3}$ corresponds
to the upper crossing between the blue and green dashed spinodal lines,
which also coincides with the upper spinodal of the unstrained Potts
transition. 

The hysteresis curves for the four representative $a$ values marked
in \ref{fig:phase_diagram_w_spinodals} are shown in Fig. \ref{fig:hysteresis}.
In this figure, we present the hysteresis curves for the nematic magnitude
$\phi\left(\tilde{\varepsilon}\right)$, the nematic director angle
$\theta\left(\tilde{\varepsilon}\right)$, and the nematic component
projected along the strain direction, $\phi_{1}=\phi\cos2\theta$.
Panels (a1)-(a3) show the case $a=0.16<a_{1}$. The system starts
deep inside the $\theta_{\pm}^{*}$ green phase (piezoelectric phase)
when $\tilde{\varepsilon}$ is large and positive. Upon decreasing
$\tilde{\varepsilon}$ (from right to left in Fig. \ref{fig:hysteresis}(a1)-(a3)),
the system remains in the metastable $\theta_{\pm}^{*}$ phase until
it reaches the dashed green spinodal line, where $\tilde{\varepsilon}<0$.
At this point, the only available minimum is the $\theta_{-}=\pi/2$
blue phase below the meta-nematic transition. Once we reverse $\tilde{\varepsilon}$
and start increasing it (from left to right in Fig. \ref{fig:hysteresis}(a1)-(a3)),
the system remains in the $\theta_{-}=\pi/2$ phase until the spinodal
blue dashed line is crossed in the $\tilde{\varepsilon}>0$ side of
the phase diagram, at which point the system moves back to the $\theta_{\pm}^{*}$
green phase. In terms of the $\phi_{1}$ component, the hysteresis
curve is a rather standard one, albeit not symmetric with respect
to either the $\phi_{1}$ or the $\tilde{\varepsilon}$ axes. 

Fig. \ref{fig:hysteresis}(b1)-(b3) shows the hysteresis curves for
the case $a_{1}<a=0.225<a_{2}$. Starting deep from the $\tilde{\varepsilon}>0$
side of the phase diagram and then decreasing $\tilde{\varepsilon}$
(i.e. going from right to left in the plots), the situation is the
same as in panels (a1)-(a3), namely, the system remains in the $\theta_{\pm}^{*}$
green phase until the green spinodal line is crossed on the $\tilde{\varepsilon}<0$
side of the phase diagram. However, upon reversing $\tilde{\varepsilon}$
and increasing it (i.e. going from left to right in the plots), the
situation changes. Once the blue dashed spinodal line is crossed,
on the $\tilde{\varepsilon}>0$ side of the phase diagram, there are
two local minima available: the global minimum corresponding to the
$\theta_{\pm}^{*}$ green phase and the metastable minimum corresponding
to the $\theta_{+}=0$ red phase. By solving Eq. (\ref{eq:langevin}),
we find that the system moves to the $\theta_{+}=0$ red phase and
remains at this local minimum until the red dashed spinodal line is
crossed, at which point the system finally moves back to the $\theta_{\pm}^{*}$
green phase. This behavior results in multi-loop hysteresis curves.

The case $a_{2}<a=0.24<a_{3}$ is depicted in Fig. \ref{fig:hysteresis}(c1)-(c3).
The behavior upon increasing $\tilde{\varepsilon}$ (i.e. going from
left to right in the plots) is the same as in panels (b1)-(b3). On
the other hand, the sequence of spinodals crossed upon decreasing
$\tilde{\varepsilon}$ (i.e. going from right to left in the plots)
is different: once the green dashed spinodal line is crossed, there
are now two local minima available, corresponding to the two $\theta_{-}=\pi/2$
blue phases associated with the two sides of the meta-nematic transition.
The solution of Eq. (\ref{eq:langevin}) shows that the system moves
to the global minimum, where it remains as $\tilde{\varepsilon}$
continues being decreased. Therefore, although the sequence of spinodals
crossed is different from the case of panels (b1)-(b3), the sequence
of metastable phases probed is the same. 

Finally, Fig. \ref{fig:hysteresis}(d1)-(d3) shows the case $a=0.26>a_{3}$.
Upon decreasing $\tilde{\varepsilon}$ (right to left in the plots),
the green dashed spinodal line is now crossed on the $\tilde{\varepsilon}>0$
side of the phase diagram. The only available minimum is the $\theta_{+}=0$
phase, which however ceases to be a solution once the $\tilde{\varepsilon}=0$
line is crossed. At this axis, $\phi\rightarrow0$, which is a consequence
of the fact that the system is above the upper spinodal $a_{3}$ of
the unstrained Potts-nematic transition. The system then moves to
the $\theta_{-}=\pi/2$ blue phase above the meta-nematic transition,
where it remains until the lower blue dashed spinodal line is crossed.
At this point, the system moves to the $\theta_{-}=\pi/2$ blue phase
below the meta-nematic transition. The behavior upon increasing $\tilde{\varepsilon}$
(left to right in the plots) can be understood in a similar manner.
The resulting hysteresis curves display multiple loops, which however
do not cross the origin, since $\phi=0$ when $\tilde{\varepsilon}=0$. 

\section{Non-zero-temperature phase diagram \label{sec:Non-zero-temperature-phase}}

\begin{figure*}[t]
\begin{raggedright}
\includegraphics[width=0.8\paperwidth]{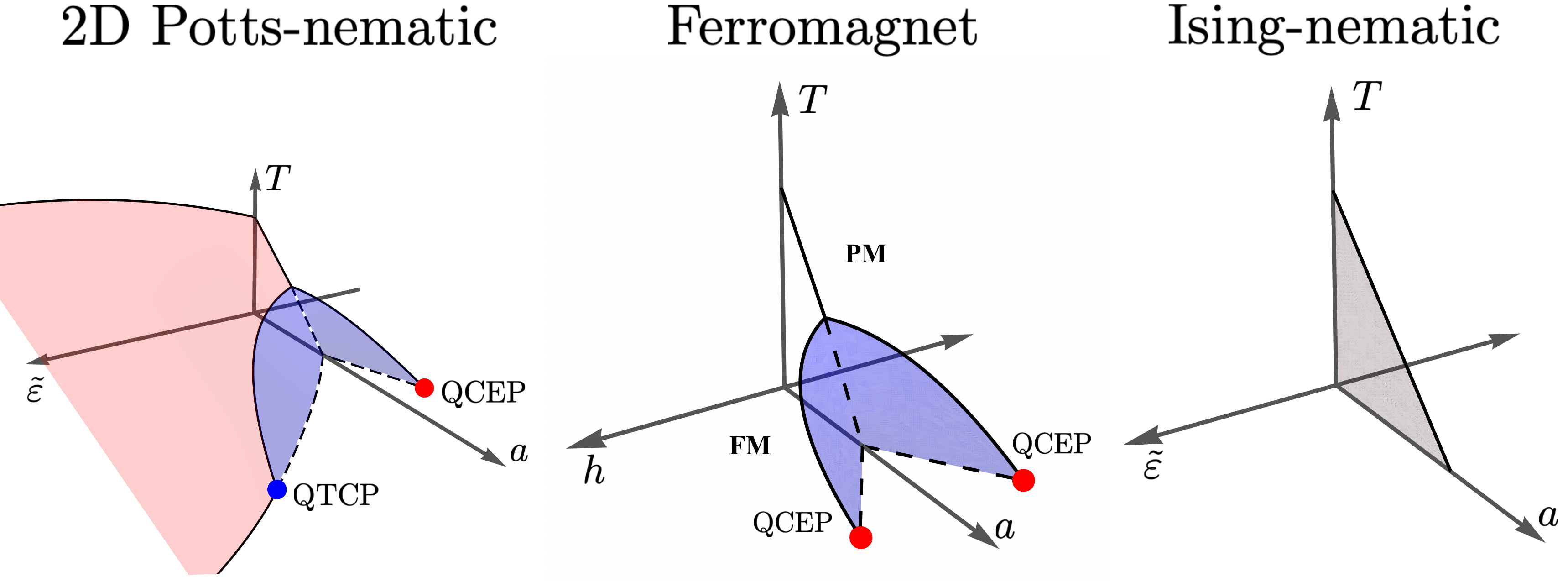}\caption{Left panel:\textbf{ }Qualitative $(\tilde{\varepsilon},a,T)$ phase
diagram of a 2D Potts-nematic system (same as Fig. \ref{fig_phase_diagram_T}).
Middle panel: $(h,a,T)$ phase diagram of an itinerant ferromagnet,
where $h$ is the magnetic field (see Refs. \citep{Belitz2005,Brando2016}).
The first-order wings are symmetric with respect to $h$ and are bounded
by a line of classical critical end-points terminating at QCEPs. Right
panel: \textbf{$(\tilde{\varepsilon},a,T)$ }phase diagram of an Ising-nematic
system. There is a second-order transition only along the $\tilde{\varepsilon}=0$
plane. Any non-zero strain along the nematic director directions smears
the transition completely. \label{fig:comparative_phase_diagrams}}
\par\end{raggedright}
\raggedright{}
\end{figure*}

At $T=0$, the Potts-nematic phase diagram is expected to be the same
for both 2D and 3D systems, since in either case the effective dimensionality
$d+z$ is larger than the upper critical dimension of the 3-state
Potts model, $d_{u}^{\mathrm{Potts}}\lesssim3$, such that a mean-field
analysis is warranted. At larger temperatures, where the bosonic quantum
dynamics can be neglected, the situation is different. Since $d=3>d_{u}^{\mathrm{Potts}}$,
in the 3D case the $\left(\tilde{\varepsilon},a,T\right)$ phase diagram
consists essentially of a sequence of ``copies'' of the phase diagram
shown in Fig. \ref{fig:phase_diagram_T=00003D0}. Similarly, for a
fixed $a$, the $\left(\tilde{\varepsilon},T\right)$ phase diagram
has the same form as the one obtained at $T=0$, but with the $y$-axis
representing $T-T_{c}$.

The situation at non-zero temperatures is more interesting in the
2D case. The fact that $d=2<d_{u}^{\mathrm{Potts}}$ implies that
the mean-field solution is not applicable. Surprisingly, despite the
presence of a cubic invariant in the free energy expansion, the 2D
3-state Potts model undergoes a second-order transition characterized
by the critical exponents $\alpha=1/3$ and $\beta=1/9$, which are
in the universality class of the hard hexagon lattice gas model \citep{review_Potts}.
Of course, one cannot exclude the possibility that for particular
microscopic models the quartic Landau coefficient is negative, rendering
the transition first-order. But, in the general case, we expect that,
in the absence of strain and at high enough temperatures, the Potts-nematic
transition is second-order. As a result, since at $T=0$ the Potts-nematic
transition is first-order, the $(a,T)$ phase diagram with fixed $\tilde{\varepsilon}=0$
should display a (classical) tricritical point. It is difficult to
estimate the position of this tricritical point, since standard perturbative
approaches such as renormalization-group calculations do not capture
the second-order character of the transition in 2D. It would be interesting
to perform Monte Carlo simulations to pinpoint the position of the
classical tricritical point.

Having established the $\left(\tilde{\varepsilon},a\right)$ phase
diagram at $T=0$ and the $(a,T)$ phase diagram at $\tilde{\varepsilon}=0$,
we can conjecture the full qualitative three-dimensional $\left(\tilde{\varepsilon},a,T\right)$
phase diagram for the 2D Potts-nematic model by continuously connecting
the QTCP and the QCEP at $T=0$ with the classical tricritical point
at $\tilde{\varepsilon}=0$. The result, shown in Fig. \ref{fig_phase_diagram_T}
and repeated for convenience in the left panel of Fig. \ref{fig:comparative_phase_diagrams},
consists of two wings (in blue) inside which the transition is first-order.
On the $\tilde{\varepsilon}<0$ side, the wing is isolated and the
transition is a symmetry-preserving meta-nematic one. On the other
hand, on the $\tilde{\varepsilon}>0$ side, the wing is connected
to a larger surface (in red) that signals a second-order transition.
Regardless of the character of the transition in the $\tilde{\varepsilon}>0$
region, it is associated with the spontaneous breaking of an in-plane
twofold rotational symmetry, which is manifested as a piezoelectric
phase in the case of twisted moiré systems.

It is interesting to compare this $\left(\tilde{\varepsilon},a,T\right)$
phase diagram with that expected for an Ising-nematic order parameter,
as realized in tetragonal lattices. As shown in the right panel of
Fig. \ref{fig:comparative_phase_diagrams}, in the Ising-nematic case
the system is generically expected to undergo a second-order transition
only along the $\tilde{\varepsilon}=0$ plane. Any strain applied
along the directions of the nematic director completely smears the
Ising-nematic phase transition. This is what renders it difficult
to unambiguously distinguish spontaneous Ising-nematic order from
strain-induced anisotropies in experimental settings, where residual
strain is invariably present. In contrast, meta-nematic and piezoelectric
transitions persist for a wide range of strain values in the case
of Potts-nematic order. Experimental observation of these effects
would provide direct evidence for spontaneous Potts-nematic order.

The wings in the $\left(\tilde{\varepsilon},a,T\right)$ phase diagram
of the Potts-nematic order, bounded by tricritical points and critical
end-points, are reminiscent of the wings generally expected in the
$\left(h,a,T\right)$ phase diagram of a metallic (Heisenberg) ferromagnet,
which is schematically shown in the middle panel of Fig. \ref{fig:comparative_phase_diagrams}
(see Refs. \citep{Belitz2005,Brando2016}). Note that here $h$ denotes
a magnetic field. At first sight, this analogy may seem unsurprising,
since both $\tilde{\varepsilon}$ and $h$ act as conjugate fields
to the nematic and ferromagnetic order parameters, respectively. However,
there are crucial qualitative differences. First, because the nematic
order parameter is 3-state Potts-like rather than continuous, there
is a fundamental asymmetry between the effects of compressive strain
and tensile strain, whereas in the ferromagnetic case the phase diagram
is symmetric with respect to the sign of the magnetic field. Second,
the mechanisms behind the first-order $T=0$ transitions are completely
different in the two cases. In the Potts-nematic case, the first-order
nature of the quantum phase transition is an intrinsic property of
the bosonic model, as it is a direct consequence of it being above
its upper critical dimension. In contrast, in the metallic ferromagnetic
case, the $T=0$ transition is rendered first-order due to the coupling
between the ferromagnetic Goldstone modes and the gapless electron-hole
excitations of the metal \citep{Belitz1999,Chubukov2004,Maslov2009}. 

\section{Conclusions \label{sec:Conclusions}}

In this paper, we used a phenomenological approach to establish the
$\left(\tilde{\varepsilon},a,T\right)$ phase diagram of the electronic
3-state Potts-nematic model in the presence of uniaxial strain applied
along one of the high-symmetry directions of a lattice that possesses
out-of-plane threefold rotational symmetry and in-plane twofold rotational
symmetry. While $a$ is a non-thermal tuning parameter, such as doping,
the parameter $\tilde{\varepsilon}$ is linearly proportional to the
applied strain. Whether compressive or tensile strain gives $\tilde{\varepsilon}<0$
or $\tilde{\varepsilon}>0$ depends on the signs of the cubic nematic
coefficient and the nemato-elastic coupling. At zero temperature and
zero strain, the mean-field approach is justified due to the reduced
upper critical dimension of the 3-state Potts model, $d_{u}^{\mathrm{Potts}}\lesssim3$.
Then, because the Potts-nematic action contains a cubic invariant,
there is no Potts-nematic QCP, but rather a first-order Potts-nematic
quantum phase transition. Upon increasing the temperature, but keeping
the strain zero, the mean-field solution ceases to be valid in the
case of a 2D lattice, and the Potts-nematic transition becomes second-order.
Thus, a classical tricritical nematic point is generally expected
in an unstrained 2D system, whereas in a 3D system the Potts-nematic
transition should be always first-order.

Notwithstanding the absence of a QCP in an unstrained 2D or 3D system,
application of strain can tune the system across a a meta-nematic
QCEP, for $\tilde{\varepsilon}<0$, and a QTCP followed by a line
of QCPs for $\tilde{\varepsilon}>0$. The former transition is symmetry-preserving,
whereas the latter spontaneously breaks the in-plane twofold rotational
symmetry of the lattice. Note that a non-zero $\tilde{\varepsilon}$
explicitly breaks the out-of-plane threefold rotational symmetry.
In lattices with $\mathrm{D_{3}}$ and $\mathrm{D_{6}}$ point-group
symmetries, which is the case for instance of twisted bilayer graphene
(TBG) and twisted double-bilayer graphene (TDBG), the transition in
the $\tilde{\varepsilon}>0$ side of the phase diagram leads to the
emergence of a non-zero electric polarization, resulting in what we
dubbed a piezoelectric phase -- since the ferroelectric order requires
the presence of external strain. Connecting the phase diagrams at
zero strain and at zero temperature, we proposed the $\left(\tilde{\varepsilon},a,T\right)$
phase diagram for a 2D Potts-nematic system shown in Fig. \ref{fig_phase_diagram_T}.
One of its key features is the existence of two first-order transition
wings, bounded by a line of tricritical points in the $\tilde{\varepsilon}>0$
side of the phase diagram, and by a line of critical end-points in
the $\tilde{\varepsilon}<0$ side. While the latter wing is isolated,
the former is connected to an open surface of second-order phase transitions
towards the piezoelectric phase. The recent observation of electronic
nematicity in TBG \citep{Kerelsky2019,Jiang2019,Choi2019,Cao2021},
TDBG \citep{Rubio2022}, and twisted trilayer graphene \citep{Zhang_Vafek2022},
which are 2D materials, indicate not only that the phase diagram of
Fig. \ref{fig_phase_diagram_T} may be realized in moiré superlattices,
but also that strain can be used to move the system towards nematic
quantum criticality. Note that this is a different mechanism from
that proposed in Ref. \citep{Parker2021} to strain-tune TBG across
a quantum phase transition.

It is interesting to contrast the results obtained here for the Potts-nematic
phase with those for an Ising-nematic phase, which is realized in
lattices with fourfold rotational symmetry. In the Ising-nematic case,
``longitudinal'' strain applied along either of the two allowed
nematic director directions smears the second-order phase transition.
However, ``transverse'' strain applied along the other two high-symmetry
directions not encompassed by the nematic director can tune the system
towards an Ising-nematic QCP \citep{Maharaj2017}. This offers an
interesting insight into why strain is capable of tuning the system
across a Potts-nematic QCP. For the lattices considered here, the
nematic director can point along any of the high-symmetry lattice
directions. Thus, uniaxial strain applied along these directions can
have either a ``longitudinal'' or a ``transverse'' character,
depending on whether strain is compressive or tensile. This asymmetry
between compressive and tensile strain traces back to the well-understood
inequivalence between positive and negative conjugate fields in the
mean-field solution of the 3-state Potts model \citep{Straley1973ThreestatePM,blankschtein1980effects}. 

The Potts-nematic QCPs that emerge in the presence of strain behave
analogously to an Ising-nematic QCP in the absence of strain. In both
the $\tilde{\varepsilon}>0$ and $\tilde{\varepsilon}<0$ sides of
the phase diagram, the two-component Potts-nematic order parameter
is effectively reduced to a single-component one by the external strain,
either because the nematic amplitude jumps between two non-zero values
while the nematic director angle is pinned by the strain ($\tilde{\varepsilon}<0$),
or because the nematic director unlocks from the strain direction
by rotating along the clockwise or the counterclockwise direction
($\tilde{\varepsilon}>0$). In fact, under these conditions, the Potts-nematic
electronic form factor reduces to the well-known ``$\mathrm{B_{1g}}$''
Ising-nematic form factor for $\tilde{\varepsilon}<0$ and ``$\mathrm{B_{2g}}$''
Ising-nematic form factor for $\tilde{\varepsilon}>0$. Consequently,
while the QCPs on the two sides of the phase diagram have the same
Hertz-Millis dynamical critical exponent $z=3$ except for a few cold
spots, for which $z=2$, these cold spots are at different locations
depending on the sign of $\tilde{\varepsilon}$. More broadly, the
strain-induced Potts-nematic QCPs should support the same phenomena
expected for the Ising-nematic QCP, such as superconductivity and
non-Fermi liquid behavior \citep{Metlitski2010,Metlitski2015,Schattner2016,Lederer2017,Klein2018,SSLee2018}.

Our results provide valuable criteria to experimentally identify intrinsic
Potts-nematic order and distinguish it from extrinsic effects via
a controlled application of uniaxial strain. Observation of the characteristic
multi-loop hysteresis curves shown in Fig. \ref{fig:hysteresis} would
be a direct confirmation not only of long-range nematic order, but
also of the Potts-like character of the order parameter. Experimentally,
$\phi_{1}$ can be probed via resistivity anisotropy measurements
similarly to those carried out in the pnictides \citep{Chu2012}.
In this regard, as pointed out in Ref. \citep{Vafek2022}, the geometry
used to measured the resistivity plays an important role in extracting
the anisotropic component of the resistivity tensor (see also Ref.
\citep{XiaoyuWang2022}). Moreover, the observation of a piezoelectric
effect in twisted moiré systems that only emerges for one type of
strain (compressive or tensile) would provide unambiguous evidence
for an intrinsic Potts-nematic instability. Interestingly, ferroelectricity
has been recently observed in a moiré heterostructure \citep{Zheng2020}.
While in this paper we focused only on externally-applied uniform
strain, any crystalline system will invariably be subjected to internal
random strain \citep{Carlson2006,Carlson2011,Meese2022}. Given the
non-trivial impact of uniform strain on the Potts-nematicity, it will
be interesting for future studies to shed light on the properties
of the 3-state Potts-nematic model in the presence of both random
strain and uniform strain.
\begin{acknowledgments}
We thank H. Ochoa and J. Venderbos for fruitful discussions. This
work was supported by the U. S. Department of Energy, Office of Science,
Basic Energy Sciences, Materials Sciences and Engineering Division,
under Award No. DE-SC0020045.
\end{acknowledgments}

\bibliographystyle{apsrev4-2}
\bibliography{references}

\begin{thebibliography}{88}%
\makeatletter
\providecommand \@ifxundefined [1]{%
 \@ifx{#1\undefined}
}%
\providecommand \@ifnum [1]{%
 \ifnum #1\expandafter \@firstoftwo
 \else \expandafter \@secondoftwo
 \fi
}%
\providecommand \@ifx [1]{%
 \ifx #1\expandafter \@firstoftwo
 \else \expandafter \@secondoftwo
 \fi
}%
\providecommand \natexlab [1]{#1}%
\providecommand \enquote  [1]{``#1''}%
\providecommand \bibnamefont  [1]{#1}%
\providecommand \bibfnamefont [1]{#1}%
\providecommand \citenamefont [1]{#1}%
\providecommand \href@noop [0]{\@secondoftwo}%
\providecommand \href [0]{\begingroup \@sanitize@url \@href}%
\providecommand \@href[1]{\@@startlink{#1}\@@href}%
\providecommand \@@href[1]{\endgroup#1\@@endlink}%
\providecommand \@sanitize@url [0]{\catcode `\\12\catcode `\$12\catcode
  `\&12\catcode `\#12\catcode `\^12\catcode `\_12\catcode `\%12\relax}%
\providecommand \@@startlink[1]{}%
\providecommand \@@endlink[0]{}%
\providecommand \url  [0]{\begingroup\@sanitize@url \@url }%
\providecommand \@url [1]{\endgroup\@href {#1}{\urlprefix }}%
\providecommand \urlprefix  [0]{URL }%
\providecommand \Eprint [0]{\href }%
\providecommand \doibase [0]{https://doi.org/}%
\providecommand \selectlanguage [0]{\@gobble}%
\providecommand \bibinfo  [0]{\@secondoftwo}%
\providecommand \bibfield  [0]{\@secondoftwo}%
\providecommand \translation [1]{[#1]}%
\providecommand \BibitemOpen [0]{}%
\providecommand \bibitemStop [0]{}%
\providecommand \bibitemNoStop [0]{.\EOS\space}%
\providecommand \EOS [0]{\spacefactor3000\relax}%
\providecommand \BibitemShut  [1]{\csname bibitem#1\endcsname}%
\let\auto@bib@innerbib\@empty
\bibitem [{\citenamefont {Kivelson}\ \emph {et~al.}(1998)\citenamefont
  {Kivelson}, \citenamefont {Fradkin},\ and\ \citenamefont
  {Emery}}]{Kivelson1998}%
  \BibitemOpen
  \bibfield  {author} {\bibinfo {author} {\bibfnamefont {S.~A.}\ \bibnamefont
  {Kivelson}}, \bibinfo {author} {\bibfnamefont {E.}~\bibnamefont {Fradkin}},\
  and\ \bibinfo {author} {\bibfnamefont {V.~J.}\ \bibnamefont {Emery}},\
  }\href@noop {} {\bibfield  {journal} {\bibinfo  {journal} {Nature}\ }\textbf
  {\bibinfo {volume} {393}},\ \bibinfo {pages} {550} (\bibinfo {year}
  {1998})}\BibitemShut {NoStop}%
\bibitem [{\citenamefont {Kivelson}\ \emph {et~al.}(2003)\citenamefont
  {Kivelson}, \citenamefont {Bindloss}, \citenamefont {Fradkin}, \citenamefont
  {Oganesyan}, \citenamefont {Tranquada}, \citenamefont {Kapitulnik},\ and\
  \citenamefont {Howald}}]{Kivelson2003}%
  \BibitemOpen
  \bibfield  {author} {\bibinfo {author} {\bibfnamefont {S.~A.}\ \bibnamefont
  {Kivelson}}, \bibinfo {author} {\bibfnamefont {I.~P.}\ \bibnamefont
  {Bindloss}}, \bibinfo {author} {\bibfnamefont {E.}~\bibnamefont {Fradkin}},
  \bibinfo {author} {\bibfnamefont {V.}~\bibnamefont {Oganesyan}}, \bibinfo
  {author} {\bibfnamefont {J.~M.}\ \bibnamefont {Tranquada}}, \bibinfo {author}
  {\bibfnamefont {A.}~\bibnamefont {Kapitulnik}},\ and\ \bibinfo {author}
  {\bibfnamefont {C.}~\bibnamefont {Howald}},\ }\href
  {https://doi.org/10.1103/RevModPhys.75.1201} {\bibfield  {journal} {\bibinfo
  {journal} {Rev. Mod. Phys.}\ }\textbf {\bibinfo {volume} {75}},\ \bibinfo
  {pages} {1201} (\bibinfo {year} {2003})}\BibitemShut {NoStop}%
\bibitem [{\citenamefont {Hinkov}\ \emph {et~al.}(2008)\citenamefont {Hinkov},
  \citenamefont {Haug}, \citenamefont {Fauqu{\'e}}, \citenamefont {Bourges},
  \citenamefont {Sidis}, \citenamefont {Ivanov}, \citenamefont {Bernhard},
  \citenamefont {Lin},\ and\ \citenamefont {Keimer}}]{Hinkov2008}%
  \BibitemOpen
  \bibfield  {author} {\bibinfo {author} {\bibfnamefont {V.}~\bibnamefont
  {Hinkov}}, \bibinfo {author} {\bibfnamefont {D.}~\bibnamefont {Haug}},
  \bibinfo {author} {\bibfnamefont {B.}~\bibnamefont {Fauqu{\'e}}}, \bibinfo
  {author} {\bibfnamefont {P.}~\bibnamefont {Bourges}}, \bibinfo {author}
  {\bibfnamefont {Y.}~\bibnamefont {Sidis}}, \bibinfo {author} {\bibfnamefont
  {A.}~\bibnamefont {Ivanov}}, \bibinfo {author} {\bibfnamefont
  {C.}~\bibnamefont {Bernhard}}, \bibinfo {author} {\bibfnamefont
  {C.}~\bibnamefont {Lin}},\ and\ \bibinfo {author} {\bibfnamefont
  {B.}~\bibnamefont {Keimer}},\ }\href@noop {} {\bibfield  {journal} {\bibinfo
  {journal} {Science}\ }\textbf {\bibinfo {volume} {319}},\ \bibinfo {pages}
  {597} (\bibinfo {year} {2008})}\BibitemShut {NoStop}%
\bibitem [{\citenamefont {Vojta}(2009)}]{Vojta2009}%
  \BibitemOpen
  \bibfield  {author} {\bibinfo {author} {\bibfnamefont {M.}~\bibnamefont
  {Vojta}},\ }\href@noop {} {\bibfield  {journal} {\bibinfo  {journal}
  {Advances in Physics}\ }\textbf {\bibinfo {volume} {58}},\ \bibinfo {pages}
  {699} (\bibinfo {year} {2009})}\BibitemShut {NoStop}%
\bibitem [{\citenamefont {Okazaki}\ \emph {et~al.}(2011)\citenamefont
  {Okazaki}, \citenamefont {Shibauchi}, \citenamefont {Shi}, \citenamefont
  {Haga}, \citenamefont {Matsuda}, \citenamefont {Yamamoto}, \citenamefont
  {Onuki}, \citenamefont {Ikeda},\ and\ \citenamefont {Matsuda}}]{Okazaki2011}%
  \BibitemOpen
  \bibfield  {author} {\bibinfo {author} {\bibfnamefont {R.}~\bibnamefont
  {Okazaki}}, \bibinfo {author} {\bibfnamefont {T.}~\bibnamefont {Shibauchi}},
  \bibinfo {author} {\bibfnamefont {H.}~\bibnamefont {Shi}}, \bibinfo {author}
  {\bibfnamefont {Y.}~\bibnamefont {Haga}}, \bibinfo {author} {\bibfnamefont
  {T.}~\bibnamefont {Matsuda}}, \bibinfo {author} {\bibfnamefont
  {E.}~\bibnamefont {Yamamoto}}, \bibinfo {author} {\bibfnamefont
  {Y.}~\bibnamefont {Onuki}}, \bibinfo {author} {\bibfnamefont
  {H.}~\bibnamefont {Ikeda}},\ and\ \bibinfo {author} {\bibfnamefont
  {Y.}~\bibnamefont {Matsuda}},\ }\href@noop {} {\bibfield  {journal} {\bibinfo
   {journal} {Science}\ }\textbf {\bibinfo {volume} {331}},\ \bibinfo {pages}
  {439} (\bibinfo {year} {2011})}\BibitemShut {NoStop}%
\bibitem [{\citenamefont {Ronning}\ \emph {et~al.}(2017)\citenamefont
  {Ronning}, \citenamefont {Helm}, \citenamefont {Shirer}, \citenamefont
  {Bachmann}, \citenamefont {Balicas}, \citenamefont {Chan}, \citenamefont
  {Ramshaw}, \citenamefont {Mcdonald}, \citenamefont {Balakirev}, \citenamefont
  {Jaime} \emph {et~al.}}]{Ronning2017}%
  \BibitemOpen
  \bibfield  {author} {\bibinfo {author} {\bibfnamefont {F.}~\bibnamefont
  {Ronning}}, \bibinfo {author} {\bibfnamefont {T.}~\bibnamefont {Helm}},
  \bibinfo {author} {\bibfnamefont {K.}~\bibnamefont {Shirer}}, \bibinfo
  {author} {\bibfnamefont {M.}~\bibnamefont {Bachmann}}, \bibinfo {author}
  {\bibfnamefont {L.}~\bibnamefont {Balicas}}, \bibinfo {author} {\bibfnamefont
  {M.~K.}\ \bibnamefont {Chan}}, \bibinfo {author} {\bibfnamefont
  {B.}~\bibnamefont {Ramshaw}}, \bibinfo {author} {\bibfnamefont {R.~D.}\
  \bibnamefont {Mcdonald}}, \bibinfo {author} {\bibfnamefont {F.~F.}\
  \bibnamefont {Balakirev}}, \bibinfo {author} {\bibfnamefont {M.}~\bibnamefont
  {Jaime}}, \emph {et~al.},\ }\href@noop {} {\bibfield  {journal} {\bibinfo
  {journal} {Nature}\ }\textbf {\bibinfo {volume} {548}},\ \bibinfo {pages}
  {313} (\bibinfo {year} {2017})}\BibitemShut {NoStop}%
\bibitem [{\citenamefont {Seo}\ \emph {et~al.}(2020)\citenamefont {Seo},
  \citenamefont {Wang}, \citenamefont {Thomas}, \citenamefont {Rahn},
  \citenamefont {Carmo}, \citenamefont {Ronning}, \citenamefont {Bauer},
  \citenamefont {dos Reis}, \citenamefont {Janoschek}, \citenamefont
  {Thompson}, \citenamefont {Fernandes},\ and\ \citenamefont {Rosa}}]{Seo2020}%
  \BibitemOpen
  \bibfield  {author} {\bibinfo {author} {\bibfnamefont {S.}~\bibnamefont
  {Seo}}, \bibinfo {author} {\bibfnamefont {X.}~\bibnamefont {Wang}}, \bibinfo
  {author} {\bibfnamefont {S.~M.}\ \bibnamefont {Thomas}}, \bibinfo {author}
  {\bibfnamefont {M.~C.}\ \bibnamefont {Rahn}}, \bibinfo {author}
  {\bibfnamefont {D.}~\bibnamefont {Carmo}}, \bibinfo {author} {\bibfnamefont
  {F.}~\bibnamefont {Ronning}}, \bibinfo {author} {\bibfnamefont {E.~D.}\
  \bibnamefont {Bauer}}, \bibinfo {author} {\bibfnamefont {R.~D.}\ \bibnamefont
  {dos Reis}}, \bibinfo {author} {\bibfnamefont {M.}~\bibnamefont {Janoschek}},
  \bibinfo {author} {\bibfnamefont {J.~D.}\ \bibnamefont {Thompson}}, \bibinfo
  {author} {\bibfnamefont {R.~M.}\ \bibnamefont {Fernandes}},\ and\ \bibinfo
  {author} {\bibfnamefont {P.~F.~S.}\ \bibnamefont {Rosa}},\ }\href
  {https://doi.org/10.1103/PhysRevX.10.011035} {\bibfield  {journal} {\bibinfo
  {journal} {Phys. Rev. X}\ }\textbf {\bibinfo {volume} {10}},\ \bibinfo
  {pages} {011035} (\bibinfo {year} {2020})}\BibitemShut {NoStop}%
\bibitem [{\citenamefont {Chu}\ \emph {et~al.}(2012)\citenamefont {Chu},
  \citenamefont {Kuo}, \citenamefont {Analytis},\ and\ \citenamefont
  {Fisher}}]{Chu2012}%
  \BibitemOpen
  \bibfield  {author} {\bibinfo {author} {\bibfnamefont {J.-H.}\ \bibnamefont
  {Chu}}, \bibinfo {author} {\bibfnamefont {H.-H.}\ \bibnamefont {Kuo}},
  \bibinfo {author} {\bibfnamefont {J.~G.}\ \bibnamefont {Analytis}},\ and\
  \bibinfo {author} {\bibfnamefont {I.~R.}\ \bibnamefont {Fisher}},\
  }\href@noop {} {\bibfield  {journal} {\bibinfo  {journal} {Science}\ }\textbf
  {\bibinfo {volume} {337}},\ \bibinfo {pages} {710} (\bibinfo {year}
  {2012})}\BibitemShut {NoStop}%
\bibitem [{\citenamefont {Fernandes}\ \emph {et~al.}(2014)\citenamefont
  {Fernandes}, \citenamefont {Chubukov},\ and\ \citenamefont
  {Schmalian}}]{Fernandes2014}%
  \BibitemOpen
  \bibfield  {author} {\bibinfo {author} {\bibfnamefont {R.}~\bibnamefont
  {Fernandes}}, \bibinfo {author} {\bibfnamefont {A.}~\bibnamefont
  {Chubukov}},\ and\ \bibinfo {author} {\bibfnamefont {J.}~\bibnamefont
  {Schmalian}},\ }\href@noop {} {\bibfield  {journal} {\bibinfo  {journal}
  {Nature physics}\ }\textbf {\bibinfo {volume} {10}},\ \bibinfo {pages} {97}
  (\bibinfo {year} {2014})}\BibitemShut {NoStop}%
\bibitem [{\citenamefont {B{\"o}hmer}\ and\ \citenamefont
  {Meingast}(2016)}]{Bohmer2016}%
  \BibitemOpen
  \bibfield  {author} {\bibinfo {author} {\bibfnamefont {A.~E.}\ \bibnamefont
  {B{\"o}hmer}}\ and\ \bibinfo {author} {\bibfnamefont {C.}~\bibnamefont
  {Meingast}},\ }\href@noop {} {\bibfield  {journal} {\bibinfo  {journal}
  {Comptes Rendus Physique}\ }\textbf {\bibinfo {volume} {17}},\ \bibinfo
  {pages} {90} (\bibinfo {year} {2016})}\BibitemShut {NoStop}%
\bibitem [{\citenamefont {B{\"o}hmer}\ \emph {et~al.}(2022)\citenamefont
  {B{\"o}hmer}, \citenamefont {Chu}, \citenamefont {Lederer},\ and\
  \citenamefont {Yi}}]{Bohmer2022}%
  \BibitemOpen
  \bibfield  {author} {\bibinfo {author} {\bibfnamefont {A.~E.}\ \bibnamefont
  {B{\"o}hmer}}, \bibinfo {author} {\bibfnamefont {J.-H.}\ \bibnamefont {Chu}},
  \bibinfo {author} {\bibfnamefont {S.}~\bibnamefont {Lederer}},\ and\ \bibinfo
  {author} {\bibfnamefont {M.}~\bibnamefont {Yi}},\ }\href@noop {} {\bibfield
  {journal} {\bibinfo  {journal} {Nature Physics}\ }\textbf {\bibinfo {volume}
  {18}},\ \bibinfo {pages} {1412} (\bibinfo {year} {2022})}\BibitemShut
  {NoStop}%
\bibitem [{\citenamefont {Fradkin}\ \emph {et~al.}(2010)\citenamefont
  {Fradkin}, \citenamefont {Kivelson}, \citenamefont {Lawler}, \citenamefont
  {Eisenstein},\ and\ \citenamefont {Mackenzie}}]{Fradkin2010}%
  \BibitemOpen
  \bibfield  {author} {\bibinfo {author} {\bibfnamefont {E.}~\bibnamefont
  {Fradkin}}, \bibinfo {author} {\bibfnamefont {S.~A.}\ \bibnamefont
  {Kivelson}}, \bibinfo {author} {\bibfnamefont {M.~J.}\ \bibnamefont
  {Lawler}}, \bibinfo {author} {\bibfnamefont {J.~P.}\ \bibnamefont
  {Eisenstein}},\ and\ \bibinfo {author} {\bibfnamefont {A.~P.}\ \bibnamefont
  {Mackenzie}},\ }\href@noop {} {\bibfield  {journal} {\bibinfo  {journal}
  {Annu. Rev. Condens. Matter Phys.}\ }\textbf {\bibinfo {volume} {1}},\
  \bibinfo {pages} {153} (\bibinfo {year} {2010})}\BibitemShut {NoStop}%
\bibitem [{\citenamefont {Carlson}\ \emph {et~al.}(2006)\citenamefont
  {Carlson}, \citenamefont {Dahmen}, \citenamefont {Fradkin},\ and\
  \citenamefont {Kivelson}}]{Carlson2006}%
  \BibitemOpen
  \bibfield  {author} {\bibinfo {author} {\bibfnamefont {E.~W.}\ \bibnamefont
  {Carlson}}, \bibinfo {author} {\bibfnamefont {K.~A.}\ \bibnamefont {Dahmen}},
  \bibinfo {author} {\bibfnamefont {E.}~\bibnamefont {Fradkin}},\ and\ \bibinfo
  {author} {\bibfnamefont {S.~A.}\ \bibnamefont {Kivelson}},\ }\href
  {https://doi.org/10.1103/PhysRevLett.96.097003} {\bibfield  {journal}
  {\bibinfo  {journal} {Phys. Rev. Lett.}\ }\textbf {\bibinfo {volume} {96}},\
  \bibinfo {pages} {097003} (\bibinfo {year} {2006})}\BibitemShut {NoStop}%
\bibitem [{\citenamefont {Carlson}\ and\ \citenamefont
  {Dahmen}(2011)}]{Carlson2011}%
  \BibitemOpen
  \bibfield  {author} {\bibinfo {author} {\bibfnamefont {E.}~\bibnamefont
  {Carlson}}\ and\ \bibinfo {author} {\bibfnamefont {K.}~\bibnamefont
  {Dahmen}},\ }\href@noop {} {\bibfield  {journal} {\bibinfo  {journal} {Nature
  communications}\ }\textbf {\bibinfo {volume} {2}},\ \bibinfo {pages} {379}
  (\bibinfo {year} {2011})}\BibitemShut {NoStop}%
\bibitem [{\citenamefont {Meese}\ \emph {et~al.}(2022)\citenamefont {Meese},
  \citenamefont {Vojta},\ and\ \citenamefont {Fernandes}}]{Meese2022}%
  \BibitemOpen
  \bibfield  {author} {\bibinfo {author} {\bibfnamefont {W.~J.}\ \bibnamefont
  {Meese}}, \bibinfo {author} {\bibfnamefont {T.}~\bibnamefont {Vojta}},\ and\
  \bibinfo {author} {\bibfnamefont {R.~M.}\ \bibnamefont {Fernandes}},\ }\href
  {https://doi.org/10.1103/PhysRevB.106.115134} {\bibfield  {journal} {\bibinfo
   {journal} {Phys. Rev. B}\ }\textbf {\bibinfo {volume} {106}},\ \bibinfo
  {pages} {115134} (\bibinfo {year} {2022})}\BibitemShut {NoStop}%
\bibitem [{\citenamefont {Wang}\ \emph
  {et~al.}(2022{\natexlab{a}})\citenamefont {Wang}, \citenamefont {Finney},
  \citenamefont {Sharpe}, \citenamefont {Rodenbach}, \citenamefont {Hsueh},
  \citenamefont {Watanabe}, \citenamefont {Taniguchi}, \citenamefont {Kastner},
  \citenamefont {Vafek},\ and\ \citenamefont
  {Goldhaber-Gordon}}]{XiaoyuWang2022}%
  \BibitemOpen
  \bibfield  {author} {\bibinfo {author} {\bibfnamefont {X.}~\bibnamefont
  {Wang}}, \bibinfo {author} {\bibfnamefont {J.}~\bibnamefont {Finney}},
  \bibinfo {author} {\bibfnamefont {A.~L.}\ \bibnamefont {Sharpe}}, \bibinfo
  {author} {\bibfnamefont {L.~K.}\ \bibnamefont {Rodenbach}}, \bibinfo {author}
  {\bibfnamefont {C.~L.}\ \bibnamefont {Hsueh}}, \bibinfo {author}
  {\bibfnamefont {K.}~\bibnamefont {Watanabe}}, \bibinfo {author}
  {\bibfnamefont {T.}~\bibnamefont {Taniguchi}}, \bibinfo {author}
  {\bibfnamefont {M.}~\bibnamefont {Kastner}}, \bibinfo {author} {\bibfnamefont
  {O.}~\bibnamefont {Vafek}},\ and\ \bibinfo {author} {\bibfnamefont
  {D.}~\bibnamefont {Goldhaber-Gordon}},\ }\href@noop {} {\bibfield  {journal}
  {\bibinfo  {journal} {arXiv:2209.08204}\ } (\bibinfo {year}
  {2022}{\natexlab{a}})}\BibitemShut {NoStop}%
\bibitem [{\citenamefont {Karahasanovic}\ and\ \citenamefont
  {Schmalian}(2016)}]{Schmalian2016}%
  \BibitemOpen
  \bibfield  {author} {\bibinfo {author} {\bibfnamefont {U.}~\bibnamefont
  {Karahasanovic}}\ and\ \bibinfo {author} {\bibfnamefont {J.}~\bibnamefont
  {Schmalian}},\ }\href {https://doi.org/10.1103/PhysRevB.93.064520} {\bibfield
   {journal} {\bibinfo  {journal} {Phys. Rev. B}\ }\textbf {\bibinfo {volume}
  {93}},\ \bibinfo {pages} {064520} (\bibinfo {year} {2016})}\BibitemShut
  {NoStop}%
\bibitem [{\citenamefont {Paul}\ and\ \citenamefont {Garst}(2017)}]{Paul2017}%
  \BibitemOpen
  \bibfield  {author} {\bibinfo {author} {\bibfnamefont {I.}~\bibnamefont
  {Paul}}\ and\ \bibinfo {author} {\bibfnamefont {M.}~\bibnamefont {Garst}},\
  }\href {https://doi.org/10.1103/PhysRevLett.118.227601} {\bibfield  {journal}
  {\bibinfo  {journal} {Phys. Rev. Lett.}\ }\textbf {\bibinfo {volume} {118}},\
  \bibinfo {pages} {227601} (\bibinfo {year} {2017})}\BibitemShut {NoStop}%
\bibitem [{\citenamefont {de~Carvalho}\ and\ \citenamefont
  {Fernandes}(2019)}]{Carvalho2019}%
  \BibitemOpen
  \bibfield  {author} {\bibinfo {author} {\bibfnamefont {V.~S.}\ \bibnamefont
  {de~Carvalho}}\ and\ \bibinfo {author} {\bibfnamefont {R.~M.}\ \bibnamefont
  {Fernandes}},\ }\href {https://doi.org/10.1103/PhysRevB.100.115103}
  {\bibfield  {journal} {\bibinfo  {journal} {Phys. Rev. B}\ }\textbf {\bibinfo
  {volume} {100}},\ \bibinfo {pages} {115103} (\bibinfo {year}
  {2019})}\BibitemShut {NoStop}%
\bibitem [{\citenamefont {Massat}\ \emph {et~al.}(2022)\citenamefont {Massat},
  \citenamefont {Wen}, \citenamefont {Jiang}, \citenamefont {Hristov},
  \citenamefont {Liu}, \citenamefont {Smaha}, \citenamefont {Feigelson},
  \citenamefont {Lee}, \citenamefont {Fernandes},\ and\ \citenamefont
  {Fisher}}]{Massat2022}%
  \BibitemOpen
  \bibfield  {author} {\bibinfo {author} {\bibfnamefont {P.}~\bibnamefont
  {Massat}}, \bibinfo {author} {\bibfnamefont {J.}~\bibnamefont {Wen}},
  \bibinfo {author} {\bibfnamefont {J.~M.}\ \bibnamefont {Jiang}}, \bibinfo
  {author} {\bibfnamefont {A.~T.}\ \bibnamefont {Hristov}}, \bibinfo {author}
  {\bibfnamefont {Y.}~\bibnamefont {Liu}}, \bibinfo {author} {\bibfnamefont
  {R.~W.}\ \bibnamefont {Smaha}}, \bibinfo {author} {\bibfnamefont {R.~S.}\
  \bibnamefont {Feigelson}}, \bibinfo {author} {\bibfnamefont {Y.~S.}\
  \bibnamefont {Lee}}, \bibinfo {author} {\bibfnamefont {R.~M.}\ \bibnamefont
  {Fernandes}},\ and\ \bibinfo {author} {\bibfnamefont {I.~R.}\ \bibnamefont
  {Fisher}},\ }\href@noop {} {\bibfield  {journal} {\bibinfo  {journal}
  {Proceedings of the National Academy of Sciences}\ }\textbf {\bibinfo
  {volume} {119}},\ \bibinfo {pages} {e2119942119} (\bibinfo {year}
  {2022})}\BibitemShut {NoStop}%
\bibitem [{\citenamefont {Feldman}\ \emph {et~al.}(2016)\citenamefont
  {Feldman}, \citenamefont {Randeria}, \citenamefont {Gyenis}, \citenamefont
  {Wu}, \citenamefont {Ji}, \citenamefont {Cava}, \citenamefont {MacDonald},\
  and\ \citenamefont {Yazdani}}]{Feldman2016}%
  \BibitemOpen
  \bibfield  {author} {\bibinfo {author} {\bibfnamefont {B.~E.}\ \bibnamefont
  {Feldman}}, \bibinfo {author} {\bibfnamefont {M.~T.}\ \bibnamefont
  {Randeria}}, \bibinfo {author} {\bibfnamefont {A.}~\bibnamefont {Gyenis}},
  \bibinfo {author} {\bibfnamefont {F.}~\bibnamefont {Wu}}, \bibinfo {author}
  {\bibfnamefont {H.}~\bibnamefont {Ji}}, \bibinfo {author} {\bibfnamefont
  {R.~J.}\ \bibnamefont {Cava}}, \bibinfo {author} {\bibfnamefont {A.~H.}\
  \bibnamefont {MacDonald}},\ and\ \bibinfo {author} {\bibfnamefont
  {A.}~\bibnamefont {Yazdani}},\ }\href@noop {} {\bibfield  {journal} {\bibinfo
   {journal} {Science}\ }\textbf {\bibinfo {volume} {354}},\ \bibinfo {pages}
  {316} (\bibinfo {year} {2016})}\BibitemShut {NoStop}%
\bibitem [{\citenamefont {Sun}\ \emph {et~al.}(2019)\citenamefont {Sun},
  \citenamefont {Kittaka}, \citenamefont {Sakakibara}, \citenamefont {Machida},
  \citenamefont {Wang}, \citenamefont {Wen}, \citenamefont {Xing},
  \citenamefont {Shi},\ and\ \citenamefont {Tamegai}}]{Tamegai2019}%
  \BibitemOpen
  \bibfield  {author} {\bibinfo {author} {\bibfnamefont {Y.}~\bibnamefont
  {Sun}}, \bibinfo {author} {\bibfnamefont {S.}~\bibnamefont {Kittaka}},
  \bibinfo {author} {\bibfnamefont {T.}~\bibnamefont {Sakakibara}}, \bibinfo
  {author} {\bibfnamefont {K.}~\bibnamefont {Machida}}, \bibinfo {author}
  {\bibfnamefont {J.}~\bibnamefont {Wang}}, \bibinfo {author} {\bibfnamefont
  {J.}~\bibnamefont {Wen}}, \bibinfo {author} {\bibfnamefont {X.}~\bibnamefont
  {Xing}}, \bibinfo {author} {\bibfnamefont {Z.}~\bibnamefont {Shi}},\ and\
  \bibinfo {author} {\bibfnamefont {T.}~\bibnamefont {Tamegai}},\ }\href
  {https://doi.org/10.1103/PhysRevLett.123.027002} {\bibfield  {journal}
  {\bibinfo  {journal} {Phys. Rev. Lett.}\ }\textbf {\bibinfo {volume} {123}},\
  \bibinfo {pages} {027002} (\bibinfo {year} {2019})}\BibitemShut {NoStop}%
\bibitem [{\citenamefont {Cho}\ \emph {et~al.}(2020)\citenamefont {Cho},
  \citenamefont {Shen}, \citenamefont {Lyu}, \citenamefont {Atanov},
  \citenamefont {Chen}, \citenamefont {Lee}, \citenamefont {Hor}, \citenamefont
  {Gawryluk}, \citenamefont {Pomjakushina}, \citenamefont {Bartkowiak} \emph
  {et~al.}}]{Cho2020}%
  \BibitemOpen
  \bibfield  {author} {\bibinfo {author} {\bibfnamefont {C.-w.}\ \bibnamefont
  {Cho}}, \bibinfo {author} {\bibfnamefont {J.}~\bibnamefont {Shen}}, \bibinfo
  {author} {\bibfnamefont {J.}~\bibnamefont {Lyu}}, \bibinfo {author}
  {\bibfnamefont {O.}~\bibnamefont {Atanov}}, \bibinfo {author} {\bibfnamefont
  {Q.}~\bibnamefont {Chen}}, \bibinfo {author} {\bibfnamefont {S.~H.}\
  \bibnamefont {Lee}}, \bibinfo {author} {\bibfnamefont {Y.~S.}\ \bibnamefont
  {Hor}}, \bibinfo {author} {\bibfnamefont {D.~J.}\ \bibnamefont {Gawryluk}},
  \bibinfo {author} {\bibfnamefont {E.}~\bibnamefont {Pomjakushina}}, \bibinfo
  {author} {\bibfnamefont {M.}~\bibnamefont {Bartkowiak}}, \emph {et~al.},\
  }\href@noop {} {\bibfield  {journal} {\bibinfo  {journal} {Nature
  Communications}\ }\textbf {\bibinfo {volume} {11}},\ \bibinfo {pages} {3056}
  (\bibinfo {year} {2020})}\BibitemShut {NoStop}%
\bibitem [{\citenamefont {Little}\ \emph {et~al.}(2020)\citenamefont {Little},
  \citenamefont {Lee}, \citenamefont {John}, \citenamefont {Doyle},
  \citenamefont {Maniv}, \citenamefont {Nair}, \citenamefont {Chen},
  \citenamefont {Rees}, \citenamefont {Venderbos}, \citenamefont {Fernandes}
  \emph {et~al.}}]{Little2020}%
  \BibitemOpen
  \bibfield  {author} {\bibinfo {author} {\bibfnamefont {A.}~\bibnamefont
  {Little}}, \bibinfo {author} {\bibfnamefont {C.}~\bibnamefont {Lee}},
  \bibinfo {author} {\bibfnamefont {C.}~\bibnamefont {John}}, \bibinfo {author}
  {\bibfnamefont {S.}~\bibnamefont {Doyle}}, \bibinfo {author} {\bibfnamefont
  {E.}~\bibnamefont {Maniv}}, \bibinfo {author} {\bibfnamefont {N.~L.}\
  \bibnamefont {Nair}}, \bibinfo {author} {\bibfnamefont {W.}~\bibnamefont
  {Chen}}, \bibinfo {author} {\bibfnamefont {D.}~\bibnamefont {Rees}}, \bibinfo
  {author} {\bibfnamefont {J.~W.}\ \bibnamefont {Venderbos}}, \bibinfo {author}
  {\bibfnamefont {R.~M.}\ \bibnamefont {Fernandes}}, \emph {et~al.},\
  }\href@noop {} {\bibfield  {journal} {\bibinfo  {journal} {Nature Materials}\
  }\textbf {\bibinfo {volume} {19}},\ \bibinfo {pages} {1062} (\bibinfo {year}
  {2020})}\BibitemShut {NoStop}%
\bibitem [{\citenamefont {Jin}\ \emph {et~al.}(2021{\natexlab{a}})\citenamefont
  {Jin}, \citenamefont {Zhang}, \citenamefont {Guo}, \citenamefont {Chen},
  \citenamefont {Zhou},\ and\ \citenamefont {Li}}]{Jin2021}%
  \BibitemOpen
  \bibfield  {author} {\bibinfo {author} {\bibfnamefont {S.}~\bibnamefont
  {Jin}}, \bibinfo {author} {\bibfnamefont {W.}~\bibnamefont {Zhang}}, \bibinfo
  {author} {\bibfnamefont {X.}~\bibnamefont {Guo}}, \bibinfo {author}
  {\bibfnamefont {X.}~\bibnamefont {Chen}}, \bibinfo {author} {\bibfnamefont
  {X.}~\bibnamefont {Zhou}},\ and\ \bibinfo {author} {\bibfnamefont
  {X.}~\bibnamefont {Li}},\ }\href
  {https://doi.org/10.1103/PhysRevLett.126.035301} {\bibfield  {journal}
  {\bibinfo  {journal} {Phys. Rev. Lett.}\ }\textbf {\bibinfo {volume} {126}},\
  \bibinfo {pages} {035301} (\bibinfo {year} {2021}{\natexlab{a}})}\BibitemShut
  {NoStop}%
\bibitem [{\citenamefont {Kerelsky}\ \emph {et~al.}(2019)\citenamefont
  {Kerelsky}, \citenamefont {McGilly}, \citenamefont {Kennes}, \citenamefont
  {Xian}, \citenamefont {Yankowitz}, \citenamefont {Chen}, \citenamefont
  {Watanabe}, \citenamefont {Taniguchi}, \citenamefont {Hone}, \citenamefont
  {Dean} \emph {et~al.}}]{Kerelsky2019}%
  \BibitemOpen
  \bibfield  {author} {\bibinfo {author} {\bibfnamefont {A.}~\bibnamefont
  {Kerelsky}}, \bibinfo {author} {\bibfnamefont {L.~J.}\ \bibnamefont
  {McGilly}}, \bibinfo {author} {\bibfnamefont {D.~M.}\ \bibnamefont {Kennes}},
  \bibinfo {author} {\bibfnamefont {L.}~\bibnamefont {Xian}}, \bibinfo {author}
  {\bibfnamefont {M.}~\bibnamefont {Yankowitz}}, \bibinfo {author}
  {\bibfnamefont {S.}~\bibnamefont {Chen}}, \bibinfo {author} {\bibfnamefont
  {K.}~\bibnamefont {Watanabe}}, \bibinfo {author} {\bibfnamefont
  {T.}~\bibnamefont {Taniguchi}}, \bibinfo {author} {\bibfnamefont
  {J.}~\bibnamefont {Hone}}, \bibinfo {author} {\bibfnamefont {C.}~\bibnamefont
  {Dean}}, \emph {et~al.},\ }\href@noop {} {\bibfield  {journal} {\bibinfo
  {journal} {Nature}\ }\textbf {\bibinfo {volume} {572}},\ \bibinfo {pages}
  {95} (\bibinfo {year} {2019})}\BibitemShut {NoStop}%
\bibitem [{\citenamefont {Jiang}\ \emph {et~al.}(2019)\citenamefont {Jiang},
  \citenamefont {Lai}, \citenamefont {Watanabe}, \citenamefont {Taniguchi},
  \citenamefont {Haule}, \citenamefont {Mao},\ and\ \citenamefont
  {Andrei}}]{Jiang2019}%
  \BibitemOpen
  \bibfield  {author} {\bibinfo {author} {\bibfnamefont {Y.}~\bibnamefont
  {Jiang}}, \bibinfo {author} {\bibfnamefont {X.}~\bibnamefont {Lai}}, \bibinfo
  {author} {\bibfnamefont {K.}~\bibnamefont {Watanabe}}, \bibinfo {author}
  {\bibfnamefont {T.}~\bibnamefont {Taniguchi}}, \bibinfo {author}
  {\bibfnamefont {K.}~\bibnamefont {Haule}}, \bibinfo {author} {\bibfnamefont
  {J.}~\bibnamefont {Mao}},\ and\ \bibinfo {author} {\bibfnamefont {E.~Y.}\
  \bibnamefont {Andrei}},\ }\href@noop {} {\bibfield  {journal} {\bibinfo
  {journal} {Nature}\ }\textbf {\bibinfo {volume} {573}},\ \bibinfo {pages}
  {91} (\bibinfo {year} {2019})}\BibitemShut {NoStop}%
\bibitem [{\citenamefont {Choi}\ \emph {et~al.}(2019)\citenamefont {Choi},
  \citenamefont {Kemmer}, \citenamefont {Peng}, \citenamefont {Thomson},
  \citenamefont {Arora}, \citenamefont {Polski}, \citenamefont {Zhang},
  \citenamefont {Ren}, \citenamefont {Alicea}, \citenamefont {Refael} \emph
  {et~al.}}]{Choi2019}%
  \BibitemOpen
  \bibfield  {author} {\bibinfo {author} {\bibfnamefont {Y.}~\bibnamefont
  {Choi}}, \bibinfo {author} {\bibfnamefont {J.}~\bibnamefont {Kemmer}},
  \bibinfo {author} {\bibfnamefont {Y.}~\bibnamefont {Peng}}, \bibinfo {author}
  {\bibfnamefont {A.}~\bibnamefont {Thomson}}, \bibinfo {author} {\bibfnamefont
  {H.}~\bibnamefont {Arora}}, \bibinfo {author} {\bibfnamefont
  {R.}~\bibnamefont {Polski}}, \bibinfo {author} {\bibfnamefont
  {Y.}~\bibnamefont {Zhang}}, \bibinfo {author} {\bibfnamefont
  {H.}~\bibnamefont {Ren}}, \bibinfo {author} {\bibfnamefont {J.}~\bibnamefont
  {Alicea}}, \bibinfo {author} {\bibfnamefont {G.}~\bibnamefont {Refael}},
  \emph {et~al.},\ }\href@noop {} {\bibfield  {journal} {\bibinfo  {journal}
  {Nature Physics}\ }\textbf {\bibinfo {volume} {15}},\ \bibinfo {pages} {1174}
  (\bibinfo {year} {2019})}\BibitemShut {NoStop}%
\bibitem [{\citenamefont {Cao}\ \emph {et~al.}(2021)\citenamefont {Cao},
  \citenamefont {Rodan-Legrain}, \citenamefont {Park}, \citenamefont {Yuan},
  \citenamefont {Watanabe}, \citenamefont {Taniguchi}, \citenamefont
  {Fernandes}, \citenamefont {Fu},\ and\ \citenamefont
  {Jarillo-Herrero}}]{Cao2021}%
  \BibitemOpen
  \bibfield  {author} {\bibinfo {author} {\bibfnamefont {Y.}~\bibnamefont
  {Cao}}, \bibinfo {author} {\bibfnamefont {D.}~\bibnamefont {Rodan-Legrain}},
  \bibinfo {author} {\bibfnamefont {J.~M.}\ \bibnamefont {Park}}, \bibinfo
  {author} {\bibfnamefont {N.~F.}\ \bibnamefont {Yuan}}, \bibinfo {author}
  {\bibfnamefont {K.}~\bibnamefont {Watanabe}}, \bibinfo {author}
  {\bibfnamefont {T.}~\bibnamefont {Taniguchi}}, \bibinfo {author}
  {\bibfnamefont {R.~M.}\ \bibnamefont {Fernandes}}, \bibinfo {author}
  {\bibfnamefont {L.}~\bibnamefont {Fu}},\ and\ \bibinfo {author}
  {\bibfnamefont {P.}~\bibnamefont {Jarillo-Herrero}},\ }\href@noop {}
  {\bibfield  {journal} {\bibinfo  {journal} {Science}\ }\textbf {\bibinfo
  {volume} {372}},\ \bibinfo {pages} {264} (\bibinfo {year}
  {2021})}\BibitemShut {NoStop}%
\bibitem [{\citenamefont {Rubio-Verd{\'u}}\ \emph {et~al.}(2022)\citenamefont
  {Rubio-Verd{\'u}}, \citenamefont {Turkel}, \citenamefont {Song},
  \citenamefont {Klebl}, \citenamefont {Samajdar}, \citenamefont {Scheurer},
  \citenamefont {Venderbos}, \citenamefont {Watanabe}, \citenamefont
  {Taniguchi}, \citenamefont {Ochoa} \emph {et~al.}}]{Rubio2022}%
  \BibitemOpen
  \bibfield  {author} {\bibinfo {author} {\bibfnamefont {C.}~\bibnamefont
  {Rubio-Verd{\'u}}}, \bibinfo {author} {\bibfnamefont {S.}~\bibnamefont
  {Turkel}}, \bibinfo {author} {\bibfnamefont {Y.}~\bibnamefont {Song}},
  \bibinfo {author} {\bibfnamefont {L.}~\bibnamefont {Klebl}}, \bibinfo
  {author} {\bibfnamefont {R.}~\bibnamefont {Samajdar}}, \bibinfo {author}
  {\bibfnamefont {M.~S.}\ \bibnamefont {Scheurer}}, \bibinfo {author}
  {\bibfnamefont {J.~W.}\ \bibnamefont {Venderbos}}, \bibinfo {author}
  {\bibfnamefont {K.}~\bibnamefont {Watanabe}}, \bibinfo {author}
  {\bibfnamefont {T.}~\bibnamefont {Taniguchi}}, \bibinfo {author}
  {\bibfnamefont {H.}~\bibnamefont {Ochoa}}, \emph {et~al.},\ }\href@noop {}
  {\bibfield  {journal} {\bibinfo  {journal} {Nature Physics}\ }\textbf
  {\bibinfo {volume} {18}},\ \bibinfo {pages} {196} (\bibinfo {year}
  {2022})}\BibitemShut {NoStop}%
\bibitem [{\citenamefont {Zhang}\ \emph {et~al.}(2022)\citenamefont {Zhang},
  \citenamefont {Wang}, \citenamefont {Watanabe}, \citenamefont {Taniguchi},
  \citenamefont {Vafek},\ and\ \citenamefont {Li}}]{Zhang_Vafek2022}%
  \BibitemOpen
  \bibfield  {author} {\bibinfo {author} {\bibfnamefont {N.~J.}\ \bibnamefont
  {Zhang}}, \bibinfo {author} {\bibfnamefont {Y.}~\bibnamefont {Wang}},
  \bibinfo {author} {\bibfnamefont {K.}~\bibnamefont {Watanabe}}, \bibinfo
  {author} {\bibfnamefont {T.}~\bibnamefont {Taniguchi}}, \bibinfo {author}
  {\bibfnamefont {O.}~\bibnamefont {Vafek}},\ and\ \bibinfo {author}
  {\bibfnamefont {J.}~\bibnamefont {Li}},\ }\href@noop {} {\bibfield  {journal}
  {\bibinfo  {journal} {arXiv:2211.01352}\ } (\bibinfo {year}
  {2022})}\BibitemShut {NoStop}%
\bibitem [{\citenamefont {Jin}\ \emph {et~al.}(2021{\natexlab{b}})\citenamefont
  {Jin}, \citenamefont {Tao}, \citenamefont {Li}, \citenamefont {Xu},
  \citenamefont {Tang}, \citenamefont {Zhu}, \citenamefont {Liu}, \citenamefont
  {Watanabe}, \citenamefont {Taniguchi}, \citenamefont {Hone} \emph
  {et~al.}}]{Jin2021_stripe}%
  \BibitemOpen
  \bibfield  {author} {\bibinfo {author} {\bibfnamefont {C.}~\bibnamefont
  {Jin}}, \bibinfo {author} {\bibfnamefont {Z.}~\bibnamefont {Tao}}, \bibinfo
  {author} {\bibfnamefont {T.}~\bibnamefont {Li}}, \bibinfo {author}
  {\bibfnamefont {Y.}~\bibnamefont {Xu}}, \bibinfo {author} {\bibfnamefont
  {Y.}~\bibnamefont {Tang}}, \bibinfo {author} {\bibfnamefont {J.}~\bibnamefont
  {Zhu}}, \bibinfo {author} {\bibfnamefont {S.}~\bibnamefont {Liu}}, \bibinfo
  {author} {\bibfnamefont {K.}~\bibnamefont {Watanabe}}, \bibinfo {author}
  {\bibfnamefont {T.}~\bibnamefont {Taniguchi}}, \bibinfo {author}
  {\bibfnamefont {J.~C.}\ \bibnamefont {Hone}}, \emph {et~al.},\ }\href@noop {}
  {\bibfield  {journal} {\bibinfo  {journal} {Nature Materials}\ }\textbf
  {\bibinfo {volume} {20}},\ \bibinfo {pages} {940} (\bibinfo {year}
  {2021}{\natexlab{b}})}\BibitemShut {NoStop}%
\bibitem [{\citenamefont {Mulder}\ \emph {et~al.}(2010)\citenamefont {Mulder},
  \citenamefont {Ganesh}, \citenamefont {Capriotti},\ and\ \citenamefont
  {Paramekanti}}]{Mulder2010}%
  \BibitemOpen
  \bibfield  {author} {\bibinfo {author} {\bibfnamefont {A.}~\bibnamefont
  {Mulder}}, \bibinfo {author} {\bibfnamefont {R.}~\bibnamefont {Ganesh}},
  \bibinfo {author} {\bibfnamefont {L.}~\bibnamefont {Capriotti}},\ and\
  \bibinfo {author} {\bibfnamefont {A.}~\bibnamefont {Paramekanti}},\ }\href
  {https://doi.org/10.1103/PhysRevB.81.214419} {\bibfield  {journal} {\bibinfo
  {journal} {Phys. Rev. B}\ }\textbf {\bibinfo {volume} {81}},\ \bibinfo
  {pages} {214419} (\bibinfo {year} {2010})}\BibitemShut {NoStop}%
\bibitem [{\citenamefont {Drouin-Touchette}\ \emph {et~al.}(2022)\citenamefont
  {Drouin-Touchette}, \citenamefont {Orth}, \citenamefont {Coleman},
  \citenamefont {Chandra},\ and\ \citenamefont {Lubensky}}]{Orth2022}%
  \BibitemOpen
  \bibfield  {author} {\bibinfo {author} {\bibfnamefont {V.}~\bibnamefont
  {Drouin-Touchette}}, \bibinfo {author} {\bibfnamefont {P.~P.}\ \bibnamefont
  {Orth}}, \bibinfo {author} {\bibfnamefont {P.}~\bibnamefont {Coleman}},
  \bibinfo {author} {\bibfnamefont {P.}~\bibnamefont {Chandra}},\ and\ \bibinfo
  {author} {\bibfnamefont {T.~C.}\ \bibnamefont {Lubensky}},\ }\href
  {https://doi.org/10.1103/PhysRevX.12.011043} {\bibfield  {journal} {\bibinfo
  {journal} {Phys. Rev. X}\ }\textbf {\bibinfo {volume} {12}},\ \bibinfo
  {pages} {011043} (\bibinfo {year} {2022})}\BibitemShut {NoStop}%
\bibitem [{\citenamefont {Li}\ and\ \citenamefont {Li}(2022)}]{Li2022}%
  \BibitemOpen
  \bibfield  {author} {\bibinfo {author} {\bibfnamefont {H.}~\bibnamefont
  {Li}}\ and\ \bibinfo {author} {\bibfnamefont {T.}~\bibnamefont {Li}},\ }\href
  {https://doi.org/10.1103/PhysRevB.106.035112} {\bibfield  {journal} {\bibinfo
   {journal} {Phys. Rev. B}\ }\textbf {\bibinfo {volume} {106}},\ \bibinfo
  {pages} {035112} (\bibinfo {year} {2022})}\BibitemShut {NoStop}%
\bibitem [{\citenamefont {Nedi{\'c}}\ \emph {et~al.}(2022)\citenamefont
  {Nedi{\'c}}, \citenamefont {Quito}, \citenamefont {Sizyuk},\ and\
  \citenamefont {Orth}}]{Nedic2022}%
  \BibitemOpen
  \bibfield  {author} {\bibinfo {author} {\bibfnamefont {A.-M.}\ \bibnamefont
  {Nedi{\'c}}}, \bibinfo {author} {\bibfnamefont {V.~L.}\ \bibnamefont
  {Quito}}, \bibinfo {author} {\bibfnamefont {Y.}~\bibnamefont {Sizyuk}},\ and\
  \bibinfo {author} {\bibfnamefont {P.~P.}\ \bibnamefont {Orth}},\ }\href@noop
  {} {\bibfield  {journal} {\bibinfo  {journal} {arXiv preprint
  arXiv:2210.04900}\ } (\bibinfo {year} {2022})}\BibitemShut {NoStop}%
\bibitem [{\citenamefont {Strockoz}\ \emph {et~al.}(2022)\citenamefont
  {Strockoz}, \citenamefont {Antonenko}, \citenamefont {LaBelle},\ and\
  \citenamefont {Venderbos}}]{Strockoz2022}%
  \BibitemOpen
  \bibfield  {author} {\bibinfo {author} {\bibfnamefont {J.}~\bibnamefont
  {Strockoz}}, \bibinfo {author} {\bibfnamefont {D.~S.}\ \bibnamefont
  {Antonenko}}, \bibinfo {author} {\bibfnamefont {D.}~\bibnamefont {LaBelle}},\
  and\ \bibinfo {author} {\bibfnamefont {J.~W.}\ \bibnamefont {Venderbos}},\
  }\href@noop {} {\bibfield  {journal} {\bibinfo  {journal} {arXiv:2211.11739}\
  } (\bibinfo {year} {2022})}\BibitemShut {NoStop}%
\bibitem [{\citenamefont {Dodaro}\ \emph {et~al.}(2018)\citenamefont {Dodaro},
  \citenamefont {Kivelson}, \citenamefont {Schattner}, \citenamefont {Sun},\
  and\ \citenamefont {Wang}}]{Dodaro2018}%
  \BibitemOpen
  \bibfield  {author} {\bibinfo {author} {\bibfnamefont {J.~F.}\ \bibnamefont
  {Dodaro}}, \bibinfo {author} {\bibfnamefont {S.~A.}\ \bibnamefont
  {Kivelson}}, \bibinfo {author} {\bibfnamefont {Y.}~\bibnamefont {Schattner}},
  \bibinfo {author} {\bibfnamefont {X.~Q.}\ \bibnamefont {Sun}},\ and\ \bibinfo
  {author} {\bibfnamefont {C.}~\bibnamefont {Wang}},\ }\href
  {https://doi.org/10.1103/PhysRevB.98.075154} {\bibfield  {journal} {\bibinfo
  {journal} {Phys. Rev. B}\ }\textbf {\bibinfo {volume} {98}},\ \bibinfo
  {pages} {075154} (\bibinfo {year} {2018})}\BibitemShut {NoStop}%
\bibitem [{\citenamefont {Venderbos}\ and\ \citenamefont
  {Fernandes}(2018)}]{Venderbos2018}%
  \BibitemOpen
  \bibfield  {author} {\bibinfo {author} {\bibfnamefont {J.~W.~F.}\
  \bibnamefont {Venderbos}}\ and\ \bibinfo {author} {\bibfnamefont {R.~M.}\
  \bibnamefont {Fernandes}},\ }\href
  {https://doi.org/10.1103/PhysRevB.98.245103} {\bibfield  {journal} {\bibinfo
  {journal} {Phys. Rev. B}\ }\textbf {\bibinfo {volume} {98}},\ \bibinfo
  {pages} {245103} (\bibinfo {year} {2018})}\BibitemShut {NoStop}%
\bibitem [{\citenamefont {Kozii}\ \emph {et~al.}(2019)\citenamefont {Kozii},
  \citenamefont {Isobe}, \citenamefont {Venderbos},\ and\ \citenamefont
  {Fu}}]{Kozii2019}%
  \BibitemOpen
  \bibfield  {author} {\bibinfo {author} {\bibfnamefont {V.}~\bibnamefont
  {Kozii}}, \bibinfo {author} {\bibfnamefont {H.}~\bibnamefont {Isobe}},
  \bibinfo {author} {\bibfnamefont {J.~W.~F.}\ \bibnamefont {Venderbos}},\ and\
  \bibinfo {author} {\bibfnamefont {L.}~\bibnamefont {Fu}},\ }\href
  {https://doi.org/10.1103/PhysRevB.99.144507} {\bibfield  {journal} {\bibinfo
  {journal} {Phys. Rev. B}\ }\textbf {\bibinfo {volume} {99}},\ \bibinfo
  {pages} {144507} (\bibinfo {year} {2019})}\BibitemShut {NoStop}%
\bibitem [{\citenamefont {Xu}\ \emph {et~al.}(2020)\citenamefont {Xu},
  \citenamefont {Wu}, \citenamefont {Jian},\ and\ \citenamefont {Xu}}]{Xu2020}%
  \BibitemOpen
  \bibfield  {author} {\bibinfo {author} {\bibfnamefont {Y.}~\bibnamefont
  {Xu}}, \bibinfo {author} {\bibfnamefont {X.-C.}\ \bibnamefont {Wu}}, \bibinfo
  {author} {\bibfnamefont {C.-M.}\ \bibnamefont {Jian}},\ and\ \bibinfo
  {author} {\bibfnamefont {C.}~\bibnamefont {Xu}},\ }\href
  {https://doi.org/10.1103/PhysRevB.101.205426} {\bibfield  {journal} {\bibinfo
   {journal} {Phys. Rev. B}\ }\textbf {\bibinfo {volume} {101}},\ \bibinfo
  {pages} {205426} (\bibinfo {year} {2020})}\BibitemShut {NoStop}%
\bibitem [{\citenamefont {Fernandes}\ and\ \citenamefont
  {Venderbos}(2020)}]{Fernandes_Venderbos}%
  \BibitemOpen
  \bibfield  {author} {\bibinfo {author} {\bibfnamefont {R.~M.}\ \bibnamefont
  {Fernandes}}\ and\ \bibinfo {author} {\bibfnamefont {J.~W.~F.}\ \bibnamefont
  {Venderbos}},\ }\href {https://doi.org/10.1126/sciadv.aba8834} {\bibfield
  {journal} {\bibinfo  {journal} {Science Advances}\ }\textbf {\bibinfo
  {volume} {6}},\ \bibinfo {pages} {eaba8834} (\bibinfo {year}
  {2020})}\BibitemShut {NoStop}%
\bibitem [{\citenamefont {Kang}\ and\ \citenamefont {Vafek}(2020)}]{Kang2020}%
  \BibitemOpen
  \bibfield  {author} {\bibinfo {author} {\bibfnamefont {J.}~\bibnamefont
  {Kang}}\ and\ \bibinfo {author} {\bibfnamefont {O.}~\bibnamefont {Vafek}},\
  }\href {https://doi.org/10.1103/PhysRevB.102.035161} {\bibfield  {journal}
  {\bibinfo  {journal} {Phys. Rev. B}\ }\textbf {\bibinfo {volume} {102}},\
  \bibinfo {pages} {035161} (\bibinfo {year} {2020})}\BibitemShut {NoStop}%
\bibitem [{\citenamefont {Xie}\ \emph {et~al.}(2021)\citenamefont {Xie},
  \citenamefont {Cowsik}, \citenamefont {Song}, \citenamefont {Lian},
  \citenamefont {Bernevig},\ and\ \citenamefont {Regnault}}]{Bernevig_TBGVI}%
  \BibitemOpen
  \bibfield  {author} {\bibinfo {author} {\bibfnamefont {F.}~\bibnamefont
  {Xie}}, \bibinfo {author} {\bibfnamefont {A.}~\bibnamefont {Cowsik}},
  \bibinfo {author} {\bibfnamefont {Z.-D.}\ \bibnamefont {Song}}, \bibinfo
  {author} {\bibfnamefont {B.}~\bibnamefont {Lian}}, \bibinfo {author}
  {\bibfnamefont {B.~A.}\ \bibnamefont {Bernevig}},\ and\ \bibinfo {author}
  {\bibfnamefont {N.}~\bibnamefont {Regnault}},\ }\href
  {https://doi.org/10.1103/PhysRevB.103.205416} {\bibfield  {journal} {\bibinfo
   {journal} {Phys. Rev. B}\ }\textbf {\bibinfo {volume} {103}},\ \bibinfo
  {pages} {205416} (\bibinfo {year} {2021})}\BibitemShut {NoStop}%
\bibitem [{\citenamefont {Chichinadze}\ \emph {et~al.}(2020)\citenamefont
  {Chichinadze}, \citenamefont {Classen},\ and\ \citenamefont
  {Chubukov}}]{Chichinadze2020}%
  \BibitemOpen
  \bibfield  {author} {\bibinfo {author} {\bibfnamefont {D.~V.}\ \bibnamefont
  {Chichinadze}}, \bibinfo {author} {\bibfnamefont {L.}~\bibnamefont
  {Classen}},\ and\ \bibinfo {author} {\bibfnamefont {A.~V.}\ \bibnamefont
  {Chubukov}},\ }\href {https://doi.org/10.1103/PhysRevB.101.224513} {\bibfield
   {journal} {\bibinfo  {journal} {Phys. Rev. B}\ }\textbf {\bibinfo {volume}
  {101}},\ \bibinfo {pages} {224513} (\bibinfo {year} {2020})}\BibitemShut
  {NoStop}%
\bibitem [{\citenamefont {Sboychakov}\ \emph {et~al.}(2020)\citenamefont
  {Sboychakov}, \citenamefont {Rozhkov}, \citenamefont {Rakhmanov},\ and\
  \citenamefont {Nori}}]{Nori2020}%
  \BibitemOpen
  \bibfield  {author} {\bibinfo {author} {\bibfnamefont {A.~O.}\ \bibnamefont
  {Sboychakov}}, \bibinfo {author} {\bibfnamefont {A.~V.}\ \bibnamefont
  {Rozhkov}}, \bibinfo {author} {\bibfnamefont {A.~L.}\ \bibnamefont
  {Rakhmanov}},\ and\ \bibinfo {author} {\bibfnamefont {F.}~\bibnamefont
  {Nori}},\ }\href {https://doi.org/10.1103/PhysRevB.102.155142} {\bibfield
  {journal} {\bibinfo  {journal} {Phys. Rev. B}\ }\textbf {\bibinfo {volume}
  {102}},\ \bibinfo {pages} {155142} (\bibinfo {year} {2020})}\BibitemShut
  {NoStop}%
\bibitem [{\citenamefont {Wang}\ \emph {et~al.}(2021)\citenamefont {Wang},
  \citenamefont {Kang},\ and\ \citenamefont {Fernandes}}]{Wang_Kang2021}%
  \BibitemOpen
  \bibfield  {author} {\bibinfo {author} {\bibfnamefont {Y.}~\bibnamefont
  {Wang}}, \bibinfo {author} {\bibfnamefont {J.}~\bibnamefont {Kang}},\ and\
  \bibinfo {author} {\bibfnamefont {R.~M.}\ \bibnamefont {Fernandes}},\ }\href
  {https://doi.org/10.1103/PhysRevB.103.024506} {\bibfield  {journal} {\bibinfo
   {journal} {Phys. Rev. B}\ }\textbf {\bibinfo {volume} {103}},\ \bibinfo
  {pages} {024506} (\bibinfo {year} {2021})}\BibitemShut {NoStop}%
\bibitem [{\citenamefont {Onari}\ and\ \citenamefont
  {Kontani}(2022)}]{Kontani2022}%
  \BibitemOpen
  \bibfield  {author} {\bibinfo {author} {\bibfnamefont {S.}~\bibnamefont
  {Onari}}\ and\ \bibinfo {author} {\bibfnamefont {H.}~\bibnamefont
  {Kontani}},\ }\href {https://doi.org/10.1103/PhysRevLett.128.066401}
  {\bibfield  {journal} {\bibinfo  {journal} {Phys. Rev. Lett.}\ }\textbf
  {\bibinfo {volume} {128}},\ \bibinfo {pages} {066401} (\bibinfo {year}
  {2022})}\BibitemShut {NoStop}%
\bibitem [{\citenamefont {Brillaux}\ \emph {et~al.}(2022)\citenamefont
  {Brillaux}, \citenamefont {Carpentier}, \citenamefont {Fedorenko},\ and\
  \citenamefont {Savary}}]{Savary2022}%
  \BibitemOpen
  \bibfield  {author} {\bibinfo {author} {\bibfnamefont {E.}~\bibnamefont
  {Brillaux}}, \bibinfo {author} {\bibfnamefont {D.}~\bibnamefont
  {Carpentier}}, \bibinfo {author} {\bibfnamefont {A.~A.}\ \bibnamefont
  {Fedorenko}},\ and\ \bibinfo {author} {\bibfnamefont {L.}~\bibnamefont
  {Savary}},\ }\href {https://doi.org/10.1103/PhysRevResearch.4.033168}
  {\bibfield  {journal} {\bibinfo  {journal} {Phys. Rev. Res.}\ }\textbf
  {\bibinfo {volume} {4}},\ \bibinfo {pages} {033168} (\bibinfo {year}
  {2022})}\BibitemShut {NoStop}%
\bibitem [{\citenamefont {Matty}\ and\ \citenamefont {Kim}(2022)}]{Matty2022}%
  \BibitemOpen
  \bibfield  {author} {\bibinfo {author} {\bibfnamefont {M.}~\bibnamefont
  {Matty}}\ and\ \bibinfo {author} {\bibfnamefont {E.-A.}\ \bibnamefont
  {Kim}},\ }\href@noop {} {\bibfield  {journal} {\bibinfo  {journal} {Nature
  Communications}\ }\textbf {\bibinfo {volume} {13}},\ \bibinfo {pages} {7098}
  (\bibinfo {year} {2022})}\BibitemShut {NoStop}%
\bibitem [{\citenamefont {Grandi}\ \emph {et~al.}(2023)\citenamefont {Grandi},
  \citenamefont {Consiglio}, \citenamefont {Sentef}, \citenamefont {Thomale},\
  and\ \citenamefont {Kennes}}]{Grandi2023}%
  \BibitemOpen
  \bibfield  {author} {\bibinfo {author} {\bibfnamefont {F.}~\bibnamefont
  {Grandi}}, \bibinfo {author} {\bibfnamefont {A.}~\bibnamefont {Consiglio}},
  \bibinfo {author} {\bibfnamefont {M.~A.}\ \bibnamefont {Sentef}}, \bibinfo
  {author} {\bibfnamefont {R.}~\bibnamefont {Thomale}},\ and\ \bibinfo {author}
  {\bibfnamefont {D.~M.}\ \bibnamefont {Kennes}},\ }\href@noop {} {\bibfield
  {journal} {\bibinfo  {journal} {arXiv:2302.01615}\ } (\bibinfo {year}
  {2023})}\BibitemShut {NoStop}%
\bibitem [{\citenamefont {Hecker}\ and\ \citenamefont
  {Schmalian}(2018)}]{Hecker2018}%
  \BibitemOpen
  \bibfield  {author} {\bibinfo {author} {\bibfnamefont {M.}~\bibnamefont
  {Hecker}}\ and\ \bibinfo {author} {\bibfnamefont {J.}~\bibnamefont
  {Schmalian}},\ }\href@noop {} {\bibfield  {journal} {\bibinfo  {journal} {npj
  Quantum Materials}\ }\textbf {\bibinfo {volume} {3}},\ \bibinfo {pages} {26}
  (\bibinfo {year} {2018})}\BibitemShut {NoStop}%
\bibitem [{\citenamefont {Fernandes}\ \emph {et~al.}(2019)\citenamefont
  {Fernandes}, \citenamefont {Orth},\ and\ \citenamefont
  {Schmalian}}]{Fernandes2019}%
  \BibitemOpen
  \bibfield  {author} {\bibinfo {author} {\bibfnamefont {R.~M.}\ \bibnamefont
  {Fernandes}}, \bibinfo {author} {\bibfnamefont {P.~P.}\ \bibnamefont
  {Orth}},\ and\ \bibinfo {author} {\bibfnamefont {J.}~\bibnamefont
  {Schmalian}},\ }\href@noop {} {\bibfield  {journal} {\bibinfo  {journal}
  {Annual Review of Condensed Matter Physics}\ }\textbf {\bibinfo {volume}
  {10}},\ \bibinfo {pages} {133} (\bibinfo {year} {2019})}\BibitemShut
  {NoStop}%
\bibitem [{\citenamefont {How}\ and\ \citenamefont {Yip}(2019)}]{How2019}%
  \BibitemOpen
  \bibfield  {author} {\bibinfo {author} {\bibfnamefont {P.~T.}\ \bibnamefont
  {How}}\ and\ \bibinfo {author} {\bibfnamefont {S.-K.}\ \bibnamefont {Yip}},\
  }\href {https://doi.org/10.1103/PhysRevB.100.134508} {\bibfield  {journal}
  {\bibinfo  {journal} {Phys. Rev. B}\ }\textbf {\bibinfo {volume} {100}},\
  \bibinfo {pages} {134508} (\bibinfo {year} {2019})}\BibitemShut {NoStop}%
\bibitem [{\citenamefont {Kuntsevich}\ \emph {et~al.}(2019)\citenamefont
  {Kuntsevich}, \citenamefont {Bryzgalov}, \citenamefont {Akzyanov},
  \citenamefont {Martovitskii}, \citenamefont {Rakhmanov},\ and\ \citenamefont
  {Selivanov}}]{Kuntsevich2019}%
  \BibitemOpen
  \bibfield  {author} {\bibinfo {author} {\bibfnamefont {A.~Y.}\ \bibnamefont
  {Kuntsevich}}, \bibinfo {author} {\bibfnamefont {M.~A.}\ \bibnamefont
  {Bryzgalov}}, \bibinfo {author} {\bibfnamefont {R.~S.}\ \bibnamefont
  {Akzyanov}}, \bibinfo {author} {\bibfnamefont {V.~P.}\ \bibnamefont
  {Martovitskii}}, \bibinfo {author} {\bibfnamefont {A.~L.}\ \bibnamefont
  {Rakhmanov}},\ and\ \bibinfo {author} {\bibfnamefont {Y.~G.}\ \bibnamefont
  {Selivanov}},\ }\href {https://doi.org/10.1103/PhysRevB.100.224509}
  {\bibfield  {journal} {\bibinfo  {journal} {Phys. Rev. B}\ }\textbf {\bibinfo
  {volume} {100}},\ \bibinfo {pages} {224509} (\bibinfo {year}
  {2019})}\BibitemShut {NoStop}%
\bibitem [{\citenamefont {Kostylev}\ \emph {et~al.}(2020)\citenamefont
  {Kostylev}, \citenamefont {Yonezawa}, \citenamefont {Wang}, \citenamefont
  {Ando},\ and\ \citenamefont {Maeno}}]{Kostylev2020}%
  \BibitemOpen
  \bibfield  {author} {\bibinfo {author} {\bibfnamefont {I.}~\bibnamefont
  {Kostylev}}, \bibinfo {author} {\bibfnamefont {S.}~\bibnamefont {Yonezawa}},
  \bibinfo {author} {\bibfnamefont {Z.}~\bibnamefont {Wang}}, \bibinfo {author}
  {\bibfnamefont {Y.}~\bibnamefont {Ando}},\ and\ \bibinfo {author}
  {\bibfnamefont {Y.}~\bibnamefont {Maeno}},\ }\href@noop {} {\bibfield
  {journal} {\bibinfo  {journal} {Nature Communications}\ }\textbf {\bibinfo
  {volume} {11}},\ \bibinfo {pages} {4152} (\bibinfo {year}
  {2020})}\BibitemShut {NoStop}%
\bibitem [{\citenamefont {Kimura}\ \emph {et~al.}(2022)\citenamefont {Kimura},
  \citenamefont {Sigrist},\ and\ \citenamefont {Kawakami}}]{Kimura2022}%
  \BibitemOpen
  \bibfield  {author} {\bibinfo {author} {\bibfnamefont {K.}~\bibnamefont
  {Kimura}}, \bibinfo {author} {\bibfnamefont {M.}~\bibnamefont {Sigrist}},\
  and\ \bibinfo {author} {\bibfnamefont {N.}~\bibnamefont {Kawakami}},\ }\href
  {https://doi.org/10.1103/PhysRevB.105.035130} {\bibfield  {journal} {\bibinfo
   {journal} {Phys. Rev. B}\ }\textbf {\bibinfo {volume} {105}},\ \bibinfo
  {pages} {035130} (\bibinfo {year} {2022})}\BibitemShut {NoStop}%
\bibitem [{\citenamefont {Hecker}\ and\ \citenamefont
  {Fernandes}(2022)}]{Hecker2022}%
  \BibitemOpen
  \bibfield  {author} {\bibinfo {author} {\bibfnamefont {M.}~\bibnamefont
  {Hecker}}\ and\ \bibinfo {author} {\bibfnamefont {R.~M.}\ \bibnamefont
  {Fernandes}},\ }\href {https://doi.org/10.1103/PhysRevB.105.174504}
  {\bibfield  {journal} {\bibinfo  {journal} {Phys. Rev. B}\ }\textbf {\bibinfo
  {volume} {105}},\ \bibinfo {pages} {174504} (\bibinfo {year}
  {2022})}\BibitemShut {NoStop}%
\bibitem [{\citenamefont {Straley}\ and\ \citenamefont
  {Fisher}(1973)}]{Straley1973ThreestatePM}%
  \BibitemOpen
  \bibfield  {author} {\bibinfo {author} {\bibfnamefont {J.~P.}\ \bibnamefont
  {Straley}}\ and\ \bibinfo {author} {\bibfnamefont {M.~E.}\ \bibnamefont
  {Fisher}},\ }\href@noop {} {\bibfield  {journal} {\bibinfo  {journal}
  {Journal of Physics A: Mathematical, Nuclear and General}\ }\textbf {\bibinfo
  {volume} {6}},\ \bibinfo {pages} {1310} (\bibinfo {year} {1973})}\BibitemShut
  {NoStop}%
\bibitem [{\citenamefont {Blankschtein}\ and\ \citenamefont
  {Aharony}(1980)}]{blankschtein1980effects}%
  \BibitemOpen
  \bibfield  {author} {\bibinfo {author} {\bibfnamefont {D.}~\bibnamefont
  {Blankschtein}}\ and\ \bibinfo {author} {\bibfnamefont {A.}~\bibnamefont
  {Aharony}},\ }\href@noop {} {\bibfield  {journal} {\bibinfo  {journal}
  {Journal of Physics C: Solid State Physics}\ }\textbf {\bibinfo {volume}
  {13}},\ \bibinfo {pages} {4635} (\bibinfo {year} {1980})}\BibitemShut
  {NoStop}%
\bibitem [{\citenamefont {Wu}(1982)}]{review_Potts}%
  \BibitemOpen
  \bibfield  {author} {\bibinfo {author} {\bibfnamefont {F.~Y.}\ \bibnamefont
  {Wu}},\ }\href {https://doi.org/10.1103/RevModPhys.54.235} {\bibfield
  {journal} {\bibinfo  {journal} {Rev. Mod. Phys.}\ }\textbf {\bibinfo {volume}
  {54}},\ \bibinfo {pages} {235} (\bibinfo {year} {1982})}\BibitemShut
  {NoStop}%
\bibitem [{\citenamefont {L\"ohneysen}\ \emph {et~al.}(2007)\citenamefont
  {L\"ohneysen}, \citenamefont {Rosch}, \citenamefont {Vojta},\ and\
  \citenamefont {W\"olfle}}]{Lohneysen2007}%
  \BibitemOpen
  \bibfield  {author} {\bibinfo {author} {\bibfnamefont {H.~v.}\ \bibnamefont
  {L\"ohneysen}}, \bibinfo {author} {\bibfnamefont {A.}~\bibnamefont {Rosch}},
  \bibinfo {author} {\bibfnamefont {M.}~\bibnamefont {Vojta}},\ and\ \bibinfo
  {author} {\bibfnamefont {P.}~\bibnamefont {W\"olfle}},\ }\href
  {https://doi.org/10.1103/RevModPhys.79.1015} {\bibfield  {journal} {\bibinfo
  {journal} {Rev. Mod. Phys.}\ }\textbf {\bibinfo {volume} {79}},\ \bibinfo
  {pages} {1015} (\bibinfo {year} {2007})}\BibitemShut {NoStop}%
\bibitem [{\citenamefont {Oganesyan}\ \emph {et~al.}(2001)\citenamefont
  {Oganesyan}, \citenamefont {Kivelson},\ and\ \citenamefont
  {Fradkin}}]{Oganesyan2001}%
  \BibitemOpen
  \bibfield  {author} {\bibinfo {author} {\bibfnamefont {V.}~\bibnamefont
  {Oganesyan}}, \bibinfo {author} {\bibfnamefont {S.~A.}\ \bibnamefont
  {Kivelson}},\ and\ \bibinfo {author} {\bibfnamefont {E.}~\bibnamefont
  {Fradkin}},\ }\href {https://doi.org/10.1103/PhysRevB.64.195109} {\bibfield
  {journal} {\bibinfo  {journal} {Phys. Rev. B}\ }\textbf {\bibinfo {volume}
  {64}},\ \bibinfo {pages} {195109} (\bibinfo {year} {2001})}\BibitemShut
  {NoStop}%
\bibitem [{\citenamefont {Metzner}\ \emph {et~al.}(2003)\citenamefont
  {Metzner}, \citenamefont {Rohe},\ and\ \citenamefont
  {Andergassen}}]{Metzner2003}%
  \BibitemOpen
  \bibfield  {author} {\bibinfo {author} {\bibfnamefont {W.}~\bibnamefont
  {Metzner}}, \bibinfo {author} {\bibfnamefont {D.}~\bibnamefont {Rohe}},\ and\
  \bibinfo {author} {\bibfnamefont {S.}~\bibnamefont {Andergassen}},\ }\href
  {https://doi.org/10.1103/PhysRevLett.91.066402} {\bibfield  {journal}
  {\bibinfo  {journal} {Phys. Rev. Lett.}\ }\textbf {\bibinfo {volume} {91}},\
  \bibinfo {pages} {066402} (\bibinfo {year} {2003})}\BibitemShut {NoStop}%
\bibitem [{\citenamefont {Garst}\ and\ \citenamefont
  {Chubukov}(2010)}]{Garst2010}%
  \BibitemOpen
  \bibfield  {author} {\bibinfo {author} {\bibfnamefont {M.}~\bibnamefont
  {Garst}}\ and\ \bibinfo {author} {\bibfnamefont {A.~V.}\ \bibnamefont
  {Chubukov}},\ }\href {https://doi.org/10.1103/PhysRevB.81.235105} {\bibfield
  {journal} {\bibinfo  {journal} {Phys. Rev. B}\ }\textbf {\bibinfo {volume}
  {81}},\ \bibinfo {pages} {235105} (\bibinfo {year} {2010})}\BibitemShut
  {NoStop}%
\bibitem [{\citenamefont {Metlitski}\ and\ \citenamefont
  {Sachdev}(2010)}]{Metlitski2010}%
  \BibitemOpen
  \bibfield  {author} {\bibinfo {author} {\bibfnamefont {M.~A.}\ \bibnamefont
  {Metlitski}}\ and\ \bibinfo {author} {\bibfnamefont {S.}~\bibnamefont
  {Sachdev}},\ }\href {https://doi.org/10.1103/PhysRevB.82.075127} {\bibfield
  {journal} {\bibinfo  {journal} {Phys. Rev. B}\ }\textbf {\bibinfo {volume}
  {82}},\ \bibinfo {pages} {075127} (\bibinfo {year} {2010})}\BibitemShut
  {NoStop}%
\bibitem [{\citenamefont {Schattner}\ \emph {et~al.}(2016)\citenamefont
  {Schattner}, \citenamefont {Lederer}, \citenamefont {Kivelson},\ and\
  \citenamefont {Berg}}]{Schattner2016}%
  \BibitemOpen
  \bibfield  {author} {\bibinfo {author} {\bibfnamefont {Y.}~\bibnamefont
  {Schattner}}, \bibinfo {author} {\bibfnamefont {S.}~\bibnamefont {Lederer}},
  \bibinfo {author} {\bibfnamefont {S.~A.}\ \bibnamefont {Kivelson}},\ and\
  \bibinfo {author} {\bibfnamefont {E.}~\bibnamefont {Berg}},\ }\href
  {https://doi.org/10.1103/PhysRevX.6.031028} {\bibfield  {journal} {\bibinfo
  {journal} {Phys. Rev. X}\ }\textbf {\bibinfo {volume} {6}},\ \bibinfo {pages}
  {031028} (\bibinfo {year} {2016})}\BibitemShut {NoStop}%
\bibitem [{\citenamefont {Lederer}\ \emph {et~al.}(2017)\citenamefont
  {Lederer}, \citenamefont {Schattner}, \citenamefont {Berg},\ and\
  \citenamefont {Kivelson}}]{Lederer2017}%
  \BibitemOpen
  \bibfield  {author} {\bibinfo {author} {\bibfnamefont {S.}~\bibnamefont
  {Lederer}}, \bibinfo {author} {\bibfnamefont {Y.}~\bibnamefont {Schattner}},
  \bibinfo {author} {\bibfnamefont {E.}~\bibnamefont {Berg}},\ and\ \bibinfo
  {author} {\bibfnamefont {S.~A.}\ \bibnamefont {Kivelson}},\ }\href@noop {}
  {\bibfield  {journal} {\bibinfo  {journal} {Proceedings of the National
  Academy of Sciences}\ }\textbf {\bibinfo {volume} {114}},\ \bibinfo {pages}
  {4905} (\bibinfo {year} {2017})}\BibitemShut {NoStop}%
\bibitem [{\citenamefont {Klein}\ and\ \citenamefont
  {Chubukov}(2018)}]{Klein_Chubukov2018}%
  \BibitemOpen
  \bibfield  {author} {\bibinfo {author} {\bibfnamefont {A.}~\bibnamefont
  {Klein}}\ and\ \bibinfo {author} {\bibfnamefont {A.}~\bibnamefont
  {Chubukov}},\ }\href {https://doi.org/10.1103/PhysRevB.98.220501} {\bibfield
  {journal} {\bibinfo  {journal} {Phys. Rev. B}\ }\textbf {\bibinfo {volume}
  {98}},\ \bibinfo {pages} {220501} (\bibinfo {year} {2018})}\BibitemShut
  {NoStop}%
\bibitem [{\citenamefont {Stoner}\ and\ \citenamefont
  {Wohlfarth}(1948)}]{Stoner1948}%
  \BibitemOpen
  \bibfield  {author} {\bibinfo {author} {\bibfnamefont {E.~C.}\ \bibnamefont
  {Stoner}}\ and\ \bibinfo {author} {\bibfnamefont {E.}~\bibnamefont
  {Wohlfarth}},\ }\href@noop {} {\bibfield  {journal} {\bibinfo  {journal}
  {Philosophical Transactions of the Royal Society of London. Series A,
  Mathematical and Physical Sciences}\ }\textbf {\bibinfo {volume} {240}},\
  \bibinfo {pages} {599} (\bibinfo {year} {1948})}\BibitemShut {NoStop}%
\bibitem [{\citenamefont {Belitz}\ \emph {et~al.}(2005)\citenamefont {Belitz},
  \citenamefont {Kirkpatrick},\ and\ \citenamefont
  {Rollb\"uhler}}]{Belitz2005}%
  \BibitemOpen
  \bibfield  {author} {\bibinfo {author} {\bibfnamefont {D.}~\bibnamefont
  {Belitz}}, \bibinfo {author} {\bibfnamefont {T.~R.}\ \bibnamefont
  {Kirkpatrick}},\ and\ \bibinfo {author} {\bibfnamefont {J.}~\bibnamefont
  {Rollb\"uhler}},\ }\href {https://doi.org/10.1103/PhysRevLett.94.247205}
  {\bibfield  {journal} {\bibinfo  {journal} {Phys. Rev. Lett.}\ }\textbf
  {\bibinfo {volume} {94}},\ \bibinfo {pages} {247205} (\bibinfo {year}
  {2005})}\BibitemShut {NoStop}%
\bibitem [{\citenamefont {Brando}\ \emph {et~al.}(2016)\citenamefont {Brando},
  \citenamefont {Belitz}, \citenamefont {Grosche},\ and\ \citenamefont
  {Kirkpatrick}}]{Brando2016}%
  \BibitemOpen
  \bibfield  {author} {\bibinfo {author} {\bibfnamefont {M.}~\bibnamefont
  {Brando}}, \bibinfo {author} {\bibfnamefont {D.}~\bibnamefont {Belitz}},
  \bibinfo {author} {\bibfnamefont {F.~M.}\ \bibnamefont {Grosche}},\ and\
  \bibinfo {author} {\bibfnamefont {T.~R.}\ \bibnamefont {Kirkpatrick}},\
  }\href {https://doi.org/10.1103/RevModPhys.88.025006} {\bibfield  {journal}
  {\bibinfo  {journal} {Rev. Mod. Phys.}\ }\textbf {\bibinfo {volume} {88}},\
  \bibinfo {pages} {025006} (\bibinfo {year} {2016})}\BibitemShut {NoStop}%
\bibitem [{\citenamefont {Belitz}\ \emph {et~al.}(1999)\citenamefont {Belitz},
  \citenamefont {Kirkpatrick},\ and\ \citenamefont {Vojta}}]{Belitz1999}%
  \BibitemOpen
  \bibfield  {author} {\bibinfo {author} {\bibfnamefont {D.}~\bibnamefont
  {Belitz}}, \bibinfo {author} {\bibfnamefont {T.~R.}\ \bibnamefont
  {Kirkpatrick}},\ and\ \bibinfo {author} {\bibfnamefont {T.}~\bibnamefont
  {Vojta}},\ }\href {https://doi.org/10.1103/PhysRevLett.82.4707} {\bibfield
  {journal} {\bibinfo  {journal} {Phys. Rev. Lett.}\ }\textbf {\bibinfo
  {volume} {82}},\ \bibinfo {pages} {4707} (\bibinfo {year}
  {1999})}\BibitemShut {NoStop}%
\bibitem [{\citenamefont {Chubukov}\ \emph {et~al.}(2004)\citenamefont
  {Chubukov}, \citenamefont {P\'epin},\ and\ \citenamefont
  {Rech}}]{Chubukov2004}%
  \BibitemOpen
  \bibfield  {author} {\bibinfo {author} {\bibfnamefont {A.~V.}\ \bibnamefont
  {Chubukov}}, \bibinfo {author} {\bibfnamefont {C.}~\bibnamefont {P\'epin}},\
  and\ \bibinfo {author} {\bibfnamefont {J.}~\bibnamefont {Rech}},\ }\href
  {https://doi.org/10.1103/PhysRevLett.92.147003} {\bibfield  {journal}
  {\bibinfo  {journal} {Phys. Rev. Lett.}\ }\textbf {\bibinfo {volume} {92}},\
  \bibinfo {pages} {147003} (\bibinfo {year} {2004})}\BibitemShut {NoStop}%
\bibitem [{\citenamefont {Maslov}\ and\ \citenamefont
  {Chubukov}(2009)}]{Maslov2009}%
  \BibitemOpen
  \bibfield  {author} {\bibinfo {author} {\bibfnamefont {D.~L.}\ \bibnamefont
  {Maslov}}\ and\ \bibinfo {author} {\bibfnamefont {A.~V.}\ \bibnamefont
  {Chubukov}},\ }\href {https://doi.org/10.1103/PhysRevB.79.075112} {\bibfield
  {journal} {\bibinfo  {journal} {Phys. Rev. B}\ }\textbf {\bibinfo {volume}
  {79}},\ \bibinfo {pages} {075112} (\bibinfo {year} {2009})}\BibitemShut
  {NoStop}%
\bibitem [{\citenamefont {Terletska}\ \emph {et~al.}(2011)\citenamefont
  {Terletska}, \citenamefont {Vu\ifmmode \check{c}\else \v{c}\fi{}i\ifmmode
  \check{c}\else \v{c}\fi{}evi\ifmmode~\acute{c}\else \'{c}\fi{}},
  \citenamefont {Tanaskovi\ifmmode~\acute{c}\else \'{c}\fi{}},\ and\
  \citenamefont {Dobrosavljevi\ifmmode~\acute{c}\else
  \'{c}\fi{}}}]{Terletska2011}%
  \BibitemOpen
  \bibfield  {author} {\bibinfo {author} {\bibfnamefont {H.}~\bibnamefont
  {Terletska}}, \bibinfo {author} {\bibfnamefont {J.}~\bibnamefont {Vu\ifmmode
  \check{c}\else \v{c}\fi{}i\ifmmode \check{c}\else
  \v{c}\fi{}evi\ifmmode~\acute{c}\else \'{c}\fi{}}}, \bibinfo {author}
  {\bibfnamefont {D.}~\bibnamefont {Tanaskovi\ifmmode~\acute{c}\else
  \'{c}\fi{}}},\ and\ \bibinfo {author} {\bibfnamefont {V.}~\bibnamefont
  {Dobrosavljevi\ifmmode~\acute{c}\else \'{c}\fi{}}},\ }\href
  {https://doi.org/10.1103/PhysRevLett.107.026401} {\bibfield  {journal}
  {\bibinfo  {journal} {Phys. Rev. Lett.}\ }\textbf {\bibinfo {volume} {107}},\
  \bibinfo {pages} {026401} (\bibinfo {year} {2011})}\BibitemShut {NoStop}%
\bibitem [{\citenamefont {Furukawa}\ \emph {et~al.}(2015)\citenamefont
  {Furukawa}, \citenamefont {Miyagawa}, \citenamefont {Taniguchi},
  \citenamefont {Kato},\ and\ \citenamefont {Kanoda}}]{Furukawa2015}%
  \BibitemOpen
  \bibfield  {author} {\bibinfo {author} {\bibfnamefont {T.}~\bibnamefont
  {Furukawa}}, \bibinfo {author} {\bibfnamefont {K.}~\bibnamefont {Miyagawa}},
  \bibinfo {author} {\bibfnamefont {H.}~\bibnamefont {Taniguchi}}, \bibinfo
  {author} {\bibfnamefont {R.}~\bibnamefont {Kato}},\ and\ \bibinfo {author}
  {\bibfnamefont {K.}~\bibnamefont {Kanoda}},\ }\href@noop {} {\bibfield
  {journal} {\bibinfo  {journal} {Nature Physics}\ }\textbf {\bibinfo {volume}
  {11}},\ \bibinfo {pages} {221} (\bibinfo {year} {2015})}\BibitemShut
  {NoStop}%
\bibitem [{\citenamefont {Wang}\ \emph
  {et~al.}(2022{\natexlab{b}})\citenamefont {Wang}, \citenamefont {Gautreau},
  \citenamefont {Birol},\ and\ \citenamefont {Fernandes}}]{Wang_Fernandes2022}%
  \BibitemOpen
  \bibfield  {author} {\bibinfo {author} {\bibfnamefont {Z.}~\bibnamefont
  {Wang}}, \bibinfo {author} {\bibfnamefont {D.}~\bibnamefont {Gautreau}},
  \bibinfo {author} {\bibfnamefont {T.}~\bibnamefont {Birol}},\ and\ \bibinfo
  {author} {\bibfnamefont {R.~M.}\ \bibnamefont {Fernandes}},\ }\href
  {https://doi.org/10.1103/PhysRevB.105.144404} {\bibfield  {journal} {\bibinfo
   {journal} {Phys. Rev. B}\ }\textbf {\bibinfo {volume} {105}},\ \bibinfo
  {pages} {144404} (\bibinfo {year} {2022}{\natexlab{b}})}\BibitemShut
  {NoStop}%
\bibitem [{\citenamefont {Hertz}(1976)}]{Hertz1976}%
  \BibitemOpen
  \bibfield  {author} {\bibinfo {author} {\bibfnamefont {J.~A.}\ \bibnamefont
  {Hertz}},\ }\href {https://doi.org/10.1103/PhysRevB.14.1165} {\bibfield
  {journal} {\bibinfo  {journal} {Phys. Rev. B}\ }\textbf {\bibinfo {volume}
  {14}},\ \bibinfo {pages} {1165} (\bibinfo {year} {1976})}\BibitemShut
  {NoStop}%
\bibitem [{\citenamefont {Millis}\ \emph {et~al.}(2002)\citenamefont {Millis},
  \citenamefont {Schofield}, \citenamefont {Lonzarich},\ and\ \citenamefont
  {Grigera}}]{Millis2002}%
  \BibitemOpen
  \bibfield  {author} {\bibinfo {author} {\bibfnamefont {A.~J.}\ \bibnamefont
  {Millis}}, \bibinfo {author} {\bibfnamefont {A.~J.}\ \bibnamefont
  {Schofield}}, \bibinfo {author} {\bibfnamefont {G.~G.}\ \bibnamefont
  {Lonzarich}},\ and\ \bibinfo {author} {\bibfnamefont {S.~A.}\ \bibnamefont
  {Grigera}},\ }\href {https://doi.org/10.1103/PhysRevLett.88.217204}
  {\bibfield  {journal} {\bibinfo  {journal} {Phys. Rev. Lett.}\ }\textbf
  {\bibinfo {volume} {88}},\ \bibinfo {pages} {217204} (\bibinfo {year}
  {2002})}\BibitemShut {NoStop}%
\bibitem [{\citenamefont {Samajdar}\ \emph {et~al.}(2021)\citenamefont
  {Samajdar}, \citenamefont {Scheurer}, \citenamefont {Turkel}, \citenamefont
  {Rubio-Verd{\'u}}, \citenamefont {Pasupathy}, \citenamefont {Venderbos},\
  and\ \citenamefont {Fernandes}}]{Samajdar2021}%
  \BibitemOpen
  \bibfield  {author} {\bibinfo {author} {\bibfnamefont {R.}~\bibnamefont
  {Samajdar}}, \bibinfo {author} {\bibfnamefont {M.~S.}\ \bibnamefont
  {Scheurer}}, \bibinfo {author} {\bibfnamefont {S.}~\bibnamefont {Turkel}},
  \bibinfo {author} {\bibfnamefont {C.}~\bibnamefont {Rubio-Verd{\'u}}},
  \bibinfo {author} {\bibfnamefont {A.~N.}\ \bibnamefont {Pasupathy}}, \bibinfo
  {author} {\bibfnamefont {J.~W.}\ \bibnamefont {Venderbos}},\ and\ \bibinfo
  {author} {\bibfnamefont {R.~M.}\ \bibnamefont {Fernandes}},\ }\href@noop {}
  {\bibfield  {journal} {\bibinfo  {journal} {2D Materials}\ }\textbf {\bibinfo
  {volume} {8}},\ \bibinfo {pages} {034005} (\bibinfo {year}
  {2021})}\BibitemShut {NoStop}%
\bibitem [{\citenamefont {Parker}\ \emph {et~al.}(2021)\citenamefont {Parker},
  \citenamefont {Soejima}, \citenamefont {Hauschild}, \citenamefont {Zaletel},\
  and\ \citenamefont {Bultinck}}]{Parker2021}%
  \BibitemOpen
  \bibfield  {author} {\bibinfo {author} {\bibfnamefont {D.~E.}\ \bibnamefont
  {Parker}}, \bibinfo {author} {\bibfnamefont {T.}~\bibnamefont {Soejima}},
  \bibinfo {author} {\bibfnamefont {J.}~\bibnamefont {Hauschild}}, \bibinfo
  {author} {\bibfnamefont {M.~P.}\ \bibnamefont {Zaletel}},\ and\ \bibinfo
  {author} {\bibfnamefont {N.}~\bibnamefont {Bultinck}},\ }\href
  {https://doi.org/10.1103/PhysRevLett.127.027601} {\bibfield  {journal}
  {\bibinfo  {journal} {Phys. Rev. Lett.}\ }\textbf {\bibinfo {volume} {127}},\
  \bibinfo {pages} {027601} (\bibinfo {year} {2021})}\BibitemShut {NoStop}%
\bibitem [{\citenamefont {Maharaj}\ \emph {et~al.}(2017)\citenamefont
  {Maharaj}, \citenamefont {Rosenberg}, \citenamefont {Hristov}, \citenamefont
  {Berg}, \citenamefont {Fernandes}, \citenamefont {Fisher},\ and\
  \citenamefont {Kivelson}}]{Maharaj2017}%
  \BibitemOpen
  \bibfield  {author} {\bibinfo {author} {\bibfnamefont {A.~V.}\ \bibnamefont
  {Maharaj}}, \bibinfo {author} {\bibfnamefont {E.~W.}\ \bibnamefont
  {Rosenberg}}, \bibinfo {author} {\bibfnamefont {A.~T.}\ \bibnamefont
  {Hristov}}, \bibinfo {author} {\bibfnamefont {E.}~\bibnamefont {Berg}},
  \bibinfo {author} {\bibfnamefont {R.~M.}\ \bibnamefont {Fernandes}}, \bibinfo
  {author} {\bibfnamefont {I.~R.}\ \bibnamefont {Fisher}},\ and\ \bibinfo
  {author} {\bibfnamefont {S.~A.}\ \bibnamefont {Kivelson}},\ }\href@noop {}
  {\bibfield  {journal} {\bibinfo  {journal} {Proceedings of the National
  Academy of Sciences}\ }\textbf {\bibinfo {volume} {114}},\ \bibinfo {pages}
  {13430} (\bibinfo {year} {2017})}\BibitemShut {NoStop}%
\bibitem [{\citenamefont {Metlitski}\ \emph {et~al.}(2015)\citenamefont
  {Metlitski}, \citenamefont {Mross}, \citenamefont {Sachdev},\ and\
  \citenamefont {Senthil}}]{Metlitski2015}%
  \BibitemOpen
  \bibfield  {author} {\bibinfo {author} {\bibfnamefont {M.~A.}\ \bibnamefont
  {Metlitski}}, \bibinfo {author} {\bibfnamefont {D.~F.}\ \bibnamefont
  {Mross}}, \bibinfo {author} {\bibfnamefont {S.}~\bibnamefont {Sachdev}},\
  and\ \bibinfo {author} {\bibfnamefont {T.}~\bibnamefont {Senthil}},\ }\href
  {https://doi.org/10.1103/PhysRevB.91.115111} {\bibfield  {journal} {\bibinfo
  {journal} {Phys. Rev. B}\ }\textbf {\bibinfo {volume} {91}},\ \bibinfo
  {pages} {115111} (\bibinfo {year} {2015})}\BibitemShut {NoStop}%
\bibitem [{\citenamefont {Klein}\ \emph {et~al.}(2018)\citenamefont {Klein},
  \citenamefont {Lederer}, \citenamefont {Chowdhury}, \citenamefont {Berg},\
  and\ \citenamefont {Chubukov}}]{Klein2018}%
  \BibitemOpen
  \bibfield  {author} {\bibinfo {author} {\bibfnamefont {A.}~\bibnamefont
  {Klein}}, \bibinfo {author} {\bibfnamefont {S.}~\bibnamefont {Lederer}},
  \bibinfo {author} {\bibfnamefont {D.}~\bibnamefont {Chowdhury}}, \bibinfo
  {author} {\bibfnamefont {E.}~\bibnamefont {Berg}},\ and\ \bibinfo {author}
  {\bibfnamefont {A.}~\bibnamefont {Chubukov}},\ }\href
  {https://doi.org/10.1103/PhysRevB.97.155115} {\bibfield  {journal} {\bibinfo
  {journal} {Phys. Rev. B}\ }\textbf {\bibinfo {volume} {97}},\ \bibinfo
  {pages} {155115} (\bibinfo {year} {2018})}\BibitemShut {NoStop}%
\bibitem [{\citenamefont {Lee}(2018)}]{SSLee2018}%
  \BibitemOpen
  \bibfield  {author} {\bibinfo {author} {\bibfnamefont {S.-S.}\ \bibnamefont
  {Lee}},\ }\href@noop {} {\bibfield  {journal} {\bibinfo  {journal} {Annual
  Review of Condensed Matter Physics}\ }\textbf {\bibinfo {volume} {9}},\
  \bibinfo {pages} {227} (\bibinfo {year} {2018})}\BibitemShut {NoStop}%
\bibitem [{\citenamefont {Vafek}(2022)}]{Vafek2022}%
  \BibitemOpen
  \bibfield  {author} {\bibinfo {author} {\bibfnamefont {O.}~\bibnamefont
  {Vafek}},\ }\href@noop {} {\bibfield  {journal} {\bibinfo  {journal}
  {arXiv:2209.08208}\ } (\bibinfo {year} {2022})}\BibitemShut {NoStop}%
\bibitem [{\citenamefont {Zheng}\ \emph {et~al.}(2020)\citenamefont {Zheng},
  \citenamefont {Ma}, \citenamefont {Bi}, \citenamefont {de~La~Barrera},
  \citenamefont {Liu}, \citenamefont {Mao}, \citenamefont {Zhang},
  \citenamefont {Kiper}, \citenamefont {Watanabe}, \citenamefont {Taniguchi}
  \emph {et~al.}}]{Zheng2020}%
  \BibitemOpen
  \bibfield  {author} {\bibinfo {author} {\bibfnamefont {Z.}~\bibnamefont
  {Zheng}}, \bibinfo {author} {\bibfnamefont {Q.}~\bibnamefont {Ma}}, \bibinfo
  {author} {\bibfnamefont {Z.}~\bibnamefont {Bi}}, \bibinfo {author}
  {\bibfnamefont {S.}~\bibnamefont {de~La~Barrera}}, \bibinfo {author}
  {\bibfnamefont {M.-H.}\ \bibnamefont {Liu}}, \bibinfo {author} {\bibfnamefont
  {N.}~\bibnamefont {Mao}}, \bibinfo {author} {\bibfnamefont {Y.}~\bibnamefont
  {Zhang}}, \bibinfo {author} {\bibfnamefont {N.}~\bibnamefont {Kiper}},
  \bibinfo {author} {\bibfnamefont {K.}~\bibnamefont {Watanabe}}, \bibinfo
  {author} {\bibfnamefont {T.}~\bibnamefont {Taniguchi}}, \emph {et~al.},\
  }\href@noop {} {\bibfield  {journal} {\bibinfo  {journal} {Nature}\ }\textbf
  {\bibinfo {volume} {588}},\ \bibinfo {pages} {71} (\bibinfo {year}
  {2020})}\BibitemShut {NoStop}%
\end{thebibliography}%

\end{document}